\documentclass[12pt,oneside]{book}
\usepackage[a4paper, top=3cm, bottom=3cm]{geometry}
\usepackage{fancyhdr}
\makeatletter
\def\cleardoublepage{\clearpage\if@twoside \ifodd\c@page\else%
    \hbox{}%
    \thispagestyle{empty}
    \newpage%
    \if@twocolumn\hbox{}\newpage\fi\fi\fi}
\makeatother

\usepackage{fancyhdr}
\usepackage{tocloft}
\pdfoutput =1
\usepackage{graphicx}
\DeclareGraphicsExtensions{.pdf}
\usepackage{float} 
\usepackage{amsmath}
\usepackage{amsfonts}   
\usepackage{amssymb}
\usepackage{subfloat}

\begin{document}
\date{}

\begin{titlepage}
\begin{center}

 {\bf \uppercase{\huge P\Large HASE \huge T\Large RANSITION \vspace{0.3cm}\\ IN \\
\vspace{0.3cm}\huge B\Large LACK \huge H\Large OLES\\
}}
\vfill

\normalsize
{\Large Thesis Submitted For The Degree Of}\\[2.2ex]
\textbf{\Large Doctor of Philosophy (Science)}\\[2ex]
In\\[2ex] 
{\Large \textbf{Physics (Theoretical)} }\\[2ex]
By\\[2ex]
\textbf{\Large Dibakar  Roychowdhury}

\vfill

\vfill

\textbf{{\Large \textbf{\large University of Calcutta, India} }}\\[2ex]
{\large 2013}

\end{center}
\end{titlepage}

\newpage
\thispagestyle{empty}
\vspace*{6.5 cm}
{\centering
\hspace*{8.6 cm} $\mathcal{TO}$\\ \hspace*{10 cm}$\mathcal{MY \ PARENTS}$}


\newpage

\chapter*{Acknowledgments}

This thesis is the outcome of the cumulative efforts that has been paid for the last three and half years during my stay at S. N. Bose National Centre for Basic Sciences, Kolkata. I feel myself to be quite lucky in the sense that from the very first day of this journey I found many people around me who constantly assisted me in various occasions. I would like to convey my sincere thanks to all of them.

First of all I would like to convey my gratitude to Prof. Rabin Banerjee for giving me this nice opportunity to work under his supervision. The entire research work that I did for the last three and half years is based on the platform provided by Prof. Banerjee. 

I am specially thankful to Dr. Amitabha Mukhopadhyay of North Bengal University and Prof. Subir Ghosh of ISI, Kolkata for their generous help and guidance during the early stages of my research career. 

I am also thankful to my collaborators Dr. Bibhas Ranjan Majhi, Dr. Sujoy Kumar Modak, Dr. Kuldeep Kumar, Dr. Sunandan Gangopadhyay and Mr. Sudipta Das, who have actively collaborated with me in various occasions and helped me to accomplish various projects in different courses of time.

I am specially thankful to Mr. Srijit Bhattacharjee, with whom I have discussed lots of physics in various occasions.

I would also like to convey my gratitude to all of my friends with whom I have shared many colorful moments during these years. In this regard I would like to convey my sincere thanks to Mr. Ambika Prasad Jena, Mr. Soumyadipta Pal, Mr. Debmalya Mukhopadhyay and Dr. Sudhakar Upadhyay for their helpful assistance whenever it was required for me.

I am particularly thankful to my group mates Mr. Arindam Lala, Mr. Shirsendu Dey, Mr. Arpan Krishna Mitra and Ms. Poulami Chakraborty and for their active participation in various academic discussions.

I am specially thankful to Mr. Biswajit Paul with whom not only I have discussed physics in various occasions but also shared most of my memorable times in the S. N. Bose Centre.

I am also grateful to Mr. Debraj Roy for his generous help and assistance whenever it was required for me. 

Finally, I would like to convey my sincere thanks to Ms. Debashree Chowdhury who has been always there for me to provide a constant mental support.


\chapter*{List of publications}
\thispagestyle{empty}


\begin{enumerate} 
\item \textit{``AdS/CFT superconductors with Power Maxwell electrodynamics: reminiscent of the Meissner effect''}\\
\textbf{Dibakar Roychowdhury}\\
Published in \textbf{Phys. Lett. B 718 (2013)1089-1094 }\\
e-Print: arXiv:1211.1612 [hep-th].\\

\item \textit{``Effect of external magnetic field on holographic superconductors in presence of nonlinear corrections''}\\
\textbf{Dibakar Roychowdhury}\\
Published in \textbf{Phys.Rev. D 86, 106009 (2012) }\\
e-Print: arXiv:1211.0904 [hep-th].\\

\item \textit{``Analytic study of properties of holographic p-wave superconductors''}\\
Sunandan Gangopadhyay, \textbf{Dibakar Roychowdhury}\\
Published in \textbf{JHEP 1208 (2012) 104 }\\
e-Print: arXiv:1207.5605 [hep-th].\\

\item \textit{``Phase transition and scaling behavior of topological charged black holes in Horava-Lifshitz gravity''}\\
Bibhas Ranjan Majhi, \textbf{Dibakar Roychowdhury}\\
Published in \textbf{Class. Quant. Grav 29 (2012) 245012  }\\
e-Print: arXiv:1205.0146 [gr-qc].\\

\item \textit{``Analytic study of Gauss-Bonnet holographic superconductors in Born-Infeld electrodynamics''}\\ 
Sunandan Gangopadhyay, \textbf{Dibakar Roychowdhury}\\ 
Published in \textbf{JHEP 1205 (2012) 156 }\\
e-Print: arXiv:1204.0673 [hep-th].\\

\item \textit{``Critical behavior of Born Infeld AdS black holes in higher dimensions''}\\  
Rabin Banerjee, \textbf{Dibakar Roychowdhury}\\  
Published in \textbf{Phys.Rev. D85 (2012) 104043 }\\
e-Print: arXiv:1203.0118 [gr-qc].\\

\item \textit{``Analytic study of properties of holographic superconductors in Born-Infeld electrodynamics''}\\  
Sunandan Gangopadhyay, \textbf{Dibakar Roychowdhury}\\  
Published in \textbf{JHEP 1205 (2012) 002 }\\
e-Print: arXiv:1201.6520 [hep-th].\\

\item \textit{``Ehrenfest's scheme and thermodynamic geometry in Born-Infeld AdS black holes''}\\ 
Arindam Lala, \textbf{Dibakar Roychowdhury}\\  
Published in \textbf{Phys.Rev. D86 (2012) 084027 }\\
e-Print: arXiv:1111.5991 [gr-qc].\\

\item \textit{``Critical phenomena in Born-Infeld AdS black holes''}\\  
Rabin Banerjee, \textbf{Dibakar Roychowdhury}\\  
Published in \textbf{Phys.Rev. D85 (2012) 044040}\\ 
e-Print: arXiv:1111.0147 [gr-qc].\\

\item \textit{``Thermodynamics of phase transition in higher dimensional AdS black holes''}\\  
Rabin Banerjee, \textbf{Dibakar Roychowdhury}\\  
Published in \textbf{JHEP 1111 (2011) 004}\\
e-Print: arXiv:1109.2433 [gr-qc].\\

\item \textit{``A unified picture of phase transition: from liquid-vapour systems to AdS black holes''}\\ 
Rabin Banerjee, Sujoy Kumar Modak,  \textbf{Dibakar Roychowdhury}\\  
Published in \textbf{JHEP 1210 (2012) 125}\\
e-Print: arXiv:1106.3877 [gr-qc].\\

\item \textit{``Symmetries of Snyder-de Sitter space and relativistic particle dynamics''}\\ 
Rabin Banerjee, Kuldeep Kumar, \textbf{Dibakar Roychowdhury} \\
Published in \textbf{JHEP 1103 (2011) 060 }\\
e-Print: arXiv:1101.2021 [hep-th].\\

\item \textit{``Corrected area law and Komar energy for noncommutative inspired Reissner-Nordstroem black hole'' }\\
Sunandan Gangopadhyay,  \textbf{Dibakar Roychowdhury} \\
Published in \textbf{Int.J.Mod.Phys. A27 (2012) 1250041 } \\
e-Print: arXiv:1012.4611 [hep-th].\\

\item \textit{``New type of phase transition in Reissner Nordstrom - AdS black hole and its thermodynamic geometry''}\\ 
Rabin Banerjee, Sumit Ghosh, \textbf{Dibakar Roychowdhury} \\
Published in \textbf{Phys.Lett. B696 (2011) 156-162 }\\
e-Print: arXiv:1008.2644 [gr-qc].\\

\item \textit{``Thermodynamics of Photon Gas with an Invariant Energy Scale''}\\
Sudipta Das,  \textbf{Dibakar Roychowdhury} \\
Published in \textbf{Phys.Rev. D81 (2010) 085039 }\\
e-Print: arXiv:1002.0192 [hep-th].\\

\item \textit{``Relativistic Thermodynamics with an Invariant Energy Scale''}\\ 
Sudipta Das, Subir Ghosh,  \textbf{Dibakar Roychowdhury} \\
Published in \textbf{ Phys.Rev. D80 (2009) 125036 }\\
e-Print: arXiv:0908.0413 [hep-th]. \\ \\ \\

 This thesis is based on the papers numbered by [ 2, 4, 5, 6, 7, 9, 10, 11] whose reprints are
attached at the end of the thesis.
\end{enumerate}
\newpage
\thispagestyle{empty}

~~~~~~~~~~~~~~~~~ ~~~~~~~~~~~~~~~~~~~~ ~~~~~~~~~~~~~

~~~~~~~~~~~~~~~~~~~~~~~~ ~~~~~~~~~~~~~~~~~~~~~~~~~~~~~~

~~~~~~~~~~~~~~~~~~~~~~~~~~~~~~~ ~~~~~~~~~~~~~~~~~~~~~~~~

~~~~~~~~~~~~~~~~~~~~~~~~ ~~~~~~~~~~~~~~~~~~~~~~~~~~~~~~~~~

~~~~~~~~~~~~~~~~~~~~~~~~~~~~~~~~~~~~~~~~~~~~ ~~~~~~~~~~~~~~~~~~

~~~~~~~~~~~~~~~~~~~~~~~~~~~~~~~~~~~~~~~~~~~~~~~~~~~~~~~~~~~~~~

~~~~~~~~~~~~~~~~~~~~~~~~~~~~~~~~~~~~~~~~~~~ ~~~~~~~~~~~~~~~~~~~

~~~~~~~~~~~~~~~~~~~~~~~~~~~~~~~~~~~~~ ~~~~~~~~~~~~~~~~~~~~~~~~

\begin{center}

 {\bf \uppercase{\huge P\Large HASE \huge T\Large RANSITION \vspace{0.3cm}\\ IN \\
\vspace{0.3cm}\huge B\Large LACK \huge H\Large OLES\\
}}
\end{center}

\newpage
\renewcommand{\cftchapdotsep}{\cftdotsep}
\tableofcontents

\chapter{Introduction}

\section{Overview}

Gravity is the weakest among the all four fundamental interactions in nature and could be clearly distinguished from the other three forces through its distinguished property known as the \textit{principle of equivalence} which states that the trajectories of all freely moving particles get affected in an identical way in presence of an external gravitational field. Mathematically this equivalence could be understood as saying that gravity is always coupled to the energy momentum tensor of some matter field. In spite of such a remarkable feature, gravity is still believed to be one of the mysterious forces of nature. The reason for this is quite clear : In spite of numerous attempts for the past several decades till date we do not have a fully satisfactory and consistent theory of \textit{Quantum Gravity}. The classical \textit{General Relativity} is still considered to be the most successful theory describing gravity. 

In the classical description of gravity we have a classical \textit{metric} that describes our space time and which may be obtained as a solution of Einstein equations of General Relativity namely\footnote{$  G_N$ is the gravitational constant and $ c $ is the speed of light.},
\begin{equation}
G_{\mu\nu} + \Lambda g_{\mu\nu} = \frac{8\pi G_{N}}{c^{4}} T_{\mu\nu}
\end{equation} 
where $ \Lambda $ is the cosmological constant and $ T_{\mu\nu} $ is the energy momentum tensor of some classical matter field. For the past several decades, many people have tried to develop various theories with gravity that basically treat the metric classically while, on the other hand, consider the matter fields quantum mechanically. Presumably such a scheme may be considered as a good approximation of the full quantum theory of gravity.

Since its birth in 1916, the theory of \textit{General Relativity} has gained renewed attention due to its several remarkable predictions about nature, among which the existence of \textit{black holes} may be regarded as the most significant one. According to \textit{General Relativity}, a sufficiently high concentration of mass in certain region of space time could produce a strong gravitational field so that even light can not escape from that region. Such a region of space time is known as \textit{black hole}.

Black holes are perhaps the most tantalizing objects of Einstein's \textit{General Relativity}. For space time containing black holes one must encounter a space time singularity provided \textit{General Relativity} is correct and the energy momentum tensor satisfies certain positive definite inequality. This singularity, which is known as black hole singularity, may be regarded as the place where all the known laws of physics formulated on a classical back ground break down. In a space-time containing black hole, there exists a boundary known as the \textit{event horizon}, which behaves as a \textit{trapped surface} such that all future directed null geodesics orthogonal to it are converging. Interestingly black hole singularity is not visible to an observer sitting outside the event horizon. Literally  speaking, the event horizon acts as a one way membrane such that an in-falling observer can easily pass through the horizon from out side to inside, whereas on the other hand, nothing can escape out of it. Due to this spatial behavior classical black holes do not radiate and therefore they possess absolute zero temperature \cite{ref1}.

The discovery of a close connection between the laws of black hole  physics to that with the laws of ordinary thermodynamics is considered to be one of the remarkable achievements of theoretical physics to have occurred during the last forty years. The existence of such a deep analogy has been found to be playing the central role towards our understanding of the theory of \textit{Quantum Gravity}. On top of it, this also illuminates the role of black holes in order to develop a consistent quantum theory of gravity as well as some fundamental aspects regarding the nature of ordinary thermodynamics itself\footnote{For excellent reviews on this subject see \cite{ref2}}. 

A possible identification of black holes as thermodynamic objects was first realized by Bardeen, Carter and Hawking \cite{ref3} during their remarkable derivation of the four laws of black hole mechanics. Based on the notion of the geometry of space time, in their calculations the authors have found that the change in mass ($ \delta M $) of a rotating black hole could be related to the change in horizon area ($ \delta A $) and the change in angular momentum ($ \delta J $) by the following relation,
\begin{equation}
\delta M = \frac{\kappa}{8\pi}\delta A + \Omega_{H} \delta J\label{neqn1}
\end{equation}  
which is known as the \textit{first law of black hole mechanics}.
Here $ \kappa $ is the \textit{surface gravity} which is constant over the event horizon of the stationary black hole and $ \Omega_{H} $ is the angular velocity at the event horizon.
There could be some additional terms present on the R.H.S of (\ref{neqn1}) when matter fields are present.  The relation (\ref{neqn1}) exactly looks like the first law of ordinary thermodynamics which states that the difference in energy ($ E $), entropy ($ S $) and the other state variables of two nearby equilibrium states may be related as, 
\begin{equation}
\delta E = T \delta S + ~~ `` work ~~ terms".
\end{equation}
Thus in order to fit the laws of black hole mechanics to that with the laws of ordinary thermodynamics we have to have the following analogies namely,
\begin{equation}
E\leftrightarrow M,~~ T\leftrightarrow \frac{\kappa}{2\pi} ~~,S\leftrightarrow \frac{A}{4}\label{neqn2}
\end{equation}
where we have identified the entity $ \frac{\kappa}{2\pi} $ as the temperature ($ T $) of the black hole along with the identification of $ \frac{A}{4} $ as the entropy (S) of the black hole\footnote{For an alternative derivation based on exact differentials, see \cite{Banerjee}.} \cite{ref4}-\cite{ref5}. 

However, in the early seventies soon after the discovery of the four laws of black hole mechanics, most of the researchers at that time viewed this analogy merely as a mathematical analogy. The reason for this was clear: Black holes in the framework of classical \textit{General Relativity} do not radiate. It took time  for people to realize that there indeed lies a deep physical meaning behind the above identifications (\ref{neqn2}) until Hawking came out with his dramatic discovery \cite{ref6} which unleashed the fact that due to quantum particle creation effects a black hole is capable of radiating to infinity all species of particles with a perfect black body spectrum at temperature,
\begin{equation}
T = \frac{\kappa}{2\pi}.
\end{equation}
Soon after the discovery of Hawking it became clear that the quantity $ \frac{\kappa}{2\pi} $ actually plays the role of the physical temperature for a black hole. Thus the discovery of Hawking firmly established the fact that black holes may be identified as thermodynamic objects once quantum mechanical effects are incorporated in the framework of classical general relativity. This is usually known as \textit{semi-classical} regime of black hole physics. Under this \textit{semi-classical} framework one can therefore study various thermodynamic aspects of black holes, like thermal stability, phase transition, black hole evaporation etc. 

The identification of black holes as thermodynamic objects with physical temperature and entropy opened a gate to a new realm. Once black holes are identified as thermodynamic objects, it is very natural to ask whether they also behave as familiar thermodynamic systems. Studying \textit{phase transition} in black holes would be naturally one of the fascinating topics in this regard. Since the birth of black hole thermodynamics, the issue of phase transition in black holes has been found to be playing a significant role towards the understanding of several crucial properties of black holes including the statistical origin of its entropy. 

Occurrence of phase transition in black holes had been first investigated by Davies and Hut in the late seventies \cite{ref7}-\cite{ref8}. Their analysis revealed the fact that Schwarzschild black holes in asymptotically flat space could be found in thermal equilibrium with the surrounding thermal radiation at some equilibrium temperature ($ T $) which is precisely the Hawking temperature. However this equilibrium is unstable as can be seen through the computation of the heat capacity ($ C= \partial M/\partial T $) which comes out to be negative. This means that if the black hole absorbs some mass from outside its temperature would go down and as a result the rate of absorption would be more than the rate of emission and the black hole would continue to grow. Therefore a canonical ensemble can not be defined for a Schwarzschild black hole in asymptotically flat space. Moreover, in course of time it also became clear that for rotating and charged black holes the heat capacity at constant volume ($ C_v $) exhibits various types of discontinuities which resemble \textit{phase transitions} \cite{refpavon}. It was found that for rotating (or charged) black holes the heat capacity $ C_v $ is not necessarily always negative. For sufficiently highly charged or rapidly rotating black holes one could find a region of stable equilibrium with $ C_v>0 $. 

Although the very first attempt to understand the thermodynamic behavior of black holes was commenced for space times that are asymptotically flat, still this could be regarded merely as a theoretical model. The reason for this lies in the fact that the concept of asymptotic flatness is not always the correct theoretical idealization and is hard to satisfy in reality. Therefore in order to develop a systematic theoretical description for black holes, in the early nineties people tried to construct gravitational theories within a bounded and finite spatial region of space time \cite{ref9}-\cite{ref14}.  It was indeed found that for a finite temperature at the spatial boundary the heat capacity of the black hole comes out to be positive and therefore there is no inconsistency in defining the black hole partition function \cite{ref9}.  
From these analysis soon it became clear that instead of studying the thermodynamic behavior of black holes in asymptotically flat spaces, one should actually consider space-times that are asymptotically curved, in other words space-times which admit a cosmological constant ($ \Lambda $).

If $ \Lambda > 0 $, then the resulting space-time is an asymptotically \textit{de Sitter} (dS) space. Like in the case of an asymptotically flat space, black holes in an asymptotically \textit{de Sitter} space also emit particles at a temperature determined by its surface gravity. On top of it, in case of an asymptotically \textit{de Sitter} space we have a cosmological horizon which also emits particles determined by its own surface gravity. Therefore a thermal equilibrium is possible only when these two surface gravities are equal, i.e; when the temperatures of both the event horizon as well the cosmological horizon are same. Later on Carlip and Vaidya had performed a detailed analysis of the thermal stability of a charged black hole in an asymptotically \textit{de Sitter} space in \cite{ref15}. For a super-cooled black hole in an asymptotically \textit{de Sitter} space they observed an interesting phase structure that has a line of first order transition which terminates on a \textit{second} order point.   
  
If $ \Lambda < 0 $, then the resulting space time is known as the \textit{anti de Sitter} (AdS) space. In an AdS space the gravitational potential relative to any origin increases as one moves away from the origin. This means that an asymptotically AdS space acts as a confining box. As a result the thermal radiation remains confined close to the black hole and cannot escape to infinity. Although zero rest mass particles can escape to infinity but the incoming and outgoing fluxes at infinity are equal. Therefore one can always consider a canonical ensemble description for black holes at any given temperature ($ T $). 

Investigating thermodynamic behavior of black holes in an asymptotically AdS space had been initiated due to Hawking and Page \cite{ref16}. According to their observations there exists a phase transition (so called \textit{Hawking Page (HP) transition}) between thermal AdS and AdS black holes in four dimensions. Later on the phase transition of AdS black holes in five dimensions attained renewed attention in the context of AdS/CFT duality, which shows that HP transition in the bulk may be identified as a confinement de-confinement transition in the boundary field theory \cite{ref17}.  In the conventional HP transition one generally starts with the thermal radiation in an AdS space. If the temperature of the thermal radiation is below certain minimum ($ T<T_0 $) then the only possible equilibrium is the thermal radiation with out black holes. On the other hand if the temperature is higher than this minimum then there could be two possible black hole solutions which could be in equilibrium with the thermal radiation. According to the findings of Hawking and Page, there exists a critical mass value ($ M_0 $) for black holes in AdS space. If the mass of the black hole is below this critical value ($ M<M_0 $) then it appears with a negative heat capacity. Which means that the lower mass black hole is unstable. Therefore it may either decay completely to thermal radiation or to a larger mass ($ M>M_0 $) black hole with a positive heat capacity which corresponds a locally stable phase.  In the second case the heat capacity changes from negative infinity to positive infinity at the minimum temperature $ T_0 $. Also there is a change in the dominance from thermal AdS to black holes at some temperature $ T_1 $ which corresponds to a change in sign in the free energy of the system. Since the discovery of such a nice phenomenon in AdS black holes, till date a number of investigations have been made regarding various thermodynamic aspects of AdS black holes \cite{ref18}-\cite{ref44}.

Studying phase transition in black holes from a completely new perspective has been initiated recently in \cite{ref31} where the idea of \textit{Clausius-Clapeyron-Ehrenfest's} equations \cite{ref107} have been successfully implemented in the context of black hole thermodynamics.  Subsequently this analysis has been further extended to analyze the phase transition occurring in charged (RN) AdS and Kerr AdS black holes \cite{dibkr}-\cite{ref32}. The motivation behind such an analysis is the following:  Although the discontinuity in the specific heat (during phase transition in black holes \cite{ref24},\cite{ref27}) is sufficient to indicate the onset of a \textit{continuous} higher order transition, but certainly it is not a systematic or complete way to correctly identify the nature of this phase transition, i.e; we cannot conclude uniquely whether the phase transition is a genuine \textit{second} order or any higher order transition.  Therefore we need certain prescription in order to identify the nature of this phase transition correctly. In ordinary thermodynamics similar situations are dealt within the framework of \textit{Ehrenfest's} equations.  For a genuine \textit{second} order phase transition both the \textit{Ehrenfest's} relations are found to be satisfied simultaneously \cite{ref107}. 

Although the analysis carried out in \cite{ref31}-\cite{ref32} was proved to be quite successful in order to classify the phase transition correctly, nevertheless it had some shortcomings. The reason for this was the lack in \textit{analytic} techniques while computing the \textit{Ehrenfest's} equations near the critical point. Therefore one of the major motivations of the present thesis is to fill up this gap by developing suitable \textit{analytic} scheme in order to compute the \textit{Ehrenfest's } equations.  As a matter of fact, based on exact \textit{analytic} computations, in the present thesis we show that it is indeed possible to verify both the \textit{Ehrenfest's} equations exactly at the critical point and thereby to fix the actual nature of the phase transition unambiguously. 

An alternative approach to study thermodynamics is using \textit{state-space} geometry. In 1979 George Ruppeiner \cite{ref45} proposed a geometrical way to study thermodynamics and statistical mechanics. Different aspects of thermodynamics and statistical mechanics of the system are encoded in the metric of the state space. Consequently various thermodynamic properties of the system can be obtained from the properties of this metric and curvature. It is successful in describing the behavior of classical systems \cite{ref45} and recently it has been found to play an important role in explaining black hole thermodynamics \cite{ref46}-\cite{ref47}. 

In conventional thermodynamic systems, phase transition phenomena plays an important role in order to explore thermodynamic properties of various systems near the critical point of phase transition. The main objective in the theory of phase transition is to study the behavior of a given system in the neighborhood of the critical point, which is characterized by the discontinuity in some thermodynamic variable. In usual thermodynamics various physical quantities (like heat capacity, compressibility etc.) pertaining to a system suffer from a singularity near the critical point. The effect of diverging correlation length near the critical point manifests as divergences of these thermodynamic quantities. This is known as \textit{static critical phenomena}. It is customary in usual (equilibrium) thermodynamics to express all these singularities by a set of \textit{static critical exponents} \cite{ref48}-\cite{stanley}, which determine the qualitative nature of the critical behavior of a given system near the critical point. The effect of diverging correlation length in case of non equilibrium thermodynamics results in a critical slowing down near the critical point which is parametrized by the dynamic critical exponents, and these two sets of exponents could be related to each other. 

In case of a \textit{second} order phase transition the so called static critical exponents are found to be universal in a sense that apart from a few factors, like spatial dimensionality, symmetry of the system etc. they do not depend on the details of the interaction. As a matter of fact different physical systems may belong to the same \textit{universality} class. Critical exponents are also found to satisfy certain \textit{scaling laws}, which in turn implies that not all the critical exponents are independent. These scaling relations are related to the \textit{scaling hypothesis} for thermodynamic functions. Since black holes are well defined thermodynamic objects, therefore it will be quite natural to study all these issues for black holes particularly defined in an asymptotically AdS space.  

Studying the critical behavior in black holes had been commenced long ago considering the charged and rotating black holes in asymptotically flat space time \cite{ref49}-\cite{ref52}. In \cite{ref50}, the critical exponents for the Kerr-Newmann black holes were calculated for the first time in order to check the validity of the \textit{scaling laws} for black holes near the critical point. Recently the thermodynamics of black holes in AdS space has attained much attention in the context of AdS/CFT duality, and the critical phenomena might play a crucial role in order to gain some insights regarding this duality. Like in the case of asymptotically flat space time, a number of attempts have also been made in order to explore the critical behavior of rotating and charged black holes both in the \textit{de Sitter} (dS) and \textit{anti de Sitter} (AdS) space times \cite{ref15}, \cite{ref53}. 

Recently gravity theories in AdS space has attained renewed attention in the context of \textit{AdS/CFT} (or, \textit{gauge/gravity }) duality \cite{ref54}-\cite{ref58}, where it has been realized that the physics of black holes in the bulk AdS space-time plays a crucial role in order to explore the underlying physics behind the various critical phenomena (like phase transition, scaling behavior etc.) occurring in a CFT living on the boundary of the AdS bulk. This fact has been established by Witten long back when he showed that the phase transition taking place between the thermal AdS space at low temperatures and the Schwarzschild AdS black hole at high temperatures could be realized as the confinement/deconfinement transition in the language of boundary CFT \cite{ref17}. The central idea behind such an analysis is based on a remarkable conjecture, which states that:

 \textit{With AdS boundary conditions, string theory is completely equivalent to a gauge theory living on the AdS boundary at infinity.}

In the early seventies, long before the discovery of \textit{AdS/CFT} duality, based on the structure of perturbation theory, t' Hooft first argued that $ 1/N $ expansion of certain $ SU(N) $ gauge theory could be viewed as a theory of strings \cite{ref59}. Since then it took more than twenty years to develop a solid theoretical platform for this idea to be more precise. Gradually it became clear that \textit{AdS/CFT} duality is the manifestation of the fact that the theory of \textit{Quantum Gravity} with AdS boundary conditions is \textit{holographic}: It could be completely described by degrees of freedom living in a lower dimensional space \cite{ref60}. 

Besides explaining various strongly coupled phenomena in gauge theories, people have also tried to explore the other non gravitational areas of physics using \textit{AdS/CFT} duality, for example it is found that the so called \textit{Hall Effect} and the \textit{Nernst Effect} could be explained using \textit{AdS/CFT} duality \cite{ref61}-\cite{ref62}. The basic idea that lies behind such an analysis is that a non gravitational system at thermal equilibrium could be well described by a black hole in the AdS bulk with the Hawking temperature  $ T $. Moreover it has been confirmed that one can actually compute the electrical conductivity and the other transport properties for a non gravitational system through linear perturbations in black holes. Inspired from these early success, people have further tried to extend the domain of applicability of \textit{AdS/CFT} duality by asking the following question:

\textit{Is it possible to find out a gravitational theory that describes superconductivity?} 

Before we try to answer this question, let us first have a quick review on conventional superconductors like Cu, Al, Nb etc. The conventional BCS theory \cite{bcs} for superconductors tells us that below certain critical temperature ($ T_c $) there happens to be a \textit{second} order phase transition from the ordinary conducting phase to a superconducting phase and the electrons with opposite spin are combined to form a charged boson called the \textit{Cooper} pair. Besides having its remarkable achievements, the BCS theory has also its limitations. For example it cannot provide any satisfactory theoretical explanation that will help us to understand the properties of a new class of high $ T_c $ superconductors which were first discovered in 1986 \cite{ref63}. Recently a new class of high $ T_c $ superconductors were discovered in 2008 which are found to be layered and the superconductivity is found to be associated with two dimensional planes \cite{ref64}. Although many people believe that electron pairs do form in these high $ T_c $ materials while actually we still do not have any proper theoretical explanation regarding the pairing mechanism that is responsible for the high $ T_c $ superconductivity. 

Surprisingly over the past few years, based on the fundamental principles of \textit{gauge/gravity} duality, many people have been able to establish a remarkable connection between \textit{General Relativity} and the phenomena like \textit{superconductivity}. The minimal ingredient required to build such a theory is the following: In order to build a superconductor we need the notion of a critical temperature ($ T_c $) and there must be two distinct phases which meet at $ T = T_c $. Speaking more specifically in order to have a \textit{conductor/superconductor} type phase transition occurring at $ T=T_c $ we must have a charged condensate (that results in a superconductivity) for $ T<T_c $. On the gravity side the role of temperature is played by the black hole where the (\textit{Hawking}) temperature of the black hole is precisely the temperature of the field theory living in the boundary. On the other hand, in order to have a charged condensate in the boundary field theory one must include charged scalar field in the bulk gravitational theory. Therefore all these arguments suggest that in order to describe a \textit{conductor/superconductor} like transition in the boundary field theory one must have a (charged) black hole in the bulk that possesses \textit{scalar hair} for $ T<T_c $ and no \textit{hair} for $ T\geq T_c $. This indeed suggests that in the bulk gravitational theory we must have a corresponding \textit{phase transition} occurring at $ T = T_c $ where two distinct phases meet. For $ T<T_c $ we have a charged hairy black hole configuration which corresponds to a superconducting phase in the boundary field theory where as on the other hand for $ T\geq T_c $ we have planar Reissner Nordstrom AdS solution which corresponds to an ordinary conducting phase in the boundary field theory. Under the frame work of \textit{AdS/CFT} duality it is easy to show that the phase transition occurring at $ T = T_c $ is indeed a \textit{second} order phase transition. Theory of superconductors that are built under the frame work of \textit{AdS/CFT} duality are known as \textit{Holographic Superconductors}. Interestingly enough such a holographic model of superconductivity has been found to shed some light on various mysterious issues like the pairing mechanism that is responsible for high $ T_c $ superconductivity. At this stage it must be noted that, one must use AdS boundary conditions in order to build a gravitational theory that includes (charged) \textit{hairy} black hole configuration at low temperatures. The reason for this is simple: In an AdS space charged particles cannot escape out to infinity because the negative cosmological constant acts like a confining box and as result there could be a charged condensate outside the event horizon. 

It was Gubser \cite{ref65}-\cite{ref66} who first argued that the gravity dual of a superconductor could be found through the mechanism of spontaneous $ U(1) $ symmetry breaking near the black hole event horizon which results in a condensation of scalar hair at a temperature (T) that is less than certain critical value ($ T_c $). This critical value ($ T_c $) below which the scalar hair forms may be identified as the critical temperature corresponding to a \textit{second} order phase transition from a normal phase to a superconducting phase in the dual field theory. It is in fact the local $ U(1) $ symmetry breaking in the bulk which corresponds to a global (or sometimes a weakly gauged) $ U(1) $ symmetry breaking in the dual field theory residing at the boundary of the AdS space and thereby inducing a superconductivity\footnote{ For excellent reviews on this subject one may consult \cite{ref67}-\cite{ref69}.}. Later on, this idea was further developed and systematically extended by Horowitz et al. \cite{ref70}-\cite{ref75}, who have found that such a simple gravitational dual can indeed reproduce all the standard features of the conventional superconductors \cite{ref76}. A number of important observations have been made in this regard which may be put as follows \cite{ref77}-\cite{ref86} :

$\bullet$ It has been found that the condensate\footnote{$ \langle\mathcal{O}\rangle $ is the vacuum expectation value of some operator in the field theory living on the boundary of the AdS and is \textit{dual} to the scalar field added in the bulk. Moreover $ \langle\mathcal{O}\rangle $ plays the role of an \textit{order parameter} in the boundary field theory.} ($ \langle\mathcal{O}\rangle $) is non zero only at sufficiently low temperatures and is found to be acquiring a first non trivial value other than zero at $ T=T_c $. The behavior of the condensate at low temperatures is surprisingly found to be similar to that obtained from the BCS theory. Near the critical temperature ($ T \sim T_c $) it is found that,
\begin{equation}
\langle\mathcal{O}\rangle \sim \sqrt{1-\frac{T}{T_c}}
\end{equation}  
which confirms the standard \textit{mean field} behavior as predicted by the Landau-Ginzburg theory\footnote{In chapter 4, we will derive this relation using perturbation technique. }.

$ \bullet $ From the computation of the free energy it has been further confirmed that the free energy for the \textit{hairy} configuration is always lower than that of the black hole configuration with out scalar hair which in the present case corresponds to the planar Reissner Nordstrom AdS space-time.

$ \bullet $ The optical conductivity $ \sigma (\omega) $ is found to be constant for $ T\geq T_c $. On the other hand for $ T<T_c $ a pronounced gap appears at low frequency. Most importantly it has been observed that exactly at $ \omega = 0 $ there appears to be a delta function which indeed suggests the presence of an infinite \textit{dc} conductivity that is the unique feature of all superconductors.

$ \bullet $ At low temperatures the ratio of the width of the gap in the optical conductivity ($ \omega_g $) to the critical temperature ($ T_c $) is found to be,
\begin{equation}
\frac{\omega_g}{T_c}\thickapprox 8
\end{equation}
Interestingly for certain high $ T_c $ cuprates this ratio has been found to be exactly eight \cite{ref87}.

 Apart from these issues recently there have been some attempts in the search for the magnetic response response in holographic superconductors. It was shown that holographic superconductors are type II in nature and both the vortex and droplet states have been constructed in this regard \cite{ref88}-\cite{ref95}. 
 
The holographic model of a superconductor that was initially constructed was based on $ s $- wave symmetry. Latter on it was found that the theory may be further modified which could still result in a dual description of a superconductor with a $ p $- wave symmetry.  This was obtained by replacing the original Maxwell action and the charged scalar field by a $ SU(2) $ gauge field in the bulk theory. These are known as holographic $ p $- wave superconductors where the order parameter is a vector and the conductivity is strongly anisotropic in a manner that is consistent with the $ p $- wave nodes on the fermi surface \cite{ref96}.

\section{Outline of the thesis}

The present thesis is based on the works \cite{ref97}-\cite{ref104} and aims to explore the issue of \textit{phase transition} in black holes under various circumstances. The whole content of this thesis may be divided into two parts. The first part of it (Chapter \textit{2} - Chapter \textit{3} ) is focused towards the application of basic thermodynamic principles to AdS black holes. Based on the thermodynamic analogy, during these works \cite{ref97}-\cite{ref98}, we have developed a unique method to study the thermodynamic stability of black holes in AdS space. It is shown that using the notion of thermal stability in ordinary thermodynamics one can indeed distinguish between different phases of a black hole that is undergoing a phase transition.  On top of it, incorporating the idea of \textit{Ehrenfest's scheme} \cite{ref31}-\cite{ref32} from usual thermodynamics to black hole thermodynamics, we have shown a novel way to determine the nature of this phase transition. 

The investigation of the phase transition phenomena has been further pushed forward during the works \cite{ref99}-\cite{ref101} where a general method has been prescribed in order to calculate the critical exponents for the charged AdS black holes near the critical point of the phase transition. The novelty of the present approach is that it could be used to calculate the critical exponents for any  other black hole that possesses a discontinuity in the heat capacity near the critical point. As a non trivial exercise, in \cite{ref101} this general method has been extended in order to compute the critical exponents for the charged topological black holes in the \textit{Ho\v{r}ava-Lifshitz} theory of gravity. The computation of these critical exponents are crucial in a sense that it helps us to explore the \textit{scaling} behavior of black holes near the critical point and thereby to justify the validity of the \textit{scaling laws} in the context of black hole thermodynamics.

The rest part of the thesis (Chapter $ 4 $ - Chapter $ 5 $) is devoted towards the study of the phase transition phenomena in planar AdS black holes under the framework of \textit{AdS/CFT} duality. These works \cite{ref102}-\cite{ref104} consider a gravitational theory coupled to the \textit{Born- Infeld} (BI) Lagrangian in an asymptotically AdS space where the planar BI-AdS black hole becomes unstable to develop a \textit{scalar hair} at low temperatures. This implies that in the bulk we have a phase transition
\begin{center}
BI- AdS $ \rightarrow $ Charged black hole with scalar hair
\end{center}
The mechanism that is responsible for such a transition is the \textit{spontaneous} breaking of $ U(1) $ symmetry at low temperatures. As a consequence of this in the boundary field theory there happens to be a \textit{superconducting} phase transition which is thereby termed as \textit{holographic superconductor}. During these works \cite{ref102}-\cite{ref104}, based on analytic calculations and using the \textit{AdS/CFT} dictionary several crucial properties of these \textit{holographic superconductors} have been investigated in the \textit{probe} limit, for example, various physical entities like, the critical temperature ($ T_c $), order parameter, the critical magnetic field etc. have been computed near the critical point of the phase transition.
\vskip 7mm

\textit{The whole thesis consists of 6 - chapters, including this introductory part. Chapter wise summary is given below.}
\vskip 15mm
{\underline{\bf{Chapter}} -2:  \textit{Thermodynamics of Phase Transition In AdS Black Holes}:
\vskip 2mm
In this chapter, based on the fundamental concepts of thermodynamics and considering a \textit{grand canonical} framework we present a detailed analysis of the phase transition phenomena in AdS black holes. The method is in line with
that used a long ago to understand the \textit{liquid-vapor} phase transition where the first order derivatives of \textit{Gibbs} potential are discontinuous and \textit{Clausius-Clapeyron} equation is satisfied. We have extended this idea to black holes in order to explore the vexing issue of phase transition phenomena that occurs between a \textit{lower} mass black hole with negative specific heat to a \textit{higher} mass black hole with positive specific heat. The critical point has been marked by the discontinuity in the corresponding heat capacity (i.e; \textit{second} order derivative in the \textit{Gibbs} potential). Moreover, using \textit{Ehrenfest's scheme} of standard thermodynamics, we have \textit{analytically} checked that the phase transition occurring in these black holes is indeed a \textit{second} order phase transition.
\vskip 3mm
{\underline{\bf{Chapter}} -3:  \textit{Critical Phenomena In Charged AdS Black Holes}:
\vskip 2mm
In this chapter, considering a \textit{canonical} ensemble we have further extended our analysis in order to investigate the \textit{scaling} behavior in charged AdS black holes near the critical point(s). We have explicitly computed all the \textit{static critical exponents} associated with various thermodynamic entities (like, heat capacity, isothermal compressibility etc.) and found that all these exponents indeed satisfy so called \textit{thermodynamic scaling laws} near the critical point(s). We have also checked the \textit{Generalized Homogeneous Function }(GHF) hypothesis \cite{ref48} for charged AdS black holes and found its compatibility with thermodynamic \textit{scaling laws}. Going beyond the usual framework of \textit{Einstein} gravity, we have also investigated the phase transition of the static, \textit{topological} charged black hole solution with arbitrary scalar curvature $2k$ in the \textit{Ho\v{r}ava-Lifshitz} theory of gravity at the \textit{Lifshitz} point $z=3$.  The analysis has been performed using the {\it canonical ensemble} frame work; i.e. the charge was kept fixed. We have found (a) for both $k=0$ and $k=1$, there is no phase transition, (b) while $k=-1$ case exhibits the \textit{second} order phase transition within the {\it physical region} of the black hole. The critical point of \textit{second} order phase transition was obtained by the divergence of the heat capacity at constant charge. Near the critical point, we have computed various critical exponents. It has been also observed that they satisfy the usual thermodynamic \textit{scaling laws}.
\vskip 3mm
{\underline{\bf{Chapter}} -4:  \textit{AdS/CFT Duality And Phase Transition In Black Holes: Holographic Superconductors}:
\vskip 2mm
In this chapter, based on the \textit{Sturm-Liouville} (SL) eigenvalue problem, we have \textit{analytically}
investigated various properties of \textit{holographic s-wave superconductors} in the background of
both \textit{Schwarzschild-AdS} and \textit{Gauss-Bonnet AdS} space-times in the framework of \textit{Born-Infeld} (BI) electrodynamics. Using perturbation technique, we have explicitly computed the relation between the critical temperature and the charge density and also found that both the \textit{Born-Infeld} (BI) coupling parameter as well as the \textit{Gauss Bonnet} (GB) coupling parameter indeed affect the formation of scalar hair at low temperatures. It has been observed that the condensation is harder to form for higher values of BI/GB coupling parameters. The critical exponent associated with the \textit{order parameter} has been found to be 1/2,  which is in good agreement to that with the universal mean field value. The analytic results thus obtained are found to be in good agreement to that with the existing numerical results.
\vskip 3mm
{\underline{\bf{Chapter}} -5:  \textit{Magnetic Response In Holographic Superconductors}:
\vskip 2mm
In this chapter, based on a different analytic scheme known as the \textit{Matching} technique \cite{ref78}, several properties of holographic $ s $-wave superconductors have been investigated in the presence of an external magnetic field and considering various higher derivative (\textit{non linear}) corrections to the usual Maxwell action. Explicit expressions for the critical temperature as well as the $ s $- wave \textit{order parameter} have been obtained in the probe limit. It has been found that below certain critical magnetic field strength ($ B_c $) there exists a superconducting phase. Most importantly it is observed that the value of this critical field strength ($ B_c $) indeed gets affected due to the presence of higher derivative corrections to the usual Maxwell action. 
\vskip 3mm
{\underline{\bf{Chapter}} -6:  \textit{Summary and Outlook}:
\vskip 2mm
Finally, in chapter-6 we present the summary of our findings along with some future prospects.

\chapter{Thermodynamics of Phase Transition In AdS Black Holes}

\section{The gibbsian approach}
It is well known that black holes behave as thermodynamic systems. The laws of black hole mechanics become similar to the usual laws of thermodynamics after appropriate identifications between the black hole parameters and the thermodynamical variables \cite{ref3}. Nonetheless black holes show some exotic behaviors as a thermodynamic system, such as (i) entropy of black holes is proportional to area and not the volume, (ii) their temperature diverges to infinity when mass tends to vanish due to Hawking evaporation. In fact these exotic behaviors enhance the scope of alternative studies on black holes. The thermodynamic properties of black holes were further elaborated in early eighties when Hawking and Page discovered a phase transition between the thermal AdS and the Schwarzschild AdS black hole \cite{ref16}. Thereafter this subject has been intensely studied in various viewpoints \cite{ref18}-\cite{ref44}. 

The \textit{gibbsian} approach to the phase transitions is based on the \textit{Clausius-Clapeyron-Ehrenfest's} equations \cite{ref107}. These equations allow for a classification of phase transitions as first order or continuous (higher order) transitions. For a first order transition the first order derivatives (entropy and volume) of the Gibbs free energy is discontinuous and \textit{Clausius-Clapeyron} equation is satisfied. Liquid to vapor transition is an ideal example of such a phase transition. Similarly for a \textit{second} order transition \textit{Ehrenfest's} relations are satisfied. However, this specific tool concerning the nature of the phase transition has never been systematically explored for black holes. For the past couple of years there have been a series of papers \cite{ref31}-\cite{ref32},\cite{ref97}-\cite{ref98} which have systematically applied these concepts to study phase transitions in black holes. Despite of some conceptual issues mentioned in the earlier paragraph, it is found that black holes not only allow us to implement above basic ideas but also support them. In normal physical systems a verification of the \textit{Clausius-Clapeyron} or \textit{Ehrenfest} relation is possible subjected to the design of a table top experiment. However black holes are found to be like a theorist's laboratory, where these ideas on phase transition could be proved without performing any table top experiment.

From all of these works \cite{ref31}-\cite{ref32},\cite{ref97}-\cite{ref98}, it is indeed evident that in case of black hole phase transition we do not encounter any discontinuity in the first order derivative of Gibbs energy and hence \textit{Clausius-Clapeyron} equation was not needed.  However, because of the discontinuity in \textit{second} order derivatives, \textit{Ehrenfest's} relations for black holes were first derived in \cite{ref31}. Because of infinite divergences of various physical entities at the critical point, numerical methods were adopted in order to check the validity of the \textit{Ehrenfest's} relations.  From these analyses it was found that away from the critical point, only the first \textit{Ehrenfest's} relation was satisfied \cite{ref31}-\cite{dibkr}, whereas on the other hand, close to the critical point both relations were satisfied \cite{ref32}.  Nevertheless it should be emphasized that such a numeric check has a drawback. Since in this method one assigns a numerical value for the angular velocity or electric potential beforehand and then check the \textit{Ehrenfest} relations, every time this value is changed it is necessary to repeat the numerical analysis.  This is a limitation for generalizing the result for other cases. On top of it, because of the infinite divergences in various physical entities near the critical point, it is indeed very difficult to estimate the physical quantities exactly at the critical point and therefore the whole numerical exercise loses its significance. Such limitations in the numerical analysis should provide us enough motivation towards developing certain analytic steps that would essentially capture the physics of black hole phase transition near the critical point.

Based on the papers \cite{ref97}-\cite{ref98}, in this chapter we present an elegant analytic technique to treat the singularities near the critical points of the phase transition curves. This is important for further studies on black hole phase transition using our techniques. The example of the $ (n+1) $ dimensional charged (RN) AdS and Schwarzschild AdS black holes are worked out in details. Moreover, we have also checked the validity of the \textit{Ehrenfest's} relations for the rotating (Kerr) black holes in $ (3+1) $ dimensions.  We show analytically that at the critical point both \textit{Ehrenfest's} relations are satisfied exactly. This confirms the onset of a \textit{second} order phase transition for these black holes. Explicit expressions for the critical temperature, critical mass etc. are given. This phase transition is characterized  by the divergence of the specific heat at the critical temperature. The sign of specific heat is different in two phases and they essentially separate two branches of AdS black holes with different mass/ horizon radius. The branch with lower mass (horizon radius) has a negative specific heat and thus falls in an unstable phase. The other branch with larger mass (horizon radius) is locally stable since it is associated with a positive specific heat and also positive Gibbs free energy. 

Before we proceed further, let us briefly mention about the organization of this chapter.  To start with, it will be nice to have a quick review on the conventional \textit{liquid-vapor} transition in the framework of standard \textit{Clausius-Clapeyron} equations which is done in section 2.2. In the next section we make a detail discussion on the basic thermodynamic structure of the phase transition occurring in $ (n+1) $ dimensional charged (RN) AdS black holes and as an appropriate limit we obtain the corresponding results for the Schwarzschild AdS case. In section 2.4, based on the Ehrenfest's scheme of standard thermodynamics we perform a detail analysis in order to identify the nature of the phase transition in these black holes. In section 2.5 we carryout an identical analysis for the $ (3+1) $ dimensional rotating (Kerr) AdS black holes and discuss the results in brief. Finally we conclude in section 2.6.


\section{Liquid to vapour phase transition} 
Liquid and vapour are just two phases of matter of a single constituent. When heated upto the boiling point liquids vapourise and at this point the slope of the coexistence curve (pressure $P$ versus temperature $T$ plot) obeys the Clausius-Clapeyron equation, given by
\begin{eqnarray}
\frac{d P}{dT}=\frac{\Delta S}{\Delta V},
\label{cc}
\end{eqnarray}
where $\Delta S$ and $\Delta V$ are the difference between the entropy and volume of the constituent in two different phases.

Note that the derivation of the above relation comes from the definition of the Gibbs potential $G=U-TS+PV$  ($U$ is the internal energy), where it is assumed that the first order derivatives of $G$ with respect to the intrinsic variables ($P$ and $T$) are discontinuous and it comes out that their specific values in two phases give the slope of the coexistence curve at the phase transition point. This is an example of the first order phase transition which clearly differs from the case that we will be studying next.

\section{Phase transition in higher dimensional AdS black holes}
We start our analysis with Reissner-Nordstr{\"o}m black holes in $ (n+1) $ dimensions.
Reissner-Nordstr{\"o}m black holes are charecterised by their mass (M) and charge (Q). The solution for $ (n+1) $ dimensional RN-AdS space time with a negative cosmological constant $ (\Lambda=-n(n-1)/2l^2) $ is defined by the line element \cite{ref28},
\begin{equation}
ds^2 = -\chi dt^2+\chi ^{-1}dr^2+r^2 d\Omega^{2}_{n-1}
\end{equation} 
where,
\begin{equation}
\chi(r) = 1-\frac{m}{r^{n-2}}+\frac{q^2}{r^{2n-4}}+\frac{r^2}{l^2}.
\end{equation}
Here $ m $ is related to the ADM mass ($ M $ ) of the of the black hole as, $ (G=1) $
\begin{equation}
M=\frac{(n-1)\omega_{n-1}}{16 \pi}m ~~ ;~~ ~~ \omega_{n-1} =\frac{2\pi^{n/2}}{\Gamma(n/2)} 
\label{mn1}
\end{equation}
where $ \omega_{n-1} $ is the volume of unit $ (n-1) $ sphere.
The parameter $ q $ is related to the electric charge $ Q $ as,
\begin{equation}
Q=\sqrt{2(n-1)(n-2)}\left( \frac{q}{8 \pi}\right).
\label{q1}
\end{equation}
The entropy of the system is defined as,
\begin{equation}
S=\frac{\omega_{n-1} r_{+}^{n-1}}{4}
\label{s1}
\end{equation}
where $ r_{+} $ is the radius of the outer event horizon defined by the condition $ \chi(r_+ )=0$.
The electrostatic potential difference between horizon and infinity is given by
\begin{equation}
\Phi=\frac{\sqrt{n-1}}{\sqrt{2(n-2)}}\frac{q}{r_{+}^{n-2}}.
\label{phi}
\end{equation} 
Using (\ref{q1}) and (\ref{s1}) we can further express (\ref{phi}) as,
\begin{equation}
\Phi=\frac{4 \pi Q}{(n-2)\omega^{\frac{1}{n-1}}_{n-1}(4S)^{\frac{n-2}{n-1}}}.
\label{phin1}
\end{equation}
From the condition $ \chi(r_+ )=0$ and using (\ref{mn1}), (\ref{q1}), (\ref{s1}) and (\ref{phin1}) we can express the black hole mass as,
\begin{equation}
M=\frac{(4S)^{\frac{n-2}{n-1}}\omega^{\frac{1}{n-1}}_{n-1}(n-1)}{16 \pi}\left[ 1+\frac{2n-4}{n-1}\Phi^{2}+\left(\frac{4S}{\omega_{n-1}}\right) ^{\frac{2}{n-1}}\right].
\label{Mn} 
\end{equation}
Using the first law of black hole mechanics $ dM=TdS+\Phi dQ $, the Hawking temperature may be obtained as,
\begin{eqnarray}
T&=&\left(\frac{\partial M}{\partial S}\right) _{Q}\nonumber\\
&=& \frac{4^{\frac{n-2}{n-1}}\omega^{\frac{1}{n-1}}_{n-1}(n-2)}{16 \pi S^{\frac{1}{n-1}}}\left[ 1-\frac{2n-4}{n-1}\Phi^{2}\right] + \frac{n 4^{\frac{n}{n-1}}S^{\frac{1}{n-1}}}{16 \pi \omega^{\frac{1}{n-1}}_{n-1} }.
\label{Tn} 
\end{eqnarray}
At this stage, it is customary to mention that in the following analysis the various black hole parameters $M,~Q,~J,~S,~T$ have been interpreted as $\frac{M}{l^{n-2}},~\frac{Q}{l^{n-2}},~\frac{S}{l^{n-1}},~Tl$ respectively, where the symbols have their standard meanings. Also, with this convention, the parameter $ l $  does not appear in any of the following equations.
\begin{figure}[h]
\centering
\includegraphics[angle=0,width=6cm,keepaspectratio]{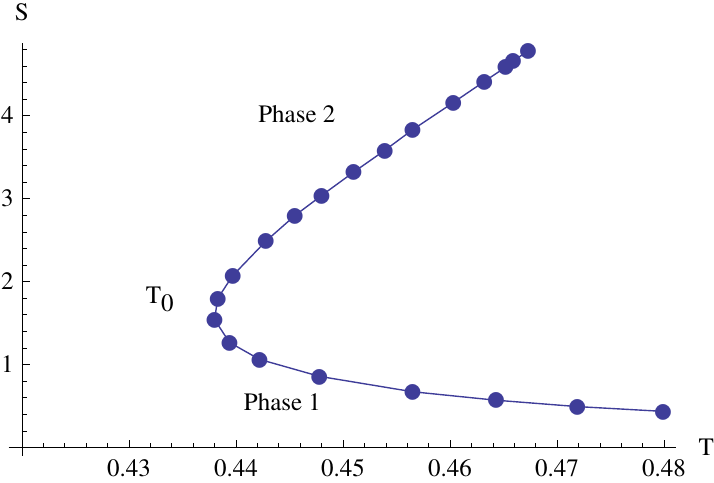}
\caption[]{\it Entropy plot ($ S $) for RN-AdS black hole with respect to temperature ($T$) for fixed $\Phi=0.2$ and $ n=4 $}
\label{figure 2a}
\end{figure} 

Using the above relation (\ref{Tn}) we plot the variation of entropy ($ S $) against temperature ($ T $) in Figure (2.1). In order to obtain an explicit expression for the temperature ($ T_0 $) corresponding to the turning point we first set $ \left( \frac{\partial T}{\partial S}\right) _{\Phi} =0$. From this condition and using (\ref{Tn}) we find
\begin{equation}
S_{0}=\left(\frac{\omega_{n-1}}{4} \right)\left(\frac{n-2}{n} \right)^{(n-1)/2}\left[1-\frac{2n-4}{n-1}\Phi^{2} \right]^{(n-1)/2}.
\label{s0}
\end{equation}
Furthermore we find that, 
\begin{equation}
\left[  \left( \frac{\partial^{2} T}{\partial S^{2}}\right) _{\Phi}\right] _{S=S_0}=\frac{n(n-2)}{8 \pi (n-1)^{3}}\left( \frac{4^{n}}{\omega_{n-1}S^{2n-3}}\right)^{1/(n-1)}.
\end{equation}
Therefore $\left( \frac{\partial^{2} T}{\partial S^{2}}\right) _{\Phi}> 0 $ for $ n>2 $. Hence the temperature $ T_0 $ corresponding to the entropy $ S=S_0 $ represents the minimum temperature for the system.
Substituting (\ref{s0}) into (\ref{Tn}) we find the corresponding minimum temperature for RN-AdS space time to be
\begin{equation}
T_{0}=\frac{\sqrt{n(n-2)}}{2\pi}\left[1-\frac{2n-4}{n-1}\Phi^{2} \right]^{1/2}.
\label{T0} 
\end{equation} 
In the charge-less limit ($\Phi=0$), we obtain the corresponding (minimum) temperature
\begin{equation}
 T_0= \frac{\sqrt{n(n-2)}}{2\pi}
 \label{t0}
\end{equation}
for Schwarzschild AdS space time in $ (n+1) $ dimension. For $ n=3 $ equation (\ref{t0}) yields
\begin{equation}
T_0=\frac{\sqrt{3}}{2 \pi}
\end{equation}
which was obtained earlier by Hawking and Page in their original analysis{\footnote{Note that $l$ does not appear since as explained before, it has been appropriately scaled out.}}  \cite{ref16}. 

It is now possible to interpret the two branches of the $ S-T $ curve, separated by the point $ (S_0,T_0) $, corresponding to two phases. Since entropy is proportional to the mass of the black hole, therefore phase 1 corresponds to a lower mass black hole and phase 2 corresponds to a higher mass black hole. We also observe that both of these phases exist above the (minimum) temperature $ T=T_0 $. To obtain an expression for the mass ($ M_0 $) of the black hole separating the two phases  we substitute (\ref{s0}) into (\ref{Mn}) which yields,
\begin{equation}
M_{0}=\frac{\omega_{n-1}(n-2)^{(n-2)/2}}{8 \pi n^{n/2}}\left[1-\frac{2n-4}{n-1}\Phi^{2} \right]^{(n-2)/2} 
\left[ (n-1)^{2}+ 2(n-2)\Phi^{2}\right]. 
\label{mc}
\end{equation}
Phase 1 corresponds to a black hole with mass $ M<M_0 $ while phase 2 corresponds to a black hole with mass $ M>M_0 $.

In the charge-less limit ($ \Phi=0 $) we obtain the corresponding critical mass for the Schwarzschild AdS black hole in    $ (n+1) $ dimension, 
\begin{equation}
M_0=\frac{\omega_{n-1}(n-2)^{(n-2)/2}(n-1)^{2}}{8 \pi n^{n/2}}.
\label{M0}
\end{equation} 
Substituting $ n=3 $ into (\ref{M0}) yields 
\begin{equation}
M_0=\frac{2}{3\sqrt{3}}
\end{equation}
which is the result obtained earlier in \cite{ref16}. 

A deeper look at the $ S-T $ plot further reveals that the slope of the curve gradually changes it's sign around $ T=T_0 $, thereby clearly indicating the discontinuity in the specific heat ($ C_{\Phi}=T(\partial S/ \partial T)_{\Phi} $) that is associated with the occurrence of a continuous higher order transition at $T=T_0 $. All these features become more prominent from the following analysis. In order to do that we first compute the specific heat at constant potential ($C_{\Phi}$), which is the analog of $C_{P}$ (specific heat at constant pressure) in conventional systems. This is found to be
\begin{eqnarray}
C_{\Phi}&=& T\left( \partial S/\partial T\right) _{\Phi}\nonumber\\ 
&=&\frac{(n-1)S\left[(n-2)\omega^{\frac{2}{n-1}}_{n-1}(1-\frac{2n-4}{n-1}\Phi^{2})+n(4S)^{\frac{2}{n-1}}\right] }{\left[ n(4S)^{\frac{2}{n-1}}-(n-2)\omega^{\frac{2}{n-1}}_{n-1}(1-\frac{2n-4}{n-1}\Phi^{2})\right] }. \label{cphi}
\end{eqnarray}

\begin{figure}[h]
\centering
\includegraphics[angle=0,width=7cm,keepaspectratio]{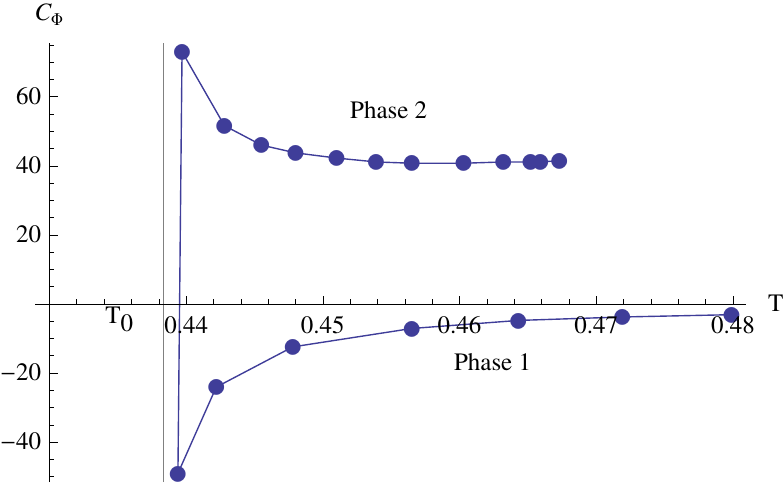}
\caption[]{\it Specific heat ($ C_{\Phi} $) plot for RN-AdS black hole with respect to temperature ($T$) for fixed $\Phi=0.2$ and $ n=4 $}
\label{figure 2a}
\end{figure} 

The critical temperature (at which the heat capacity ($ C_{\Phi} $) diverges) may be found by setting the denominator of (\ref{cphi}) equal to zero, which exactly yields the value of entropy ($ S $) as found in (\ref{s0}). Substituting this value into (\ref{Tn}) it is now trivial to show that $ T_0 $ (\ref{T0}) corresponds to the critical temperature at which $ C_{\Phi} $ diverges. This diverging nature of $C_{\Phi}$ at $ T=T_0 $ may also be observed from Figure (2.2) where we plot heat capacity ($ C_{\Phi} $) against the temperature ($ T $). From this plot we find that the heat capacity ($ C_{\Phi} $) changes from negative infinity to positive infinity at  $ T=T_0 $ \cite{ref24}. One can further note that the smaller mass black hole (phase 1) corresponds to an unstable since it posses a negative specific heat ($C_{\Phi}<0$), whereas the larger mass black hole falls into a stable phase (phase 2) as it corresponds to a positive heat capacity ($C_{\Phi}>0$).
\begin{figure}[h]
\centering
\includegraphics[angle=0,width=7cm,keepaspectratio]{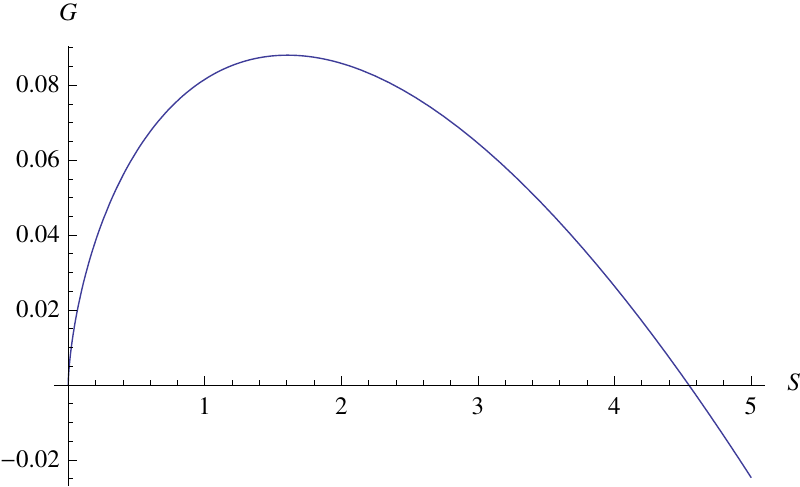}
\caption[]{\it Gibbs free energy ($G$) plot for RN-AdS black hole with respect to entropy ($S$) for fixed $\Phi=0.2$ and $ n=4 $}
\label{figure 2a}
\end{figure}

It is now customary to define grand canonical potential which is known as Gibbs free energy. For RN-AdS black hole this is defined as $G = M-TS-\Phi Q$ where the last term is the analog of $PV$ term in conventional systems. 
Using (\ref{phin1}), (\ref{Mn}) and (\ref{Tn}) one can easily write $G$ as a function of ($S,~\Phi$),  
\begin{eqnarray}
G = \frac{(4S)^{\frac{n-2}{n-1}}\omega^{\frac{1}{n-1}}_{n-1}}{16 \pi}\left[ 1-\frac{2n-4}{n-1}\Phi^{2}\right] 
- \frac{ (4S)^{\frac{n}{n-1}}}{16 \pi \omega^{\frac{1}{n-1}}_{n-1} }.
\label{grn1}
\end{eqnarray}

Since in the above relation we cannot directly substitute entropy ($ S $) in terms of temperature ($ T $) therefore in order to obtain the $ G-T $ plot we first note the variation of Gibbs free energy ($ G $) against entropy ($ S $) using the above relation (\ref{grn1}).

As a next step, (from Figure (2.3)) we note the values of $ G $ for different values of $ S $ and obtain the corresponding values of $ T $ from Figure (2.1). Finally we put all those data together in order to obtain the following numerical plot (fig. (2.4)).

\begin{figure}[h]
\centering
\includegraphics[angle=0,width=7cm,keepaspectratio]{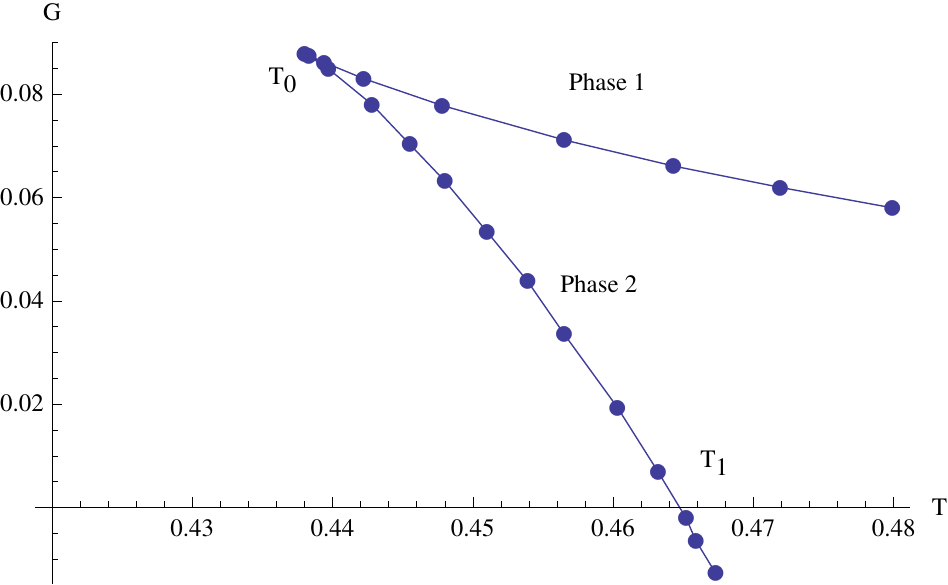}
\caption[]{\it Gibbs free energy ($G$) plot for RN-AdS black hole with respect to temperature ($T$) for fixed $\Phi=0.2$ and $ n=4 $}
\label{figure 2a}
\end{figure}

From Figure (2.4) we note that the value of $ G $ for phase 1 is greater than that of phase 2. Therefore the larger mass black hole falls into a more stable phase than the black hole with smaller mass.  We also observe that $ G $ is always positive for $ T_0<T<T_1 $ . This also corroborates the findings of \cite{ref16} for the Schwarzschild AdS black hole.

In order to calculate $T_1$ we first set $ G=0 $ (in \ref{grn1}) which yields
\begin{equation}
S_1=\left( \frac{\omega_{n-1}}{4}\right) \left[ 1-\frac{2n-4}{n-1}\Phi^{2}\right]^ \frac{(n-1)}{2}.
\label{s'}
\end{equation}
Substituting this into (\ref{T}) we finally obtain
\begin{equation}
T_1=\frac{(n-1)}{2 \pi}\left[ 1-\frac{2n-4}{n-1}\Phi^{2}\right]^ \frac{1}{2}.
\label{Tg}
\end{equation}
In the charge-less limit ($ \Phi=0 $) we obtain the corresponding temperature for the Schwarzschild AdS black hole in    $ (n+1) $ dimension 
\begin{equation}
T_1=\frac{(n-1)}{2 \pi}.
\label{T'}
\end{equation}
Substituting $ n=3 $ into (\ref{T'}) yields 
\begin{equation}
T_1=\frac{1}{\pi}
\end{equation}
which reproduces the result found earlier in \cite{ref16}. 

Furthermore from (\ref{T0}) and (\ref{Tg}) we note that 
\begin{equation}
T_0=\frac{\sqrt{n(n-2)}}{n-1}T_{1}\label{univ}
\end{equation}
which is a universal relation in the sense that it does not depend on black hole parameters.

\section{Ehrenfest's equations}
Classification of the nature of phase transition using \textit{Ehrenfest's} scheme is a very elegant technique in standard thermodynamics. In spite of its several applications to other systems \cite{ref108}- \cite{ref112}, it is still not a widely explored scheme in the context of black hole thermodynamics although some attempts have been made recently \cite{ref31}-\cite{ref32}. In the remaining part of this paper we shall analyze and classify the phase transition phenomena in RN AdS black holes by exploiting \textit{Ehrenfest's} scheme. 

Looking at the ($S-T$) graph (fig (2.1)) we find that entropy $ (S) $ is indeed a continuous function of temperature $ (T) $. Therefore the possibility of a first order phase transition is ruled out.  However, the infinite divergence of specific heat near the critical point $ (T=T_0) $ (see fig (2.3)) strongly indicates the onset of a higher order (continuous) phase transition. The nature of this divergence is further illuminated by looking at the $ C_{\Phi}-S $ graph (fig (2.5)).
\begin{figure}[h]
\centering
\includegraphics[angle=0,width=7cm,keepaspectratio]{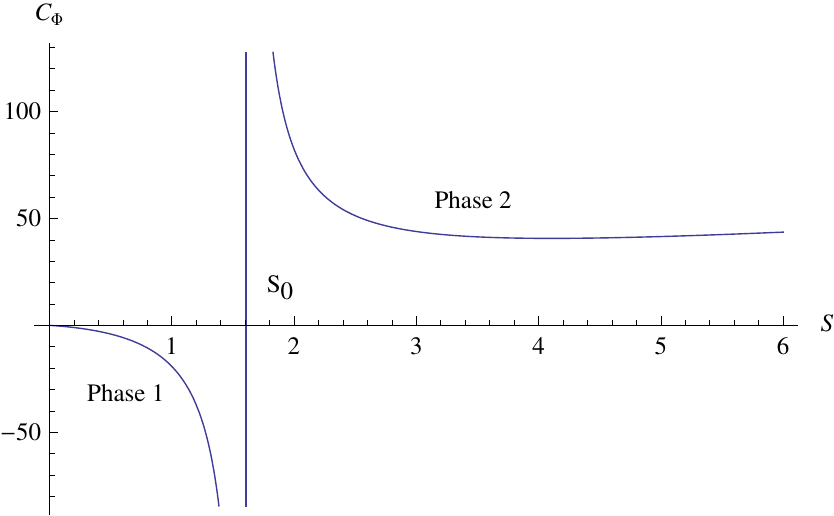}
\caption[]{\it Specific heat ($ C_{\Phi} $) plot for RN-AdS black hole with respect to entropy ($S$) for fixed $\Phi=0.2$ and $ n=4 $}
\label{figure 2a}
\end{figure}

We now exploit \textit{Ehrenfest's} scheme in order to understand the nature of the phase transition.  \textit{Ehrenfest's} scheme basically consists of a pair of equations known as \textit{Ehrenfest's} equations of first and second kind \cite{ref107, ref48}. For a standard thermodynamic system these equations may be written as \cite{ref107}
\begin{eqnarray}
&& \left(\frac{\partial P}{\partial T}\right)_{S} = \frac{C_{P_2}-C_{P_1}}{T V(\alpha_2-\alpha_1)}=\frac{\Delta C_{P}}{T V \Delta \alpha}
\label{eh1}\\
&&\left(\frac{\partial P}{\partial T}\right)_{V} = \frac{\alpha_{2}-\alpha_{1}}{k_{T_{2}}-k_{T_{1}}}=\frac{\Delta \alpha}{\Delta k_T}
\label{eh2}
\end{eqnarray} 
For a genuine \textit{second} order phase transition both of these equations have to be satisfied simultaneously.  

Let us now start by considering the RN-AdS black holes. Bringing the analogy ($ V \leftrightarrow Q, P \leftrightarrow -\Phi$) between the thermodynamic state variables and various black hole parameters, we are now in a position to write down the \textit{Ehrenfest's} equations for this system as \cite{dibkr}, 
\begin{eqnarray}
&& -\left(\frac{\partial \Phi}{\partial T}\right)_{S} = \frac{C_{\Phi_2}-C_{\Phi_1}}{T Q(\alpha_2-\alpha_1)}=\frac{\Delta C_{\Phi}}{T Q \Delta \alpha}
\label{ehf1}\\
&&-\left(\frac{\partial \Phi}{\partial T}\right)_{Q} = \frac{\alpha_{2}-\alpha_{1}}{k_{T_{2}}-k_{T_{1}}}=\frac{\Delta \alpha}{\Delta k_T}
\label{ehf2}
\end{eqnarray} 
where, $\alpha =\frac{1}{Q}\left( \frac{\partial Q}{\partial T}\right) _{\Phi}$ is the analog of volume expansion coefficient and $k_T =\frac{1}{Q}\left( \frac{\partial Q}{\partial \Phi}\right) _{T}$ is the analog of isothermal compressibility. Their explicit forms are given by,\\ 
\begin{eqnarray}
\alpha &=& \frac{16 \pi (n-2)\omega^{\frac{1}{n-1}}_{n-1}S^{\frac{1}{n-1}}}{4^{\frac{n-2}{n-1}}\left[ n(4S)^{\frac{2}{n-1}}-(n-2)\omega^{\frac{2}{n-1}}_{n-1}(1-\frac{2n-4}{n-1}\Phi^{2})\right]}\label{alphan} \\
k_T &=& \frac{\left[ n(4S)^{\frac{2}{n-1}}-(n-2)\omega^{\frac{2}{n-1}}_{n-1}(1-\frac{(4n-6)(n-2)}{n-1}\Phi^{2})\right]}{\Phi \left[ n(4S)^{\frac{2}{n-1}}-(n-2)\omega^{\frac{2}{n-1}}_{n-1}(1-\frac{2n-4}{n-1}\Phi^{2})\right]}.\label{knt}
\end{eqnarray}

Observe that the denominators of $ C_{\Phi} $ (\ref{cphi}), $ \alpha $ (\ref{alphan}), and $ k_T $ (\ref{knt}) are all identical. Hence $C_{\Phi}$, $ \alpha $ and $ k_T $ diverge at the critical point since, as already discussed, this point just corresponds to the vanishing of the denominator in $ C_{\Phi} $. However as shown below, the R.H.S. of (\ref{ehf1}) and (\ref{ehf2}) which involves the ratio of these quantities, remain finite at the critical point.

Using (\ref{Tn}), the L.H.S. of the first \textit{Ehrenfest's} equation (\ref{ehf1}) may be found as (at the critical point $S_0$)  
\begin{equation}
-\left[ \left(\frac{\partial \Phi}{\partial T}\right)_{S}\right] _{S=S_0}= \frac{\pi (n-1)}{\Phi(n-2)^2}\left( \frac{4S_0}{\omega_{n-1}}\right) ^{1/(n-1)}.
\label{eh1}
\end{equation} 
In order to calculate the right hand side, we first note that
\begin{equation}
Q\alpha=(\partial Q/\partial T)_{\Phi}=(\partial Q/ \partial S)_{\Phi} (C_{\Phi}/T).
\end{equation}
Therefore the R.H.S. of (\ref{ehf1}) becomes,
\begin{equation}
\frac{\Delta C_{\Phi}}{T_0 Q \Delta \alpha}=\left[ \left( \frac{\partial S}{\partial Q}\right) _{\Phi}\right] _{S=S_0}.
\label{Eq2}
\end{equation} 
Using (\ref{phin1}) we calculate the R.H.S. of the above equation and obtain,
\begin{equation}
\frac{\Delta C_{\Phi}}{T_0 Q \Delta \alpha}
=\frac{\pi (n-1)}{\Phi(n-2)^2}\left( \frac{4S_0}{\omega_{n-1}}\right) ^{1/(n-1)}.
\label{Eh2}
\end{equation}
Equations (\ref{eh1}) and (\ref{Eh2}) clearly reveal the validity of the first \textit{Ehrenfest's} equation. Remarkably we find that the divergence in $C_{\Phi}$ is canceled with that of $\alpha$ in the first equation.

In order to evaluate the L.H.S. of (\ref{ehf2}) we first note that, since $ T=T(S,\Phi) $, therefore,
\begin{equation}
\left(\frac{ \partial T}{\partial \Phi}\right) _{Q}=\left(\frac{ \partial T}{\partial S}\right) _{\Phi}\left(\frac{ \partial S}{\partial \Phi}\right) _{Q}+\left(\frac{ \partial T}{\partial \Phi}\right) _{S}.
\label{Eq1}
\end{equation} 
Since specific heat $ (C_{\Phi}) $ diverges at the critical point ($ S_{0} $) therefore it is evident from (\ref{cphi}) that $\left[\left(\frac{\partial T}{\partial S}\right)_{\Phi}\right]_{S = S_{0}} = 0$. Also from (\ref{phin1}) we find that, $ \left(\frac{ \partial S}{\partial \Phi}\right) _{Q} $ has a finite value when evaluated at $ S=S_{0} $. Therefore the first term on the R.H.S. of (\ref{Eq1}) vanishes. This is a very special thermodynamic feature of AdS black holes, which may not be true for other systems. Therefore under this circumstances we obtain,
 \begin{equation}
 -\left[ \left(\frac{\partial \Phi}{\partial T}\right)_{Q}\right] _{S=S_0}=-\left[ \left(\frac{\partial \Phi}{\partial T}\right)_{S}\right] _{S=S_0}.
 \label{Eq3}
 \end{equation}
Using the thermodynamic identity,
\begin{equation}
\left(\frac{\partial Q}{\partial \Phi}\right)_{T}\left(\frac{\partial \Phi}{\partial T}\right)_{Q}\left(\frac{\partial T}{\partial Q}\right)_{\Phi}=-1
\end{equation}
we find, 
\begin{equation}
Qk_T=\left(\frac{\partial Q}{\partial \Phi}\right)_{T}=-\left(\frac{\partial T}{\partial \Phi}\right)_{Q}\left(\frac{\partial Q}{\partial T}\right)_{\Phi}=-\left(\frac{\partial T}{\partial \Phi}\right)_{Q} Q\alpha.
\end{equation}
Therefore R.H.S. of (\ref{ehf2}) may be found as
\begin{equation}
\frac{\Delta \alpha}{\Delta k_T}=-\left[ \left(\frac{\partial \Phi}{\partial T}\right)_{Q}\right] _{S=S_0}.
\label{Eq4}
\end{equation}
This indeed shows the validity of the second\textit{ Ehrenfest's} equation. Also we find that divergences in $\alpha$ and $k_T$  get canceled in the second equation like in the previous case. 
Using (\ref{ehf1}), (\ref{Eq3}) and (\ref{Eq4}) the Prigogine- Defay (PD) ratio $(\Pi) $ \cite{ref112}-\cite{ref113} may be found to be,
\begin{equation}
\Pi=\frac{\Delta C_{\Phi} \Delta k_T}{T_0 Q (\Delta\alpha)^{2}}=1.
\end{equation} 
Hence the phase transition occurring at $ T=T_0 $  is a \textit{second} order equilibrium transition \cite{ref108, ref109},\cite{ref111, ref112}. This is true in spite of the fact that the phase transition curves are smeared and divergent near the critical point.

\section{Kerr AdS black hole}
In this section we discuss the phase transition for $ (3+1) $ dimensional rotating (Kerr) AdS black holes under the framework of \textit{Ehrenfest's} equations.
For the Kerr-AdS black hole one can easily perform a similar analysis and describe phase transition for that case. The graphical analysis shown above for RN-AdS black hole physically remains unchanged for the Kerr-AdS case. In particular one can calculate the inverse temperature, 
\begin{eqnarray}
\beta^{Kerr}=\frac{4\pi^{\frac{3}{2}}{[S(\pi+S)(\pi+S-S\Omega^2)]^{1/2}}}{\pi^2-2\pi S(\Omega^2-2)-3S^2(\Omega^2-1)}
\label{temp}
\end{eqnarray}
and the condition $\left[ \partial\beta^{Kerr}/\partial r_{+}\right] _{\Omega}=0$ gives, 
\begin{eqnarray}
(\pi+S)^3(3S-\pi)-6S^2(\pi+S)^2\Omega^2+S^3(4\pi+3S)\Omega^4=0.
\label{betm}
\end{eqnarray}
By substituting the positive solution of this polynomial in (\ref{temp}) one finds the minimum temperature $T^{Kerr}_0$. Furthermore for the Kerr-AdS black hole the heat capacity diverges exactly where (\ref{betm}) holds \cite{ref32} which corresponds to the minimum temperature ($T^{Kerr}_{0}$). In the irrotational limit $T^{Kerr}_0$ reproduces the known result of \cite{ref16}. Results for $M_0$ and $T_1$ can also be calculated by a similar procedure. These expressions are rather lengthy and hence omitted. Moreover there are no new insights that have not already been discussed in the RN-AdS example.

\textit{Ehrenfest's} set of equations for Kerr-AdS black hole are given by \cite{ref32},
\begin{eqnarray}
-\left(\frac{\partial \Omega}{\partial T}\right)_{S} &=& \frac{C_{\Omega_2}-C_{\Omega_1}}{TJ(\alpha_2-\alpha_1)}\label{ehrk1}\\
-\left(\frac{\partial \Omega}{\partial T}\right)_{J} &=& \frac{\alpha_{2}-\alpha_{1}}{k_{T_{2}}-k_{T_{1}}}.\label{ehrk2}
\end{eqnarray}
The expressions for specific heat ($C_{\Omega}$), analog of the volume expansion coefficient ($\alpha$) and compressibility ($k_T$) are all provided in \cite{ref32}. Once again, they all have the same denominator (like the corresponding case for RN-AdS). Considering the explicit expressions given in \cite{ref32} and using the same techniques we find that both sides of (\ref{ehrk1}) and (\ref{ehrk2}) lead to an identical result, given by, 
\begin{equation}
l.h.s=r.h.s=\frac{4\pi^{\frac{3}{2}}(\pi+S_0-S_0\Omega^2)^{\frac{3}{2}}(\pi+S_0)^{\frac{1}{2}}}{\sqrt{S_0}\Omega[3(\pi+S_0)^2-S_0\Omega^2(2\pi+3S_0)]} \label{kerr}.
\end{equation}
 This shows that the phase transition for Kerr-AdS black hole is also \textit{second} order. 
\section{Discussions}
In this chapter, based on standard thermodynamics, we have systematically developed an approach to analyze the phase transition phenomena in Reissner Nordstrom AdS black holes in arbitrary dimensions. In suitable limits the results for the Schwarzschild AdS black hole are obtained. This phase transition is characterized by divergences in specific heat, volume expansivity and compressibility near the critical point\footnote{It is important to point out the difference between the phase transition we discuss in this chapter to that with the conventional \textit{Hawking-Page} (HP) phase transition. Our analysis is strictly confined to the critical temperature ($ T_0 $) where the specific heat ($ C_{\Phi} $) acquires an infinite divergence. On the other hand \textit{Hawking-Page} transition occurs at a temperature $ T_1 $ ($ T_1> T_0  $) where Gibbs free energy ($G$) changes its sign \cite{ref26}-\cite{ref27}. While the HP transition is \textit{first} order, the one is considered here is \textit{second} order.}.  Furthermore, within the \textit{gibbsian} approach, and using a grand canonical ensemble, we have provided a unique way (which is valid in any dimension greater than two) to analyze both the quantitative and qualitative features of this phase transition. From our analysis it is clear that such a transition takes place from a black hole with smaller mass ($ M<M_0 $) to a black hole with larger mass ($ M>M_0 $), where $ M_0 $ denotes the critical mass. Also, we have found that the black hole with smaller mass falls in an unstable phase since it has negative specific heat, whereas that with the larger mass possesses a positive specific heat and thereby corresponds to a stable phase. We have provided completely generalized expressions for the critical temperature ($ T_0 $) and critical mass ($ M_0 $) that characterize this phase transition. Also, we have explicitly calculated the temperature $ T_1 $ ($ T_1>T_0 $) which is associated with the change in sign in Gibb's free energy ($ G $). Following our approach, we have established a universal relation (\ref{univ}) between $ T_0 $ and $ T_1 $ that is valid in any arbitrary dimension ($ n>2 $). 

Finally and most importantly, employing the \textit{Ehrenfest's} scheme we have resolved the vexing issue regarding the nature of phase transition in Reissner Nordstrom AdS black holes. In order to do that we have analytically checked the validity of both the \textit{Ehrenfest's} equations near the critical point. This clearly suggests that the phase transition that is associated with the divergence in the heat capacity at the critical point ($ T_0 $) is indeed a \textit{second} order equilibrium transition. A similar analysis for the $ (3+1) $ dimensional rotating (Kerr) AdS black holes has been performed as well and results identical to the charged (RN) AdS case are obtained.

From the entire analysis of this chapter it is indeed quite evident that during the black hole phase transition one should actually encounter some sort of singularity in various thermodynamic entities (like, heat capacity, isothermal compressibility etc.). 
According to the basic principles of statistical mechanics such a divergence could be regarded as the manifestation of certain diverging \textit{correlation length} near the critical point of the phase transition curve. Therefore in order to further advance the study of the phase transition, one should check the \textit{scaling} behavior of black holes near the critical point and find out the corresponding \textit{universality class}.  We shall discuss this issue in the next chapter.

\chapter{Critical Phenomena In Charged AdS Black Holes}

\section{Critical phenomena and black holes}

In ordinary thermodynamics, a phase transition occurs whenever there is a singularity in the free energy or one of its derivatives. The corresponding point of discontinuity is known as the critical point of phase transition. In \textit{Ehrenfest's} classification of phase transition, the order of the phase transition is characterized by the order of the derivative of the free energy that suffers discontinuity at the critical points. For example, if the \textit{first} derivative of the free energy is discontinuous then the corresponding phase transition is \textit{first} order in nature. On the other hand, if the \textit{first} derivatives are continuous and \textit{second} derivatives are discontinuous then the transition may be referred as higher order or \textit{continuous}.  These type of transitions correspond to a divergent heat capacity, an infinite correlation length, and a power law decay of correlations near the critical point. 

The primary aim of the theory of phase transition is to study the singular behavior of various thermodynamic entities (for example heat capacity) near the critical point. It order to do that one often expresses these singularities in terms of power laws characterized by a set of \textit{static} critical exponents \cite{ref48}-\cite{stanley}. Generally these exponents depend on a few parameters of the system, like, (1) the spatial dimensionality ($ d $) of the space in which the system is embedded, (2) the range of interactions in the system etc. To be more specific, it is observed that for systems possessing short range interactions, these exponents depend on the spatial dimensionality ($ d $). On the other hand for systems with long range interactions these exponents become independent of $ d $, which is the basic characteristic of a \textit{mean field theory} and is observed for the case $ d>4 $ in usual systems. It is interesting to note that the (static) critical exponents are also found to satisfy so called \textit{thermodynamic scaling laws} \cite{ref48}-\cite{stanley} which apply to a wide variety of thermodynamical systems, from elementary particles to turbulent fluid flow. Such studies have also been performed, albeit partially, in the context of black holes \cite{ref49}-\cite{ref53}. In this chapter we aim to provide a detailed analysis of these issues and show how the study of critical phenomena in black holes is integrated with the corresponding studies in other areas of physics adopting a \textit{mean field} approximation.

The goal of the present chapter is to carry out a detailed \textit{analytic} computation in order to explore the \textit{scaling} behavior of black holes near the critical point(s)\footnote{The critical points of the phase transition are characterized by the discontinuities in the heat capacity at constant charge $ (C_Q) $.} of the phase transition curve in various theories of gravity. To do that, in our analysis we consider examples both from the usual \textit{Einstein} gravity as well as from the \textit{Ho\v{r}ava-Lifshitz} theory of gravity. This eventually helps us to make a systematic comparison between the \textit{scaling} behavior of black holes in various theories of gravity as well as to identify the corresponding \textit{universality class}. 

The entire content of this chapter is based on the papers \cite{ref99}-\cite{ref101} and could be divided mainly into two parts. In the first part, based on a \textit{canonical} framework, we aim to study the critical behavior of charged black holes taking the particular example of Born-Infeld AdS (BI AdS) black holes in $ (n+1) $ dimensions. Results obtained in the above case smoothly translate to that of the $ (n+1) $ dimensional RN AdS case in the appropriate limit. We compute the \textit{static} critical exponents associated with this phase transition and verify the \textit{scaling} laws. Interestingly enough these laws are found to be compatible with the \textit{static scaling hypothesis} of usual statistical systems \cite{ref48}-\cite{stanley}. 

In the remaining part of this chapter we make a detail analysis on the phase transition and \textit{scaling} behavior of  charged topological black holes in the \textit{Ho\v{r}ava-Lifshitz} theory of gravity at the \textit{Lifshitz} point $z=3$. 
Based on a \textit{canonical} framework (i.e; keeping the charge ($ Q $) of the black hole fixed \cite{ref28}), we investigate the critical behavior of topological charged black holes in \textit{Ho\v{r}ava-Lifshitz} theory of gravity at the \textit{Lifshitz} point $z=3$. The black hole solution was given in \cite{ref114}. Although all the thermodynamic quantities were evaluated earlier \cite{ref114}, a detailed study of the nature of phase transition is still lacking. Particularly the issue regarding the \textit{scaling behavior} of (charged) black holes has never been investigated so far in the framework of \textit{Ho\v{r}ava-Lifshitz} theory of gravity.  This essentially gives us enough motivation to carry out a systematic  analysis of the phase transition as well as the \textit{scaling} behavior of topological black holes in the \textit{Ho\v{r}ava-Lifshitz} theory of gravity.  The present thesis therefore aims to explore the phase transition phenomena considering all the three cases taking $ k=0,\pm 1 $. We observe the following interesting features:
\vskip 1mm
\noindent
$\bullet$ There is no \textit{Hawking Page} transition \cite{ref16} for black hole with $ k=0,1 $.
\vskip 1mm
\noindent
$\bullet$  For $k=-1$, there is a upper bound in the value of the event horizon, above which the temperature becomes negative. This indicates that above this critical value of horizon, the black hole solution does not exist. We will call this valid range as the {\it physical region}. 
\vskip 1mm
\noindent
$\bullet$ Within the physical region, interesting phase structure could be observed for the \textit{hyperbolic} charged black holes ($ k=-1 $). In this particular case we observe the \textit{second} order transition. This is {\it different} from the usual one. In \textit{Einstein} gravity, the \textit{Hawking Page} transition \textit{mostly} occurs for the $k=1$ case, while \textit{usually} there is no such transition for the $k=0,-1$ case \cite{ref26,brimhm}.
\vskip 1mm
\noindent

Finally, we explicitly calculate all the \textit{static critical exponents} associated with the \textit{second} order transition, and check the validity of \textit{thermodynamic scaling laws} near the critical point. Interestingly enough it is found that the \textit{hyperbolic} charged black holes ($ k = -1 $) in the \textit{Ho\v{r}ava-Lifshitz} theory of gravity fall under the same \textit{universality class} as the black holes having spherically symmetric topology ($ k = 1 $) in the usual \textit{Einstein} gravity. The values of these critical exponents indeed suggest a universal \textit{mean field} behavior in black holes which is valid in both  \textit{Einstein} as well as \textit{Ho\v{r}ava-Lifshitz} theory of gravity.

The plan of the chapter is as follows: In section 3.2, we make a qualitative as well as quantitative analysis of the various thermodynamic entities for $ (n+1) $ dimensional BI AdS black holes in order to have a meaningful discussion on their critical behavior in the subsequent sections. In section 3.3, we explicitly compute the \textit{static} critical exponents in order to check the validity of the \textit{scaling laws} both for the BI AdS and RN AdS black holes in $ (n+1) $ dimensions. We begin in section 3.4 by giving a brief introduction of the charged topological black hole solutions in \textit{Ho\v{r}ava-Lifshitz} gravity and use it in section 3.5 to study the different thermodynamic quantities as well as the phase transition. In section 3.6, the critical exponents near the critical point(s) are being evaluated and a brief discussion on the validity of the ordinary scaling laws is presented.  Finally, we conclude in section 3.7.


\section{Charged black holes in higher dimensional AdS space}

For the past couple of decades gravity theories with Born-Infeld (BI) action have garnered considerable attention due to its several remarkable features. For example, Born-Infeld type effective actions, which arise naturally in open super strings and D branes are free from physical singularities \cite{ref115}. Also, using the Born Infeld action one can in fact study various thermodynamic features of Reissner Nordstrom AdS black holes in a suitable limit.  During the last ten years many attempts have been made in order to understand the  thermodynamics of Born-Infeld black holes in AdS space \cite{ref116}. In spite of these efforts, several significant issues remain unanswered. Studying the critical phenomena is one of them. In order to have a deeper insight regarding the underlying phase structure for these black holes  one needs to compute the critical exponents associated with the phase transition and check the validity of the  \textit{scaling laws} near the critical point. A similar remark also holds for the higher dimensional Reissner Nordstrom AdS (RN AdS) black holes.  

In the present thesis therefore we aim to fill up this gap step by step. In order to do that we first compute the essential thermodynamic entities both for the BI AdS and RN AdS black holes which will be required in the next step to calculate the critical exponents for these black holes.

The action for the \textit{Einstein- Born-Infeld} gravity in $ (n+1) $ dimensions $ (n\geq 3) $ is given by \cite{ref116},
\begin{equation}
S= \int d^{n+1}x\sqrt{-g}\left[ \frac{R-2\Lambda}{16\pi G}+L(F)\right] 
\end{equation}
where,
\begin{equation}
L(F)=\frac{b^{2}}{4\pi G}\left( 1-\sqrt{1+\frac{2F}{b^{2}}}\right) 
\end{equation}
with $ F=\frac{1}{4} F_{\mu\nu}F^{\mu\nu} $. Here  $ b $ is the Born-Infeld parameter with the dimension of mass and $ \Lambda(=-n(n-1)/2l^2) $ is the cosmological constant. It is also to be noted that for the rest of our analysis we set Newton's constant $ G=1 $.   

By solving the equations of motion, the Born-Infeld anti de sitter (BI AdS) solution may be found as,
\begin{equation}
ds^2 = -\chi dt^2+\chi ^{-1}dr^2+r^2 d\Omega^{2}\label{metricn}
\end{equation} 
where,
\begin{eqnarray}
\chi(r) = 1-\frac{m}{r^{n-2}}+\left[\frac{4b^{2}}{n(n-1)} +\frac{1}{l^{2}}\right]r^{2} -\frac{2\sqrt{2}b}{n(n-1)r^{n-3}}\sqrt{2b^{2}r^{2n-2}+(n-1)(n-2)q^{2}}\nonumber\\
+\frac{2(n-1)q^{2}}{nr^{2n-4}}H\left[\frac{n-2}{2n-2},\frac{1}{2},\frac{3n-4}{2n-2},-\frac{(n-1)(n-2)q^{2}}{2b^{2}r^{2n-2}} \right] 
\label{chi}
\end{eqnarray}
and $ H $ is a \textit{hyper-geometric function} \cite{ref117}. 

It is interesting to note that for $ n\geq 3 $ the metric (\ref{chi}) has a curvature singularity at $ r=0 $. It is in fact possible to show that this singularity is hidden behind the event horizon(s) whose location may be obtained through the condition $\chi(r) = 0 $. Therefore the above solution (\ref{metricn}) could be interpreted as a space time with black holes \cite{ref116}.

In the limit $ b\rightarrow\infty $ and $ Q\neq0 $ one obtains the corresponding solution for Reissner Nordstrom (RN) AdS black holes. Clearly this is a nonlinear generalization of the RN AdS black holes. Here $ m $ is related to the mass ($ M $) of the black hole as \cite{ref116}, 
\begin{eqnarray}
M=\frac{(n-1)}{16\pi}\omega_{n-1}m\nonumber\\
\label{mq}
\end{eqnarray}
where $ \omega_{n-1}\left(=\frac{2\pi^{n/2}}{\Gamma(n/2)}\right) $ is the volume of the unit $(n-1)$ sphere. Identical expression could also be found for the RN AdS case \cite{ref28}.

Electric charge ($ Q $) may defined as \cite{ref116},
\begin{equation}
Q=\frac{1}{4\pi}\int \ast F d\Omega 
\end{equation}
which finally yields,
\begin{equation}
Q=\frac{\sqrt{(n-1)(n-2)}}{4\pi\sqrt{2}}\omega_{n-1}q\label{charge}.
\end{equation}

 Using (\ref{mq}) one can rewrite (\ref{chi}) as,
\begin{eqnarray}
\chi(r) = 1-\frac{16\pi M}{(n-1)\omega_{n-1}r^{n-2}}+\frac{r^{2}}{l^{2}}+\frac{4b^{2}r^{2}}{n(n-1)}\left[ 1-\sqrt{1+\frac{16\pi^{2}Q^{2}}{b^{2}r^{2(n-1)}\omega_{n-1}^{2}}}\right] \nonumber\\
+\frac{64\pi^{2}Q^{2}}{n(n-2)r^{2n-4}\omega_{n-1}^{2}}H\left[\frac{n-2}{2n-2},\frac{1}{2},\frac{3n-4}{2n-2},-\frac{16\pi^{2}Q^{2}}{b^{2}r^{2n-2}\omega_{n-1}^{2}} \right] 
\label{chi1}.
\end{eqnarray}

In order to obtain an explicit expression for the mass ($ M $) of the black hole we set $\chi(r_{+})=0$, which yields, $ (G=1) $
\begin{eqnarray}
M= \frac{(n-1)\omega_{n-1}r_{+}^{n-2}}{16\pi}+\frac{\omega_{n-1}(n-1)r_{+}^{n}}{16\pi l^{2}} + \frac{b^{2}r_{+}^{n}\omega_{n-1}}{4\pi n}\left[ 1-\sqrt{1+\frac{16\pi^{2}Q^{2}}{b^{2}r_{+}^{2(n-1)}\omega_{n-1}^{2}}}\right]\nonumber\\
+\frac{4\pi Q^{2}(n-1)}{n(n-2)\omega_{n-1}r_{+}^{n-2}}\left[ 1-\frac{8\pi^{2}Q^{2}(n-2)}{b^{2}(3n-4)r_{+}^{2n-2}\omega_{n-1}^{2}}\right]+ O(1/b^{4}) 
\label{M}
\end{eqnarray}
where $ r_{+} $ is the radius of the outer event horizon. Here the parameter $ b $ is chosen in such a way so that $ \frac{1}{b^{2}}\ll 1 $ or, in other words our analysis is carried out in the large $ b $ limit. Also, such a limit is necessary for abstracting the results for RN AdS black holes obtained by taking $ b\rightarrow\infty $. Therefore all the higher order terms from $ (1/b^{2})^{2} $ onwards have been dropped out from the series expansion of $ H\left[\frac{n-2}{2n-2},\frac{1}{2},\frac{3n-4}{2n-2},-\frac{16\pi^{2}Q^{2}}{b^{2}r^{2n-2}\omega_{n-1}^{2}} \right] $. Using (\ref{charge}) one can express the electrostatic potential difference ($ \Phi $) between the horizon and infinity as,
\begin{eqnarray}
\Phi &=&\sqrt{\frac{n-1}{2n-4}}\frac{q}{r_{+}^{n-2}}H\left[\frac{n-2}{2n-2},\frac{1}{2},\frac{3n-4}{2n-2},-\frac{(n-1)(n-2)q^{2}}{2b^{2}r^{2n-2}} \right] \nonumber\\
&=&  \frac{4\pi Q}{(n-2)\omega_{n-1}r_{+}^{n-2}}\left[ 1-\frac{8\pi^{2}Q^{2}(n-2)}{b^{2}(3n-4)r_{+}^{2n-2}\omega_{n-1}^{2}}\right]+ O(1/b^{4})
\label{Phi}
\end{eqnarray}
where $ Q $ is the electric charge. Henceforth all our results are valid upto $ O(1/b^{2}) $ only.

Using (\ref{chi1}) and (\ref{M}), the Hawking temperature may be obtained as,
\begin{eqnarray}
T&=& \frac{\chi^{'}(r_{+})}{4\pi}\nonumber\\
&=&\frac{1}{4\pi}\left[ \frac{n-2}{r_{+}}+\frac{nr_{+}}{l^{2}}+\frac{4b^{2}r_{+}}{n-1}\left( 1-\sqrt{1+\frac{16\pi^{2}Q^{2}}{b^{2}r_+^{2(n-1)}\omega_{n-1}^{2}}}\right)   \right] \label{T}.
\end{eqnarray}
In the appropriate limit ($ b\rightarrow\infty $) the corresponding expression of Hawking temperature for the RN AdS black hole may be obtained as,
\begin{eqnarray}
T_{RN AdS}=\frac{1}{4\pi}\left[ \frac{n-2}{r_{+}}+\frac{nr_{+}}{l^{2}}-\frac{32\pi^{2}Q^{2}}{(n-1)r_{+}^{2n-3}\omega_{n-1}^{2}} \right] \label{Trn}.
\end{eqnarray}

Using (\ref{M}) and (\ref{T}) the entropy of the BI AdS black hole may be found as,
\begin{eqnarray}
S=\int T^{-1}\left( \frac{\partial M}{\partial r_{+}}\right)_{Q} dr_{+}= \frac{\omega_{n-1}r_{+}^{n-1}}{4}\label{S}. 
\end{eqnarray}
It is interesting to note that identical expression could also be found for the RN AdS case \cite{ref28}. 

In order to investigate the critical phenomena, it is necessary to compute the heat capacity at constant charge ($ C_{Q} $). Using (\ref{T}) and (\ref{S}) the specific heat (at constant charge) may be found as,
\begin{eqnarray}
C_{Q}=T\left(\frac{\partial S}{\partial T} \right)_{Q} = T \frac{\left( \partial S/\partial r_{+}\right)_{Q}}{\left( \partial T/\partial r_{+}\right)_{Q}}=\frac{\Im(r_+,Q)}{\Re(r_+,Q)}\label{CQ}
\end{eqnarray}
where,
\begin{eqnarray}
\Im(r_+,Q)=\frac{(n-1)\omega_{n-1}r_+^{3n-7}}{4}\sqrt{1+\frac{16\pi^{2}Q^{2}}{b^{2}r_+^{2(n-1)}\omega_{n-1}^{2}}}\nonumber\\
\times \left[(n-2)r_{+}^{2}+\frac{nr_{+}^{4}}{l^{2}}+
\frac{4b^{2}r_{+}^{4}}{n-1}\left( 1-\sqrt{1+\frac{16\pi^{2}Q^{2}}{b^{2}r_+^{2(n-1)}\omega_{n-1}^{2}}}\right)  \right] 
\end{eqnarray}
and,
\begin{eqnarray}
\Re(r_+,Q)= r_+^{2n-4}\left(\frac{nr_+^{2}}{l^{2}}-n+2 \right) \sqrt{1+\frac{16\pi^{2}Q^{2}}{b^{2}r_+^{2(n-1)}\omega_{n-1}^{2}}}+\left(\frac{n-2}{n-1} \right) \frac{64 \pi^{2}Q^{2}}{\omega_{n-1}^{2}} \nonumber\\
-\frac{4b^{2}r_{+}^{2n-2}}{n-1}\left( 1-\sqrt{1+\frac{16\pi^{2}Q^{2}}{b^{2}r_+^{2(n-1)}\omega_{n-1}^{2}}}\right)\label{d}.
\end{eqnarray}

From (\ref{CQ}) we note that in order to have a divergence in $ C_Q $ one must satisfy the following condition, 
\begin{eqnarray}
\Re(r_+,Q)= r_+^{2n-4}\left(\frac{nr_+^{2}}{l^{2}}-n+2 \right) \sqrt{1+\frac{16\pi^{2}Q^{2}}{b^{2}r_+^{2(n-1)}\omega_{n-1}^{2}}}+\left(\frac{n-2}{n-1} \right) \frac{64 \pi^{2}Q^{2}}{\omega_{n-1}^{2}} \nonumber\\
-\frac{4b^{2}r_{+}^{2n-2}}{n-1}\left( 1-\sqrt{1+\frac{16\pi^{2}Q^{2}}{b^{2}r_+^{2(n-1)}\omega_{n-1}^{2}}}\right)=0.\label{root}
\end{eqnarray}

Although it is quite difficult to solve (\ref{root}) analytically, however one can attempt to solve this equation numerically. In order to do that it is first necessary to fix the parameters ($ b $ and $ Q $) of the theory. In this regard it is always possible to give certain plausibility arguments that give a bound on the parameter space in order to have real positive roots for (\ref{root}). The boundedness of the parameter space could be achieved by demanding that a smooth \textit{extremal} limit holds. In other words, we choose the parameters in such a way so that an \textit{extremal} black hole could be found in the appropriate limit. Furthermore, once the choice of parameters has been determined in this manner, it is easy to show that meaningful results are obtained in the \textit{non extremal} case iff one is confined to this choice. 

Let us consider the \textit{extremal} BI AdS black holes. Here both $ \chi(r) $ and $ \frac{d\chi}{dr} $ vanish at the degenerate horizon ($ r_e $). From the above two conditions and using (\ref{M}) we arrive at the following equation,  
\begin{equation}
1+\left(\frac{4b^{2}}{n-1} +\frac{n}{l^{2}}\right)\frac{r_{e}^{2}}{n-2}-\frac{4b^{2}r_{e}^{2}}{(n-1)(n-2)}\sqrt{1+\frac{16\pi^{2}Q^{2}}{b^{2}r_e^{2(n-1)}\omega_{n-1}^{2}}} =0.\label{extm}
\end{equation}

In the following we discuss solutions of (\ref{extm}) corresponding to different choices in $ n $. 

\begin{center}
(I)\textbf{\textit{ The $ n=3 $ case :}}
\end{center}

For $ n = 3 $, the equation (\ref{extm}) takes the following form,
\begin{equation}
1+ \left(2b^{2}+\frac{3}{l^{2}} \right)r^{2}_{e} -2b^{2}\sqrt{r^{4}_{e}+\frac{Q^{2}}{b^{2}}}=0
\end{equation}   
whose solution is,
\begin{equation}
r^{2}_{e}=\frac{l^{2}}{6}\left( \frac{1+\frac{3}{2b^{2}l^{2}}}{1+\frac{1}{b^{2}l^{2}}}\right) \left[ -1+\sqrt{1+\frac{12(1+\frac{3}{4b^{2}l^{2}})}{b^{2}l^{2}(1+\frac{3}{4b^{2}l^{2}})^{2}}\left(b^{2}Q^{2}-\frac{1}{4} \right)} \right]. 
\end{equation}
In order to have a real root we must have $ bQ \geq 0.5 $. The \textit{forbidden} region for BI AdS black holes is thus given by $ 0\leq bQ <0.5 $ while the allowed region is,
\begin{equation}
0.5\leq bQ\leq\infty.\label{ext}
\end{equation}

Requiring that a smooth \textit{extremal} limit holds, we choose the parameter space satisfying the condition (\ref{ext}) in order to find out the roots of the equation (\ref{root}). Numerical analysis reveals that now there are two positive and two negative roots for $ r_+ $. We therefore pursue our analysis subject to the
condition (\ref{ext}).

\begin{subtables}
\begin{table}[htb]
\caption{Roots of equation (\ref{root}) for $ n=3 $, $ b=10 $ and $ l=10 $}   
\centering                          
\begin{tabular}{c c c c c c c}            
\hline\hline                        
$Q$ & $r_1$  & $r_2$ &$r_3$ & $r_4 $ &  \\ [0.05ex]
\hline
0.5 & 0.875061 & 5.70663 & -0.875061 & -5.70663 \\
0.4 & 0.696540 & 5.73116 & -0.696540 & -5.73116 \\                              
0.3 & 0.519874 & 5.74988 & -0.519874 & -5.74988 \\
0.2 & 0.344194 & 5.76306 & -0.344194 & -5.76306 \\ 
0.1 & 0.167309 & 5.77090 & -0.167309 & -5.77090 \\ 
0.05& 0.0707235& 5.77285 & -0.0707235& -5.77285 \\ [0.05ex]         
\hline                              
\end{tabular}\label{E1}  
\end{table}
\begin{table}[htb]
\caption{Roots of equation (\ref{root}) for $ n=3 $, $ b=15 $ and $ l=10 $}   
\centering                          
\begin{tabular}{c c c c c c c}            
\hline\hline                        
$Q$ & $r_1$  & $r_2$ &$r_3$ & $r_4 $ &  \\ [0.05ex]
\hline
0.5 & 0.875680 & 5.70663 & -0.875680 & -5.70663 \\
0.4 & 0.697319 & 5.73116 & -0.697319 & -5.73116 \\                              
0.3 & 0.520919 & 5.74988 & -0.520919 & -5.74988 \\
0.2 & 0.345784 & 5.76306 & -0.345784 & -5.76306 \\ 
0.1 & 0.170722 & 5.77090 & -0.170722 & -5.77090 \\ [0.05ex]         
\hline                              
\end{tabular}
\label{E2}          
\end{table}
\begin{table}[htb]
\caption{Roots of equation (\ref{root}) for $ n=3 $, $ b=20 $ and $ l=10 $}   
\centering                          
\begin{tabular}{c c c c c c c}            
\hline\hline                        
$Q$ & $r_1$  & $r_2$ &$r_3$ & $r_4 $ &  \\ [0.05ex]
\hline
0.5 & 0.875896 & 5.70663 & -0.875896 & -5.70663 \\
0.4 & 0.697590 & 5.73116 & -0.697590 & -5.73116 \\                              
0.3 & 0.521283 & 5.74988 & -0.521283 & -5.74988 \\
0.2 & 0.346335 & 5.76306 & -0.346335 & -5.76306 \\ 
0.1 & 0.171859 & 5.77090 & -0.171859 & -5.77090 \\ [0.05ex]         
\hline                              
\end{tabular}
\label{E3}          
\end{table}
\end{subtables}

In Tables (3.1a), (\ref{E2}) and (\ref{E3}) we give the solution for (\ref{root}) obtained by numerical methods subject to the condition (\ref{ext}). The value $ bQ=0.5 $ when the bound is saturated is also considered (last entry in Table ($ 3.1a $)). From the numerical solutions it is quite evident that, for a general choice of parameters ($ Q $ and $ b $), equation (\ref{root}) possesses four distinct real roots implying that the denominator of (\ref{CQ}) has only simple poles. Among these four roots there are two positive and two negative roots. The phase transition occurs at the two positive roots ($ r_1 $ and $ r_2 $). This fact has also been depicted in various Figures ((3.1) and (3.2)) for different sets of parameters. For example, for the specific choice of $ Q=0.5 $, $ b=10 $ (and $ l=10 $), the roots of the equation (\ref{root}) are $ \pm 0.875061$ and $\pm 5.70663$.  We take only the two positive roots ($ r_1=0.875061 $, $ r_2=5.70663 $) which are also marked in the Figure 3.1.

\begin{figure}[h]
\centering
\includegraphics[angle=0,width=8cm,keepaspectratio]{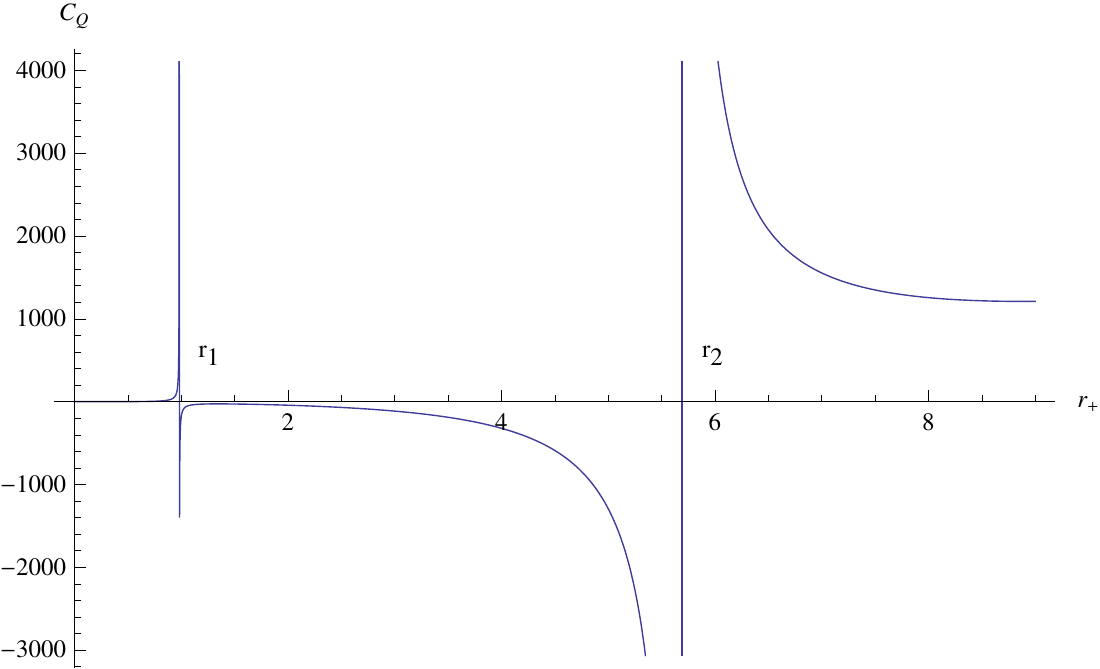}
\caption[]{\it Specific heat plot ($ C_{Q} $) for Born-Infeld AdS black hole with respect to $r_{+}$ for $Q(=Q_c)=0.5$, $ n=3 $, $ b=10 $ and $ l=10 $. The discontinuities at $ r_1 $ and $ r_2 $ are shown.}
\label{figure 2a}
\end{figure} 
\begin{figure}[h]
\centering
\includegraphics[angle=0,width=8cm,keepaspectratio]{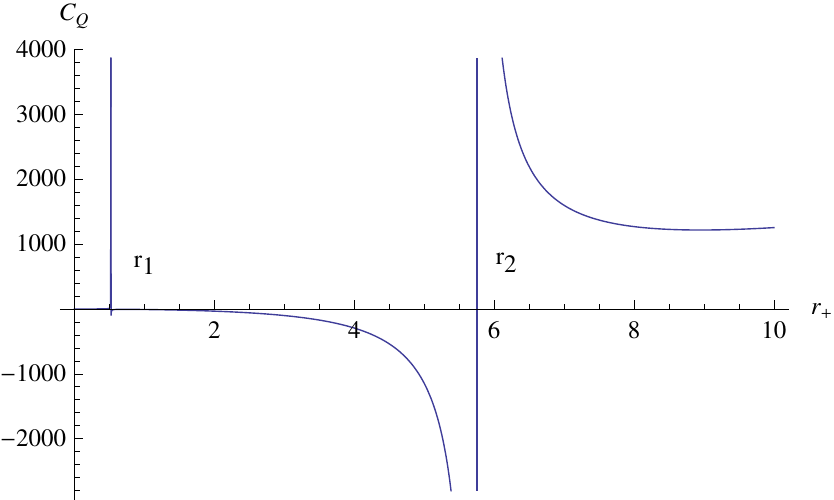}
\caption[]{\it Discontinuity of Specific heat ($ C_{Q} $) for Born-Infeld AdS black hole at $r_{+}= r_2$ for $Q(=Q_c)=0.3$, $ n =3 $, $ b=15 $ and $ l=10 $. The discontinuities at $ r_1 $ and $ r_2 $ are shown.}
\label{figure 2a}
\end{figure}
From Figures 3.1 and 3.2 we observe that $ C_{Q} $ suffers discontinuities at two different points, namely $ r_{1} $ and $ r_{2} $ (discussed in the previous paragraph), which may be identified as the critical points for the phase transition phenomena in BI AdS black holes. From the above figures it is evident that the heat capacity is positive for $ r_+ < r_1 $ and $ r_+ >r_2 $, while it is negative in the intermediate range $ r_1<r_+<r_2 $.  Since the  black hole with smaller mass possesses lesser entropy/horizon radius than the black hole with larger mass, therefore the point $ r_{1} $ corresponds to the critical point for the transition between a smaller mass black hole with positive specific heat ($ C_Q>0 $) to an intermediate unstable black hole with negative heat capacity ($ C_Q<0 $). On the other hand  $ r_{2} $ corresponds to a transition from the intermediate unstable black hole to a larger mass black hole with positive heat capacity ($ C_Q>0 $).  
\\ \\

\begin{center}
(II)\textbf{\textit{ The $ n=4 $ case :}}
\end{center}
For $ n=4 $, on the other hand, equation (\ref{extm}) turns out to be a cubic equation in the variable $ r_{e}^{2} $, which is indeed quite difficult to solve analytically. However, it is possible to solve equation (\ref{extm}) numerically for  $ r_{e}^{2} $. The solutions are provided below in a tabular form (Table 3.2) for various choice of parameters ($ Q $ and $ b $).
\begin{table}[htb]
\caption{Numerical solutions for $r_e^{2}$ for $ n=4 $ and $ l=10 $ (extremal case)}   
\centering                          
\begin{tabular}{c c c c c c c}            
\hline\hline                        
$Q$ & $b$  & $r_{e1}^{2}$ &$r_{e2}^{2}$ & $r_{e3}^{2} $   \\ [0.05ex]
\hline
0.5 & 10 & -0.187184 & 0.180352 & 50.7835 \\
0.5 & 1 & -102.266 & -0.776247 & 0.04254 \\                              
0.5 & 0.5 & -6.74452 & -4.46669 & 0.0112109 \\
0.05 & 0.05 & -0.542572-i 4.97467 & -0.542572+i 4.97467  & $1.12579\times 10^{-6}$ \\  [0.05ex]         
\hline                              
\end{tabular}\label{E1}  
\end{table}
From the roots of equation (\ref{extm}) it is quite evident that for $ n=4 $ there is as such no bound on the parameter space of BI AdS black holes as far as a smooth \textit{extremal} limit is concerned. It is also interesting to note that for $ bQ\leq 0.5 $ we have only one real positive root for $ r_e $, whereas for $ bQ>0.5 $ we have two real positive roots for $ r_e $. We are now in a position to find out the roots of the equation (\ref{root}) numerically for $ n=4 $ considering various values of the parameters ($ b $ and $ Q $), which are given below in Table 3.3. 
\begin{table}[htb]
\caption{Roots of the equation (\ref{root}) for $ n=4 $ and $ l=10 $(non extremal case)}   
\centering                          
\begin{tabular}{c c c c c c c c c c}            
\hline\hline                        
$Q$ & $b$  & $r_1$ &$r_2$ & $r_3 $ &$r_4$&$ r_{5,6} $ \\ [0.05ex]
\hline
0.5&10&0.6410&7.0708&-7.0708&-0.6410&$5.9376\times 10^{-17}\pm i 0.641$\\ 
0.5&1&0.4538&7.0708&-7.0708&-0.4538&$-3.6375\times 10^{-17}\pm i 0.735$\\
0.5&0.5&0.2139&7.0708&-7.0708&-0.2139&$3.2485\times 10^{-17}\pm i 0.875$\\
0.05&0.05&0.0021&7.0710&-7.0710&-0.0021&$ 3.3787\times10^{-17}\pm i0.860$\\
\\ [0.05ex]         
\hline                              
\end{tabular}
\label{E1}          
\end{table} 
Two crucial points are to be noted at this stage- (i) $ C_Q $ (\ref{CQ}) possesses only simple poles and (ii) there are always two real positive roots ($ r_1 $ and $ r_2 $) of (\ref{root}) for different choice of parameters. These two roots ($ r_1 $ and $ r_2 $) correspond to the critical points for the phase transition phenomena occurring in BI AdS black holes. For our detailed analysis we choose $ Q=0.5 $  and $ b=10 $, corresponding to the first row in Table 3.3. With this particular choice the critical points are found to be  $ r_1= 0.6410$ and $ r_2=7.0708 $, which are also depicted in various Figures (3.3 and 3.4).
\begin{figure}[h]
\centering
\includegraphics[angle=0,width=8cm,keepaspectratio]{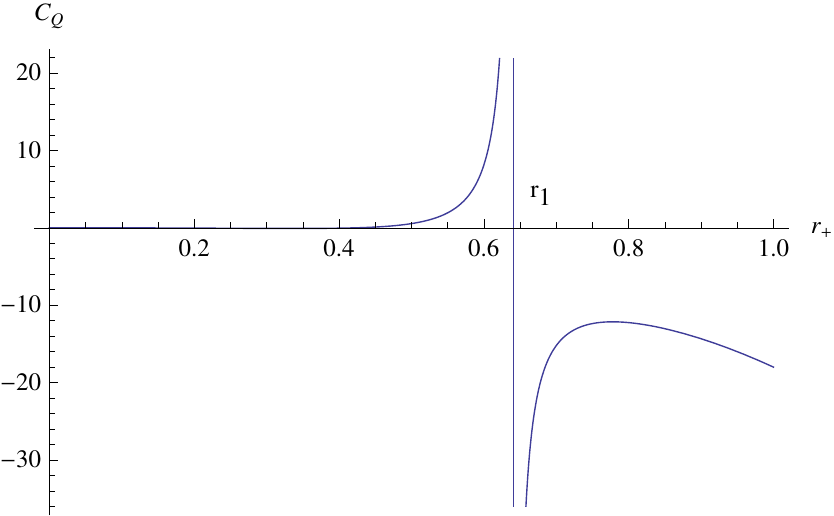}
\caption[]{\it Discontinuity of Specific heat ($ C_{Q} $) for Born-Infeld AdS black hole at $r_{+}= r_1$ for $Q(=Q_c)=0.5$, $ b=10 $, $ n=4 $ and $ l=10 $.}
\label{figure 2a}
\end{figure}
\begin{figure}[h]
\centering
\includegraphics[angle=0,width=8cm,keepaspectratio]{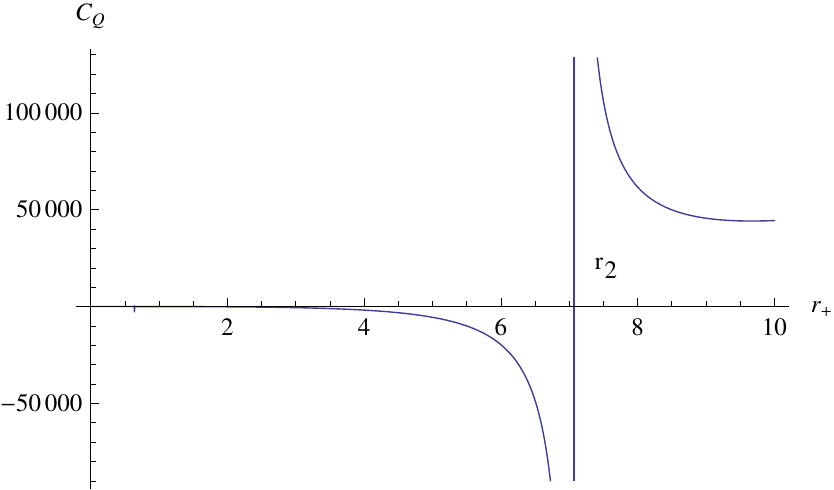}
\caption[]{\it Discontinuity of Specific heat ($ C_{Q} $) for Born-Infeld AdS black hole at $r_{+}= r_2$ for $Q(=Q_c)=0.5$, $ b=10 $, $ n=4 $ and $ l=10 $.}
\label{figure 2a}
\end{figure}
Like in the previous case with $ n = 3 $, here we also find that the heat capacity ($ C_{Q} $)  suffers discontinuities exactly at two points, namely $ r_{1} $ and $ r_{2} $, which could be identified as the critical points for the phase transition phenomena in BI AdS black holes. From these figures we note that there is a sign flip in the heat capacity around $ r_i(i=1,2) $, which indicates the onset of a \textit{continuous} higher order transition near critical points. The qualitative features of the phase transition therefore does not change as we increase the spatial dimensionality of our system.
\begin{figure}[h]
\centering
\includegraphics[angle=0,width=8cm,keepaspectratio]{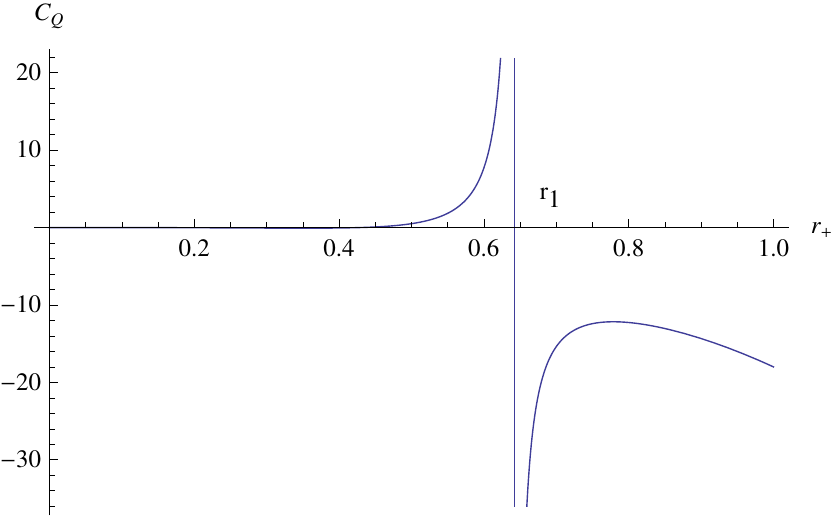}
\caption[]{\it Discontinuity of Specific heat ($ C_{Q} $) for Reissner Nordstrom AdS black hole at $r_{+}= r_1$ for $Q(=Q_c)=0.5$, $ n=4 $ and $ l=10 $.}
\label{figure 2a}
\end{figure}
\begin{figure}[h]
\centering
\includegraphics[angle=0,width=8cm,keepaspectratio]{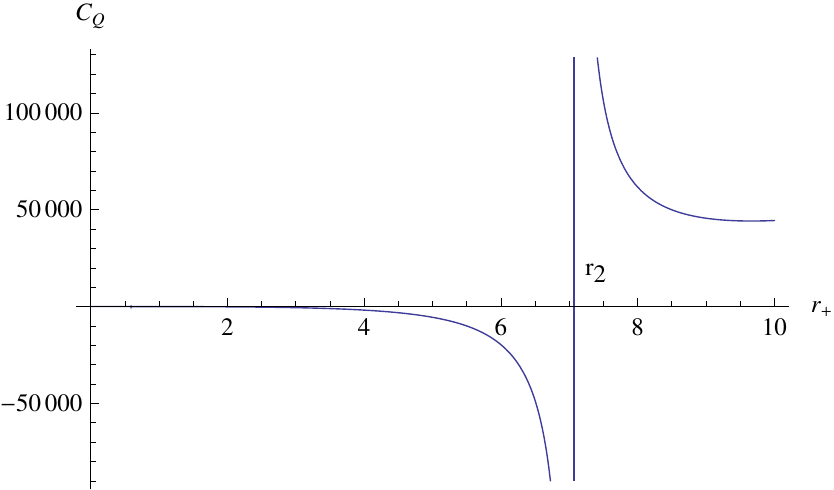}
\caption[]{\it Discontinuity of Specific heat ($ C_{Q} $) for Reissner Nordstrom AdS black hole at $r_{+}= r_2$ for $Q(=Q_c)=0.5$, $ n=4 $ and $ l=10 $.}
\label{figure 2a}
\end{figure}

Similar features may also be observed for the RN AdS case. The expression for the corresponding heat capacity is obtained by taking the $ b\rightarrow\infty $ limit of (\ref{CQ}). This finally leads to,
\begin{equation}
\left(C_Q \right)_{RN AdS}=\frac{(n-1)\omega_{n-1}r_+^{3n-7}\left[(n-2)r_{+}^{2}+\frac{nr_{+}^{4}}{l^{2}}-\frac{32\pi^{2}Q^{2}}{(n-1)r_{+}^{2n-6}\omega_{n-1}^{2}}\right]}{4\left[ r_+^{2n-4}\left(\frac{nr_+^{2}}{l^{2}}-n+2 \right)+\frac{32(2n-3) \pi^{2}Q^{2}}{(n-1)\omega_{n-1}^{2}}\right]}. 
\end{equation}
The plot of $ C_Q $ vs $ r_+ $ for the RN AdS case are given in Figures 3.5 and 3.6. The nature of the phase transition is similar to the BI AdS case.\\
 
\begin{center}
(III)\textbf{\textit{ The $ n=5 $ case :}}
\end{center}
For $ n=5 $, equation (\ref{extm}) turns out to be a quartic equation in the variable $ r_e^{2} $. The corresponding roots  ($ r_{e}^{2} $) of the equation (\ref{extm}) are provided in the following tabular form (Table 3.4). From Table 3.4 we note that, even for $ bQ<0.5 $ one can have two real positive roots of the equation (\ref{extm}) in $ (5+1) $ dimensions. However one should note that for $ bQ\rightarrow0 $ the number of real positive roots again reduces to one (see, for instance, the last row in Table 3.4). 
\begin{table}[htb]
\caption{Numerical solutions for $r_e^{2}$ for $ n=5 $ and $ l=10 $(extremal case)}   
\centering                          
\begin{tabular}{c c c c c c c c}            
\hline\hline                        
$Q$ & $b$  & $r_{e1}^{2}$ &$r_{e2}^{2}$ & $r_{e3}^{2} $ & $ r_{e4}^{2} $   \\ [0.05ex]
\hline
0.5 & 10 & -0.11104-0.183039 i & -0.11104+0.183039 i & 0.207093 & 60.06 \\
0.5 & 1 & -1.39385 & -0.0819668 & 0.0775052 & 66.0137 \\                              
0.5 & 0.5 & -4.72104 & -0.0399487 & 0.0396321 & 84.7214 \\
0.05 & 0.05 & -100 & -40  & -0.00039789 & 0.000397885 \\  [0.05ex]         
\hline                              
\end{tabular}\label{E1}  
\end{table}
\begin{figure}[h]
\centering
\includegraphics[angle=0,width=8cm,keepaspectratio]{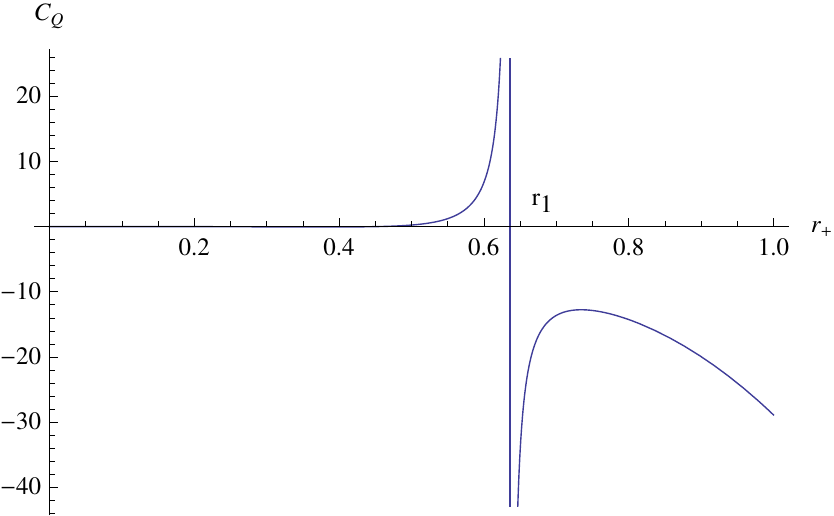}
\caption[]{\it Discontinuity of Specific heat ($ C_{Q} $) for Born-Infeld AdS black hole at $r_{+}= r_1$ for $Q(=Q_c)=0.5$, $ b=10 $, $ n=5 $ and $ l=10 $.}
\label{figure 2a}
\end{figure}
\begin{figure}[h]
\centering
\includegraphics[angle=0,width=8cm,keepaspectratio]{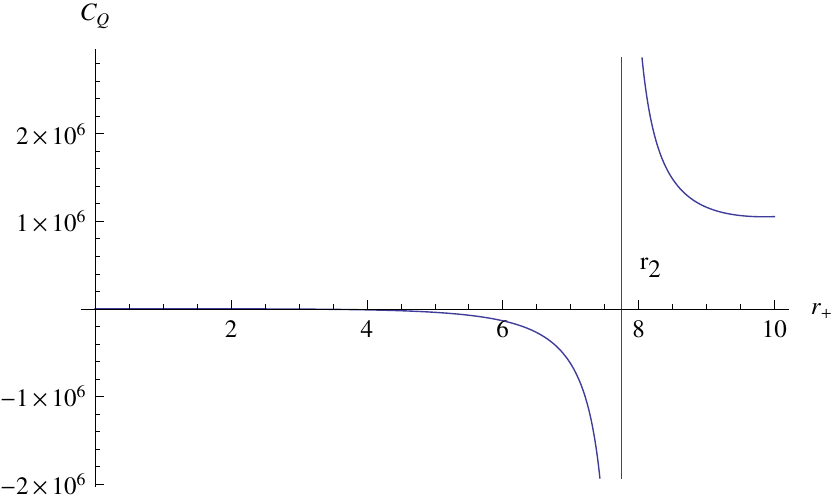}
\caption[]{\it Discontinuity of Specific heat ($ C_{Q} $) for Born-Infeld AdS black hole at $r_{+}= r_2$ for $Q(=Q_c)=0.5$, $ b=10 $, $ n=5 $ and $ l=10 $.}
\label{figure 2a}
\end{figure}

Finally, we aim to find out the roots of the equation (\ref{root}) for different choice of parameters ($ Q $ and $ b $), which essentially gives us the critical points for the phase transition in the \textit{non extremal} regime.
\begin{table}[htb]
\caption{Roots of the equation (\ref{root}) for $ n=5 $ and $ l=10 $(non extremal case)}   
\centering                          
\begin{tabular}{c c c c c c c c c c c}            
\hline\hline                        
$Q$ & $b$  & $r_1$ &$r_2$ & $r_3 $ &$r_4$&$ r_{5,6} $& $ r_{7,8} $ \\ [0.005ex]
\hline
0.5&10&0.63&7.74&-7.74&-0.63&$-0.31\pm i 0.55$ &$0.31\pm i 0.55$  \\ 
0.5&1&0.49&7.74&-7.74&-0.49&$-0.29\pm i 0.64$&$0.29\pm i 0.64$\\
0.5&0.5&0.34&7.74&-7.74&-0.34&$-0.32\pm i 0.76$&$0.32\pm i 0.76$\\
0.05&0.05&0.03&7.74&-7.74&-0.03&$ -0.31\pm i 0.76 $&$0.31\pm i 0.76 $\\
\\ [0.005ex]         
\hline                              
\end{tabular}
\label{E1}          
\end{table}

Like in the earlier case, from Table 3.5 we observe that $ C_Q $(\ref{CQ}) possesses simple poles, two of which are real positive  ($ r_1 $ and $ r_2 $) that may be regarded as the critical points corresponding to the phase transition phenomena in (non extremal) BI AdS black holes. Taking $ Q=0.5 $ and $ b=10 $, these critical points have been shown explicitly in Figures (3.7 and 3.8). Likewise, in the appropriate limit ($ b\rightarrow\infty $), one can also obtain the corresponding critical points for the RN AdS case.

Therefore, from the above analysis it is quite suggestive that the boundedness on the parameter space, which exists in $ (3+1) $ dimension, eventually disappears in higher dimensions which is also consistent with our finding of the critical points in the corresponding \textit{non extremal} case for various choice of parameters. 


\section{Critical exponents and scaling laws in higher dimensions}
In the usual theory of phase transitions it is customary to study the behavior of a given system close to its critical point by means of a set of critical exponents $ (\alpha,\beta,\gamma,\delta,\varphi,\psi,\nu,\eta) $. These critical exponents determine the qualitative nature of the critical behavior of the given system in the neighborhood of the critical point. By virtue of the so called \textit{scaling} laws, one can in fact see that only two of the eight critical exponents are actually independent. Based on the \textit{renormalization group} approach, one can also calculate these critical exponents. Systems having identical critical exponents have similar critical behavior and hence fall under the same \textit{universality class}. Here we show similar conclusions may be drawn for black holes also. We find that, apart from $ \delta $ that takes the value 2, all other critical exponents are coming out to be $ \frac{1}{2} $. Moreover, these critical exponents are found to be independent of the dimensionality of the space time. These results indeed suggest an underlying \textit{mean field} behavior for the (AdS) black holes. 

In order to calculate the critical exponent ($ \alpha $) that is associated with the divergences in the heat capacity ($ C_{Q} $), we first note that near the critical points ($ r_i $) we can write
\begin{equation}
r_{+}=r_{i}(1+\Delta), ~~~~ i=1,2 \label{tri}
\end{equation}
where $|\Delta| <<1 $.
\begin{figure}[h]
\centering
\includegraphics[angle=0,width=7cm,keepaspectratio]{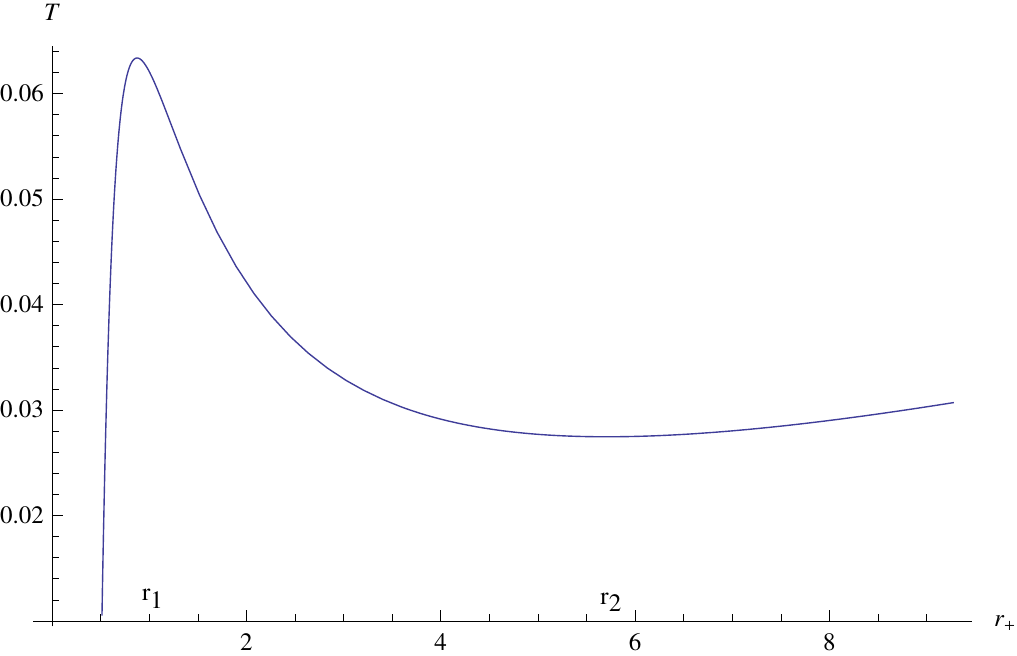}
\caption[]{\it Hawking temperature plot ($ T $) for Born-Infeld AdS black hole with respect to $r_{+}$ for $Q(=Q_c)=0.5$, $ b=10 $, $ n = 3 $ and $ l=10 $.}
\label{figure 2a}
\end{figure} 
\begin{figure}[h]
\centering
\includegraphics[angle=0,width=7cm,keepaspectratio]{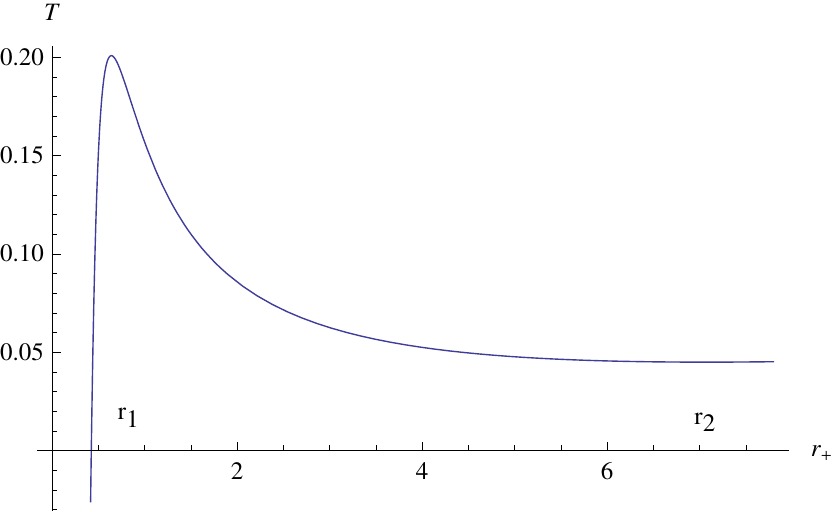}
\caption[]{\it Temperature ($T$) plot for Born-Infeld AdS black hole for $Q(=Q_c)=0.5$, $ b=10 $, $ n=4 $ and $ l=10 $.}
\label{figure 2a}
\end{figure}
\begin{figure}[h]
\centering
\includegraphics[angle=0,width=7cm,keepaspectratio]{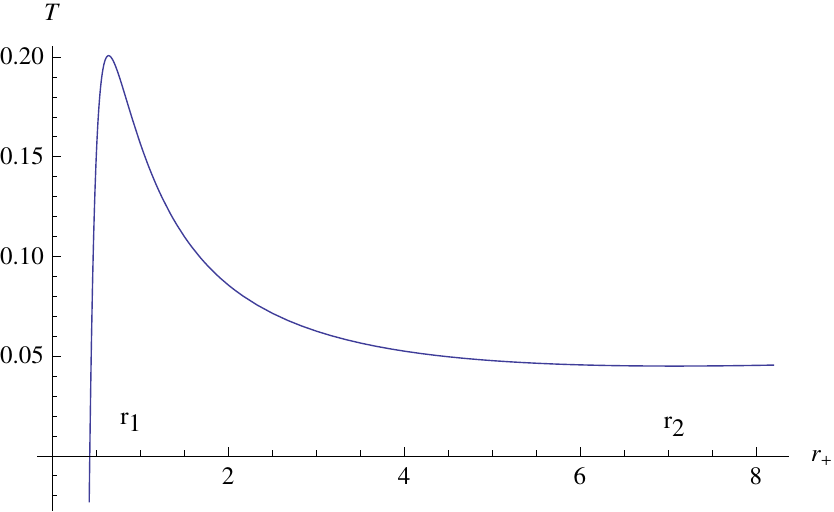}
\caption[]{\it Temperature ($T$) plot for Reissner Nordstrom AdS black hole for $Q(=Q_c)=0.5$, $ b=10 $, $ n=4 $ and $ l=10. $}
\label{figure 2a}
\end{figure}
As already discussed there are two distinct positive roots for the critical point $ r_i $ ($ r_1 $ and $ r_2 $). Also, any function of $ r_+ $, in particular the temperature $ T(r_+) $, may be expressed as,
\begin{equation}
T(r_+)=T(r_i)(1+\epsilon)
\end{equation}
where $ |\epsilon|<<1$.
As a next step, for a fixed value of the charge ($ Q $), we Taylor expand $ T(r_+) $ in a sufficiently small neighborhood of $ r_i $ which yields,
\begin{eqnarray}
T(r_+)= T(r_i)+\left[ \left( \frac{\partial T}{\partial r_+}\right)_{Q=Q_c}\right]_{r_+=r_i} (r_+-r_i)+\frac{1}{2} \left[ \left( \frac{\partial^{2} T}{\partial r^{2}_+}\right)_{Q=Q_c}\right]_{r_+=r_i} (r_+-r_i)^{2}\nonumber\\
+higher~order~terms.\label{Tr}
\end{eqnarray}
Since $ C_Q $ diverges at $ r_+=r_i $, therefore the second term on the R.H.S. of (\ref{Tr}) vanishes by virtue of equation (\ref{CQ}). This fact has also been depicted in various other figures (see Figures 3.9, 3.10 and 3.11).  
Using (\ref{tri}) we finally obtain from (\ref{Tr})
\begin{equation}
\Delta=\frac{\epsilon^{1/2}}{D_{i}^{1/2}}\label{delta1}
\end{equation} 
where\footnote{We use the notation $ T(r_i)=T_i $.},
\begin{eqnarray}
D_i =\frac{r^{2}_{i}}{2T_{i}}\left[ \left( \frac{\partial^{2} T}{\partial r^{2}_+}\right)_{Q=Q_c}\right]_{r_+=r_i}=\frac{\Sigma(r_i,Q_c)}{4\pi r^{2n-3}_{i}T_i \left(1+\frac{16 \pi^{2} Q^{2}_{c}}{{b^{2}r^{2n-2}_{i}\omega_{n-1}^{2}}}\right)^{3/2}}
\end{eqnarray}
with,
\begin{eqnarray}
\Sigma(r_i,Q_c) = (n-2)r^{2n-4}_{i}\left(1+\frac{16\pi^{2} Q^{2}_{c}}{{b^{2}r^{2n-2}_{i}\omega_{n-1}^{2}}}\right)^{3/2}+\frac{512(n-1)\pi^{4}Q^{4}_{c}}{b^{2}r^{2n-2}_{i}\omega_{n-1}^{4}}\nonumber\\
-32(2n-3)\pi^{2}Q_{c}^{2}\omega_{n-1}^{-2}\left(1+\frac{16\pi^{2} Q^{2}_{c}}{b^{2}r^{2n-2}_{i}\omega_{n-1}^{2}}\right).
\end{eqnarray}
In the limit $ b\rightarrow\infty $, the corresponding expression for RN AdS case may be obtained as,
\begin{eqnarray}
\left[D_i \right]_{RN AdS}=\frac{(n-2)r^{2n-4}_{i}-32(2n-3)\pi^{2}Q_{c}^{2}\omega_{n-1}^{-2}}{4\pi r^{2n-3}_{i}[T_i]_{RN AdS}}. 
\end{eqnarray}

A closer look at Figures 3.9, 3.10 and 3.11 reveals that in the neighborhood of $ r_+=r_2 $ we always have $ T(r_+)>T(r_2) $ so that $ \epsilon $ is positive. On the other hand, for any point close to  $ r_+=r_1 $ we have $ T(r_+)<T(r_1) $ implying that $ \epsilon $ is negative. We will exploit these observations to find $ C_Q $ near the critical points.

Let us first compute the value of $ C_Q $ near the critical point $ r_+=r_2 $ (where $ \epsilon $ is positive). As a first step, substituting $ r_+ $ from (\ref{tri}) into (\ref{CQ}) we obtain,
\begin{equation}
C_Q=\frac{\Im(r_i,Q_c)}{\Re(r_i(1+\Delta),Q_c)}.
\end{equation}
In the next step, Taylor expanding around the critical point $\Delta\approx 0$ we obtain,
\begin{eqnarray}
C_Q \simeq ~~~\left[ \frac{A_{i}}{\epsilon^{1/2}}\right] _{r_i=r_2}
\end{eqnarray}
where,
\begin{equation}
A_i=\frac{D^{1/2}_{i} \zeta(r_i,Q_c)}{\xi(r_i,Q_c)}
\end{equation}
with,
\begin{eqnarray}
\zeta(r_i,Q_c)= \frac{(n-1)\omega_{n-1}r_{i}^{3n-7}}{8}\sqrt{1+\frac{16\pi^{2}Q_{c}^{2}}{b^{2}r_{i}^{2n-2}\omega_{n-1}^{2}}}\nonumber\\
\times\left[(n-2)r_{i}^{2}+\frac{nr_{i}^{4}}{l^{2}}+\frac{4b^{2}r_{i}^{4}}{n-1}\left( 1-\sqrt{1+\frac{16\pi^{2}Q_{c}^{2}}{b^{2}r_{i}^{2n-2}\omega_{n-1}^{2}}}\right)  \right] 
\end{eqnarray}
and,

\begin{eqnarray}
\xi(r_i,Q_c)= r_{i}^{2n-4}\left[(n-1)\frac{nr_{i}^{2}}{l^{2}}-(n-2)^{2} \right]\sqrt{1+\frac{16\pi^{2}Q_{c}^{2}}{b^{2}r_{i}^{2n-2}\omega_{n-1}^{2}}}\nonumber\\
-\frac{8(n-1)\pi^{2}Q_{c}^{2}}{b^{2}r_{i}^{2}\omega_{n-1}^{2}}\left(2-n+\frac{nr_{i}^{2}}{l^{2}} \right).  
\end{eqnarray}
 On the other hand, following a similar approach, the singular behavior of $ C_Q $ near $ r_+=r_1 $ (where $ \epsilon $ is negative) may be expressed as,
\begin{equation}
C_Q \simeq  ~~~\left[ \frac{A_{i}}{(-\epsilon)^{1/2}}\right] _{r_i=r_1}.
\end{equation}
Combining both of these facts into a single expression, we may therefore express the singular behavior of the heat capacity ($ C_Q $) near the critical points as,
\begin{eqnarray}
C_Q &\simeq & ~~~ \frac{A_{i}}{|\epsilon |^{1/2}}\nonumber\\
&=&~~~\frac{A_{i}T^{1/2}_{i}}{|T-T_i|^{1/2}}\label{CQ1}.
\end{eqnarray}

Similar arguments also hold for the RN AdS case. In order to obtain the corresponding expression for the singular behavior of the heat capacity near the critical points, we set $ b\rightarrow\infty $, which finally yields,
\begin{equation}
\left[C_Q \right]_{RN AdS}\simeq \frac{[A_{i}]_{RN AdS}[T^{1/2}_{i}]_{RN AdS}}{|T-T_i|^{1/2}}\label{rnads1}
\end{equation}
where,
\begin{eqnarray}
[A_{i}]_{RN AdS}=[D_i^{1/2}]_{RN AdS}\frac{(n-1)\omega_{n-1}r_{i}^{n-3}\left[(n-2)r_{i}^{2}+\frac{nr_{i}^{4}}{l^{2}}-\frac{32\pi^{2}Q_{c}^{2}}{(n-1)r_{i}^{2n-6}\omega_{n-1}^{2}}\right]}{8\left[(n-1)\frac{nr_{i}^{2}}{l^{2}}-(n-2)^{2} \right]}.
\end{eqnarray}

Comparing (\ref{CQ1}) and (\ref{rnads1}) with the standard form \cite{ref48},
\begin{equation}
C_Q \sim |T-T_i|^{-\alpha}
\end{equation}
we find $ \alpha=1/2 $.

Next, we want to calculate the critical exponent $ \beta $ which is related to the electric potential ($ \Phi $) for a fixed value of charge as,
\begin{equation}
\Phi(r_+) - \Phi(r_i) \sim |T-T_i|^{\beta}\label{Ph}.
\end{equation}
In order to do that we Taylor expand $ \Phi(r_+) $ close to the critical point $ r_+=r_i $ which yields, 
\begin{eqnarray}
\Phi(r_+)= \Phi(r_i)+\left[ \left( \frac{\partial \Phi}{\partial r_+}\right)_{Q=Q_c}\right]_{r_+=r_i} (r_+-r_i)
+ higher~~ order~~ terms.\label{phir}
\end{eqnarray}
Ignoring all the higher order terms in (\ref{phir}) and using (\ref{Phi}) and (\ref{delta1}) we finally obtain,
\begin{equation}
\Phi(r_+)- \Phi(r_i)= - \left( \frac{4\pi Q_c}{r_i^{n-2}\omega_{n-1} T^{1/2}_i D^{1/2}_{i}}\right) \left(1-\frac{8\pi^{2}Q^{2}_{c}}{b^{2}r^{2n-2}_{i}\omega_{n-1}^{2}} \right) |T-T_i|^{1/2}\label{Pr}.
\end{equation}
For the RN AdS case ($ b\rightarrow\infty $) the corresponding expression becomes,
\begin{eqnarray}
\Phi(r_+)- \Phi(r_i)= -  \frac{4\pi Q_c}{r_i^{n-2}\omega_{n-1} [T^{1/2}_i]_{RN AdS} [D^{1/2}_{i}]_{RN AdS}}  |T-T_i|^{1/2}\label{rnads2}.
\end{eqnarray}  
Comparing (\ref{Pr}) and (\ref{rnads2}) with (\ref{Ph}) we find $ \beta=1/2 $.

We next calculate the critical exponent $ \gamma $ which is related to the singular behavior of the isothermal compressibility related derivative $ K^{-1}_{T} $ \cite{ref50}(near the critical points $ r_i $) for a fixed value of charge ($ Q=Q_c $). This is defined as,
\begin{equation}
K^{-1}_{T}\sim |T-T_i|^{-\gamma}\label{k}.
\end{equation}
In order to calculate $ K^{-1}_{T} $ we first note that,
\begin{eqnarray}
 K^{-1}_{T}=Q\left( \partial\Phi /\partial Q\right)_{T} =- Q \left(\frac{\partial \Phi}{\partial T} \right)_{Q} \left(\frac{\partial T}{\partial Q} \right)_{\Phi},
\end{eqnarray}
where we have used the thermodynamic identity $\left(\frac{\partial \Phi}{\partial T} \right)_{Q}\left(\frac{\partial T}{\partial Q} \right)_{\Phi}\left(\frac{\partial Q}{\partial \Phi} \right)_{T} =-1$. 

Finally, using (\ref{Phi}) and (\ref{T}) the expression for $ K^{-1}_{T} $ may be found as,
\begin{equation}
K^{-1}_{T}=\frac{4\pi Q}{(n-2)\omega_{n-1}r_{+}^{n-2}}\left( 1-\frac{24\pi^{2}Q^{2}(n-2)}{(3n-4)\omega_{n-1}^{2}b^{2}r_{+}^{2n-2}}\right) \frac{\wp(Q,r_+)}{\Re(Q,r_+)}\label{kt}
\end{equation}
where,
\begin{eqnarray}
\wp(Q,r_+)= r_{+}^{2n-2}\left(\frac{n}{l^{2}}-\frac{n-2}{r_{+}^{2}} \right) \sqrt{1+\frac{16\pi^{2}Q^{2}}{b^{2}r_+^{2(n-1)}\omega_{n-1}^{2}}}+ \frac{1024\pi^{4}Q^{4}(n-2)}{r_{+}^{2n-2}b^{2}\omega_{n-1}^{4}(n-1)(3n-4)}\nonumber\\
-\frac{4b^{2}r_{+}^{2n-2}}{n-1}\left( 1-\sqrt{1+\frac{16\pi^{2}Q^{2}}{b^{2}r_+^{2(n-1)}\omega_{n-1}^{2}}}\right).
\end{eqnarray}
Taking the appropriate limit ($ b\rightarrow\infty $) one can easily obtain the corresponding expression of $K^{-1}_{T}$ for the RN AdS black hole as,
\begin{equation}
\left( K^{-1}_{T}\right)_{RN AdS} =\frac{4\pi Q}{(n-2)\omega_{n-1}r_{+}^{n-2}}\frac{\left[r_{+}^{2n-2}\left(\frac{n}{l^{2}}-\frac{n-2}{r_{+}^{2}} \right)+\frac{32\pi^{2}Q^{2}}{(n-1)\omega_{n-1}^{2}} \right] }{\left[r_+^{2n-4}\left(\frac{nr_+^{2}}{l^{2}}-n+2 \right)+\frac{32(2n-3) \pi^{2}Q^{2}}{(n-1)\omega_{n-1}^{2}} \right] }.
\end{equation}
From the above expressions it is interesting to note that both $ K^{-1}_{T} $ and the heat capacity ($ C_{Q} $) posses common singularities. It is interesting to note that a similar feature may also be observed for the Kerr Newmann black hole in asymptotically flat space \cite{ref50}. It is reassuring to note that all these features are in general compatible with standard thermodynamic systems\cite{ref118}.  

Following our previous approach, we substitute $ r_+ $ from (\ref{tri}) into (\ref{kt}) and use (\ref{delta1}) which finally yields,
\begin{equation}
K^{-1}_{T}=\frac{B_i}{|\epsilon|^{1/2}}= \frac{B_i T^{1/2}_i}{|T-T_i|^{1/2}}\label{KT}
\end{equation}   
where,
\begin{equation}
B_i=\frac{2\pi Q_c}{(n-2)\omega_{n-1}r_{i}^{n-2}}\left( 1-\frac{24\pi^{2}Q_{c}^{2}(n-2)}{(3n-4)\omega_{n-1}^{2}b^{2}r_{i}^{2n-2}}\right)\frac{D^{1/2}_i\wp(Q_c,r_i)}{\xi(Q_c,r_i)}.
\end{equation}
In the appropriate limit ($ b\rightarrow\infty $) the corresponding expression for the RN AdS black hole may be obtained as,
\begin{equation}
[K^{-1}_{T}]_{RN AdS}=\frac{[B_i]_{RN AdS}}{|\epsilon|^{1/2}}= \frac{[B_i]_{RN AdS} [T^{1/2}_i]_{RN AdS}}{|T-T_i|^{1/2}}\label{rnads3}
\end{equation}
where,
\begin{equation}
[B_i]_{RN AdS}= \frac{2\pi Q_c[D_{i}^{1/2}]_{RN AdS}}{(n-2)\omega_{n-1}r_{i}^{3n-6}}\frac{\left[r_{i}^{2n-2}\left(\frac{n}{l^{2}}-\frac{n-2}{r_{i}^{2}} \right)+\frac{32\pi^{2}Q_{c}^{2}}{(n-1)\omega_{n-1}^{2}}\right]}{\left[(n-1)\frac{nr_{i}^{2}}{l^{2}}-(n-2)^{2} \right]}.
\end{equation}
Comparing (\ref{KT}) and (\ref{rnads3}) with (\ref{k}) we note that $ \gamma=1/2 $.

Let us now calculate the critical exponent ($ \delta $) which is related to the electric potential ($ \Phi $) for the fixed value of the temperature $ T=T_i $ as,
\begin{equation}
\Phi(r_+) - \Phi(r_i) \sim |Q-Q_i|^{1/\delta}\label{PQ},
\end{equation}
where $ Q_i $ is the value of charge ($ Q $) at $ r_+=r_i $. We expand $ Q(r_+) $ in a sufficiently small neighborhood of $r_+= r_i $ which yields,
 \begin{eqnarray}
Q(r_+)= Q(r_i)+\left[ \left( \frac{\partial Q}{\partial r_+}\right)_{T=T_i}\right]_{r_+=r_i} (r_+-r_i)+\frac{1}{2} \left[ \left( \frac{\partial^{2} Q}{\partial r^{2}_+}\right)_{T=T_i}\right]_{r_+=r_i} (r_+-r_i)^{2}\nonumber\\
+ higher~~ order~~ terms.\label{Q1}
\end{eqnarray}
Using the functional form,
\begin{equation}
T=T(r_+,Q)
\end{equation}
and following our previous argument we note that,
\begin{equation}
\left[ \left( \frac{\partial Q}{\partial r_+}\right)_{T}\right]_{r_+=r_i}=-\left[ \left( \frac{\partial T}{\partial r_+}\right)_{Q}\right]_{r_+=r_i} \left( \frac{\partial Q}{\partial T}\right)_{r_+=r_i}=0 \label{e2}.
\end{equation}
Also note that near the critical point we can express the charge ($ Q $) as,
\begin{equation}
Q(r_+)= Q(r_i)(1+\Pi) \label{Qr}
\end{equation}
with $ |\Pi|<<1 $.  
Finally using (\ref{tri}), (\ref{Q1}) and (\ref{Qr}) we obtain 
\begin{equation}
\Delta=\left(\frac{2Q_i}{M_i r^{2}_{i}} \right)^{1/2}\Pi^{1/2}\label{e4}
\end{equation}  
where,
\begin{eqnarray}
M_i=\left[ \left( \frac{\partial^{2} Q}{\partial r^{2}_+}\right)_{T}\right]_{r_+=r_i}
=\frac{(n-1)\omega_{n-1}^{2}\Gamma(r_i,Q_c)}{64\pi^{2}r_iQ_c\sqrt{1+\frac{16\pi^{2}Q_{c}^{2}}{b^{2}r_{i}^{2n-2}\omega_{n-1}^{2}}}} \label{Mi}
\end{eqnarray}
with,
\begin{eqnarray}
\Gamma(r_i,Q_c)=r_{i}^{2n-4}\left(1+\frac{16\pi^{2}Q_{c}^{2}}{b^{2}r_{i}^{2n-2}\omega_{n-1}^{2}} \right)\left[ \frac{n}{l^{2}}(2n-3)r_{i}^{2}-(n-2)(2n-5)+\frac{4b^{2}(2n-3)r_{i}^{2}}{n-1}\right]\nonumber\\
-\frac{64\pi^{2}Q_{c}^{2}}{\omega_{n-1}^{2}} 
- \sqrt{1+\frac{16\pi^{2}Q_{c}^{2}}{b^{2}r_{i}^{2n-2}\omega_{n-1}^{2}}}\left[ \frac{4b^{2}(2n-3)r_{i}^{2n-2}}{n-1}+\frac{64(n-2)\pi^{2}Q_{c}^{2}}{(n-1)\omega_{n-1}^{2}}\right]\nonumber\\
-r_{i}^{2n-4}\left(\frac{nr_{i}^{2}}{l^{2}}-n+2 \right)\left( 1+\frac{16\pi^{2}Q_{c}^{2}}{b^{2}r_{i}^{2n-2}\omega_{n-1}^{2}}\right).   
\end{eqnarray}
Let us now consider the functional relation
\begin{equation}
\Phi=\Phi(r_+,Q)
\end{equation}
from which we find,
\begin{equation}
\left[ \left( \frac{\partial \Phi}{\partial r_+}\right)_{T} \right]_{r_+=r_i} = \left[ \left( \frac{\partial \Phi}{\partial r_+}\right)_{Q} \right]_{r_+=r_i}+\left[ \left( \frac{\partial Q}{\partial r_+}\right)_{T}\right]_{r_+=r_i}\left( \frac{\partial \Phi}{\partial Q}\right)_{r_+=r_i}\label{e1}.
\end{equation}
By using (\ref{Phi}) and (\ref{e2}) we obtain from (\ref{e1}) the result,
\begin{equation}
\left[ \left( \frac{\partial \Phi}{\partial r_+}\right)_{T} \right]_{r_+=r_i} =- \left( \frac{4\pi Q_c}{r_i^{n-1}\omega_{n-1}}\right) \left(1-\frac{8\pi^{2}Q^{2}_{c}}{b^{2}r^{2n-2}_{i}\omega_{n-1}^{2}} \right)\label{phibi}.
\end{equation}  
As a next step, for a fixed value of the temperature ($ T $), we Taylor expand $ \Phi(r_+,T) $ close to the critical point $ r_+=r_i $, which yields,
\begin{eqnarray}
\Phi(r_+)= \Phi(r_i)+\left[ \left( \frac{\partial \Phi}{\partial r_+}\right)_{T=T_i}\right]_{r_+=r_i} (r_+-r_i)
+ higher~~ order~~ terms\label{e3}.
\end{eqnarray}
Ignoring all the higher order terms in (\ref{e3}) and using (\ref{e4}) and (\ref{phibi}) we obtain 
\begin{equation}
\Phi(r_+)-\Phi(r_i)=-\left( \frac{2}{M_i}\right) ^{1/2}\left( \frac{4\pi Q_c}{r_i^{n-1}\omega_{n-1}}\right) \left(1-\frac{8\pi^{2}Q^{2}_{c}}{b^{2}r^{2n-2}_{i}\omega_{n-1}^{2}} \right) |Q-Q_i|^{1/2}\label{e6}.
\end{equation} 
Finally, taking the limit $ b\rightarrow\infty $, we abstract the corresponding expression for the RN AdS black hole as,
\begin{equation}
\Phi(r_+)-\Phi(r_i)=-\left( \frac{2}{[M_i]_{RN AdS}}\right) ^{1/2}\left( \frac{4\pi Q_c}{r_i^{n-1}\omega_{n-1}}\right)|Q-Q_i|^{1/2}\label{rnads4}
\end{equation}
where,
\begin{eqnarray}
[M_i]_{RN AdS}= \frac{(n-1)\omega_{n-1}^{2}[\Gamma(r_i,Q_c)]_{RN AdS}}{64\pi^{2}r_iQ_c}
\end{eqnarray}
with,
\begin{eqnarray}
[\Gamma(r_i,Q_c)]_{RN AdS}=r_{i}^{2n-4}\left[ \frac{n}{l^{2}}(2n-3)r_{i}^{2}-(n-2)(2n-5)\right]-\frac{64\pi^{2}Q_{c}^{2}}{\omega_{n-1}^{2}}\nonumber\\ 
- \frac{64(n-2)\pi^{2}Q_{c}^{2}}{(n-1)\omega_{n-1}^{2}}
-r_{i}^{2n-4}\left(\frac{nr_{i}^{2}}{l^{2}}-n+2 \right).
\end{eqnarray}
Comparing (\ref{e6}) and (\ref{rnads4}) with (\ref{PQ}) we note that $ \delta=2 $.

In order to calculate the critical exponent $ \varphi $ we note from (\ref{CQ1}) that near the critical point $ r_+ =r_i$ the heat capacity behaves as,
\begin{equation}
C_Q \sim \frac{1}{\Delta}.
\end{equation}
Using (\ref{e4}) this yields our cherished form,
\begin{equation}
C_Q =\frac{A_ir_i\sqrt{M_i}}{\sqrt{2}D_i|Q-Q_i|^{1/2}} \label{e10}.
\end{equation}
For the RN AdS black hole, the corresponding expression becomes,
\begin{equation}
[C_Q]_{RN AdS}=\frac{[A_i]_{RN AdS}r_i\sqrt{[M_i]_{RN AdS}}}{\sqrt{2}[D_i]_{RN AdS}|Q-Q_i|^{1/2}}\label{rnads5}
\end{equation} 
Comparing (\ref{e10}) and (\ref{rnads5}) with the standard relation,
\begin{equation}
C_Q \sim \frac{1}{|Q-Q_i|^{\varphi}}
\end{equation}
we note that $ \varphi=1/2 $.

In an attempt to calculate the critical exponent $ \psi $,  we use (\ref{S}) and (\ref{tri}) in order to expand the entropy ($ S $) near the critical point $ r_+=r_i $. Then it follows, 
\begin{eqnarray}
S(r_+)&=&S(r_i)+\left[ \left( \frac{\partial S}{\partial r_+}\right)\right] _{r_+=r_i}(r_+-r_i)+ higher~~ order~~ terms\nonumber\\ 
&=& S(r_i)+ \frac{\omega_{n-1}}{4}(n-1)r_{i}^{n-1}\Delta \label{e7}
\end{eqnarray}
where we have ignored the higher order terms in $ \Delta $ as we did earlier.
Exploiting (\ref{e4}) we obtain,
\begin{equation}
S(r_+)-S(r_i)=\frac{\omega_{n-1}}{4}(n-1)r_{i}^{n-2}\sqrt{\frac{2}{M_i}}|Q-Q_i|^{1/2}\label{e8}.
\end{equation}
It is now quite straightforward to obtain the corresponding expression for the RN AdS black hole as,
\begin{equation}
S(r_+)-S(r_i)=\frac{\omega_{n-1}}{4}(n-1)r_{i}^{n-2}\sqrt{\frac{2}{[M_i]_{RN AdS}}}|Q-Q_i|^{1/2}\label{rnads6}.
\end{equation}
Comparing (\ref{e8}) and (\ref{rnads6}) with the standard form
\begin{equation}
S(r_+)-S(r_i) \sim |Q-Q_i|^{\psi}
\end{equation}
we note that $ \psi=1/2 $.

Let us now tabulate the various \textit{static} critical exponents obtained so far.
\begin{table}[h]
\caption{Various static critical exponents and their values}   
\centering                          
\begin{tabular}{c c c c c c c c c}            
\hline\hline                        
$Critical~Exponents  $ & $\alpha$ & $\beta$  & $\gamma$ & $ \delta $ & $\varphi $ & $ \psi $ \\ [2.0ex]
\hline
Values & 1/2 & 1/2 & 1/2 & 2 & 1/2 & 1/2\\ [2.0ex]         
\hline                              
\end{tabular}
\label{E1}          
\end{table}

In usual thermodynamics the critical exponents are found to satisfy certain \textit{scaling} relations among themselves, known as \textit{thermodynamic scaling laws} \cite{ref48}-\cite{stanley}, which may be expressed as,
\begin{eqnarray}
\alpha + 2\beta +\gamma = 2,~~~\alpha +\beta(\delta + 1) =2,~~~(2-\alpha)(\delta \psi -1)+1 =(1-\alpha)\delta ~~\nonumber\\
~~ \gamma(\delta +1) =(2-\alpha)(\delta -1),~~~\gamma =\beta (\delta - 1),~~~\varphi + 2\psi -\delta^{-1} =1 \label{scaling laws}.
\end{eqnarray}     
From the values of the critical exponents (given in Table 3.6), it is easily seen that all the above relations (\ref{scaling laws}) are indeed satisfied both for BI AdS as well as RN AdS black holes.\\\ \\\\

\textbf{\textit{Static scaling hypothesis:}}

The \textit{static scaling hypothesis} plays an important role in order to explore the singular behavior of several thermodynamic functions near the critical point. The \textit{scaling hypothesis} for the critical phenomena has been made in a variety of situations and applied to thermodynamic functions as well as static and dynamic correlation functions \cite{ref119}-\cite{ref121}. In spite of its several applications to ordinary thermodynamic systems, it is not a widely explored scheme in the context of black hole thermodynamics. In fact it has never been explored in higher dimensions so far.

For ordinary thermodynamic systems, near the critical point, the free energy may be split into two parts, one \textit{regular} at the critical point while the other contains the \textit{singular} part. The \textit{scaling hypothesis} is made for the singular part. In the subsequent sections we aim to study the \textit{scaling hypothesis} for the BI AdS black holes in higher dimensions. The corresponding discussion for the RN AdS case arises as a natural consequence of the BI AdS case.  

The \textit{static scaling hypothesis} for black holes \cite{Lousto:1993yr} states that \textit{close to the critical point, the singular part of the Helmholtz free energy $ F(T,Q)= M-TS $ is a generalized homogeneous function of its variables} i.e; there exists two numbers(known as \textit{scaling parameters}) $ p $ and $ q $ such that for all positive $ \Lambda $
\begin{equation}
F(\Lambda^{p}\epsilon,\Lambda^{q}\Pi)=\Lambda F(\epsilon,\Pi)\label{ghf}.
\end{equation}

Let us now calculate these \textit{scaling parameters} for the BI AdS black hole (and consequently for RN AdS case also) in arbitrary dimensions. In order to do that we first Taylor expand $ F(T,Q) $ close to the critical point $ r_+=r_i $ which yields,
\begin{eqnarray}
F(T,Q)= F(T,Q)|_{r_+=r_i} + \left[ \left( \frac{\partial F}{\partial T}\right)_{Q}\right]_{r_+=r_i} (T-T_i)+\left[ \left( \frac{\partial F}{\partial Q}\right)_{T}\right]_{r_+=r_i} (Q-Q_i)\nonumber\\
+\frac{1}{2} \left[ \left( \frac{\partial^{2} F}{\partial T^{2}}\right)_{Q}\right]_{r_+=r_i} (T-T_i)^{2}+\frac{1}{2} \left[ \left( \frac{\partial^{2} F}{\partial Q^{2}}\right)_{T}\right]_{r_+=r_i} (Q-Q_i)^{2} \nonumber\\
+\left[ \left( \frac{\partial^{2}F}{\partial Q \partial T}\right)\right]_{r_+=r_i} (Q-Q_i)(T-T_i)+higher~order~terms. \label{F1}
\end{eqnarray}
From our previous discussions it is by now quite evident that both 
\begin{equation}
C_Q= - T \left( \frac{\partial^{2} F}{\partial T^{2}}\right)_{Q} ~~~ and ~~~ K^{-1}_{T}= Q \left( \frac{\partial^{2} F}{\partial Q^{2}}\right)_{T}
\end{equation}
diverge at the critical point $ r_+=r_i $. Therefore from (\ref{F1}) we identify the singular part of the free energy as,
\begin{eqnarray}
F_{singular}&=&\frac{1}{2} \left[ \left( \frac{\partial^{2} F}{\partial T^{2}}\right)_{Q}\right]_{r_+=r_i} (T-T_i)^{2}+\frac{1}{2} \left[ \left( \frac{\partial^{2} F}{\partial Q^{2}}\right)_{T}\right]_{r_+=r_i} (Q-Q_i)^{2}\nonumber\\
&=& -\frac{C_Q}{2T_i}(T-T_i)^{2}+\frac{K^{-1}_{T}}{2Q_i} (Q-Q_i)^{2}.
\end{eqnarray} 
Using (\ref{delta1}), (\ref{CQ1}), (\ref{KT}) and (\ref{e4}) we finally obtain 
\begin{equation}
F_{singular}= a_i \epsilon^{3/2}+b_i \Pi^{3/2}\label{F}
\end{equation}
where,
\begin{equation}
 a_i=-\frac{A_i T_i }{2}~~~ and, ~~~ b_i=\frac{Q^{3/2}_{i}r_i M^{1/2}_i B_i}{2^{3/2}D^{1/2}_i}.
\end{equation}
Setting the limit $ b\rightarrow\infty $ one can obtain the corresponding expression for RN AdS black hole as,
\begin{equation}
[F_{singular}]_{RN AdS}= [a_i]_{RN AdS} \epsilon^{3/2}+ [b_i]_{RN AdS} \Pi^{3/2}\label{rnads7}
\end{equation}
where,
\begin{equation}
 [a_i]_{RN AdS}=-\frac{[A_i]_{RN AdS} [T_i]_{RN AdS} }{2}~~~ and, ~~~ [b_i]_{RN AdS}=\frac{Q^{3/2}_{i}r_i [M^{1/2}_i]_{RN AdS} [B_i]_{RN AdS}}{2^{3/2}[D^{1/2}_i]_{RN AdS}}.
\end{equation}

Finally, from (\ref{F}) and (\ref{rnads7}) we note that in order to satisfy (\ref{ghf}) we must have $ p=q=2/3 $. Note that although the scaling parameters $ p $ and $ q $ are in general different for a generalized homogeneous function, this is a special case  where both the numbers $ p $ and $ q $ have identical values. In other words $ F $ behaves as a usual homogeneous function.

At this stage it is important to note that, in standard thermodynamics, the various critical exponents are related to the scaling parameters as \cite{stanley},
\begin{eqnarray}
\alpha = 2-\frac{1}{p},~~~\beta =\frac{1-q}{p},~~~\delta =\frac{q}{1-q}~~\nonumber\\
~~ \gamma =\frac{2q-1}{p},~~~\psi =\frac{1-p}{q},~~~\varphi =\frac{2p-1}{q}.
\end{eqnarray}
It is reassuring to note that these relations are also satisfied both for the BI AdS and RN AdS black holes. One can further see that elimination of the two scaling parameters ($ p $ and $ q $) from the above set of relations eventually leads to (\ref{scaling laws}). This is consistent with the fact that the scaling hypothesis replaces the exponent \textit{inequalities} with the \textit{equalities} \cite{ref48}-\cite{stanley}, as found in usual systems. The fact that all the critical exponents can be expressed in terms of two \textit{scaling parameters} ($ p $ and $ q $) suggests that if two exponents are specified then all others can be determined. 

Finally, we are in a position to provide a suggestive way to find out the rest of the critical exponents $ \nu $ and $ \eta $ which are associated with the \textit{correlation length} and \textit{correlation function} respectively. \textit{Assuming} the additional scaling relations \cite{ref48},
\begin{equation}
\gamma = \nu (2-\eta),~~~ 2-\alpha = \nu d \label{addbi}
\end{equation}
to be valid, where $ d(=n) $ is the spatial dimensionality of the system, we find  
\begin{equation}
\nu =\frac{3}{2n},~~~ \eta =\frac{6-n}{3} \label{eta}.
\end{equation}
Frankly speaking, the above relations are clearly suggestive rather than definitive. Till date it is not clear a priory whether these additional relations (\ref{addbi}) are indeed valid for black holes. It will be more definitive as well as interesting if one attempts to compute the values for $ \nu $ and $ \eta $ directly from the \textit{correlation} of scalar modes in gravity theory. 

So far our analysis was strictly confined to the usual scenario of \textit{Einstein} gravity. For the rest of this chapter we aim to explore similar situations for charged topological black holes in the context of \textit{Ho\v{r}ava-Lifshitz} gravity. 

\section{Charged topological black holes in Ho\v{r}ava-Lifshitz gravity}

 Studying the thermodynamics as well as the critical behavior  of topological black hole solutions in the framework of usual \textit{Einstein} gravity \cite{ref26, brimhm}, \textit{Einstein-Gauss-Bonnet gravity} and \textit{dilaton} gravity \cite{ref30} as well as \textit{Lovelock} gravity \cite{refcai} has been a fascinating topic of research for the past few decades. Inspired by the dynamical critical phenomena in usual condensed matter systems, few years back P. Ho\v{r}ava has proposed a UV complete theory of gravity \cite{horava} that reduces to the usual \textit{Einstein} gravity at large scales. Since then a number of attempts have been made in order to understand various aspects of this theory \cite{horava1}-\cite{horava3} including different cosmological aspects \cite{Calcagni:2009ar}. Various black hole solutions were also found in \cite{ref114,Lu:2009em,Tang:2009bu}. The thermodynamics of these black holes have been studied in \cite{ref114,hl1,hl2}.  Although these attempts are self contained and rigorous, still there remains some major questions which have not yet been attempted. Studying the critical behavior of black holes is one of the important aspects, which we aim to explore for \textit{ Ho\v{r}ava-Lifshitz} theory of gravity. 

In this section a brief discussion on the black hole solution in \textit{Ho\v{r}ava-Lifshitz} gravity will be presented. For details, one can follow \cite{ref114} where the meaning of all the parameters are given explicitly. We will mainly concentrate on the solutions at the \textit{Lifshitz} point $z=3$, particularly, the static spherically symmetric topological charged black holes solution
\begin{equation}
ds^{2}=-f(r)dt^{2}+\frac{dr^{2}}{f(r)}+r^{2}d\Omega_{k}^{2}~,
\end{equation}
where, $d\Omega_{k}^2$ is the line element for a two dimensional Einstein space with constant scalar curvature $2k$. Without loss of generality, one can take $k = 0,\pm 1$ respectively. The form of the metric coefficient, for the detailed balance condition, is given by \cite{ref114},
\begin{equation}
f(r)=k+x^{2}-\sqrt{\alpha x-\frac{q^{2}}{2}}~
\label{BH}
\end{equation}
where, $x = \sqrt{-\Lambda}r$ and $ \Lambda(=-\frac{3}{l^{2}}) $ corresponds to the negative cosmological constant.
The physical mass and the charge ($ Q $) corresponding to the black hole solution are respectively given by, 
\begin{equation}
M=\frac{\kappa^{2}\mu^{2}\Omega_{k}\sqrt{-\Lambda}}{16}\alpha; \,\,\,\ Q=\frac{\kappa^{2}\mu^{2}\Omega_{k}\sqrt{-\Lambda}}{16} q
\end{equation}
where, $ \alpha $, $q$ are the integration constants, $ \Omega_{k} $ is the volume of the two dimensional Einstein space and $ \kappa $, $ \mu $ are the constant parameters of the theory.
The event horizon is the solution of the equation $f(r_+) = 0$.

\section{Phase transition}
 In ordinary thermodynamics the phase transition is studied by the divergence of relevant thermodynamic quantities. Here, the same technique will be adopted.  We shall first calculate the Hawking temperature, entropy and specific heat of the black hole using a \textit{canonical ensemble} frame work, which means that we shall carry out our analysis keeping the total charge ($ Q $) of the black hole fixed.  Finally, a graphical analysis will be given to study the phase transition. Before we proceed further, let us mention that in the following analysis we re-scale our variables as $ M\rightarrow \frac{M}{\Omega_{k}} $, $ S\rightarrow\frac{S}{\Omega_{k}} $, $ Q\rightarrow \frac{Q}{A \Omega_{k}\sqrt{-\Lambda}} $, where we have set $ 16\pi A=1 $ with $ A= \frac{\kappa^{2}\mu^{2}}{16}$. 

The Hawking temperature is calculated as,
\begin{equation}
T=\frac{f^{'}(r_+)}{4\pi}=\frac{\sqrt{-\Lambda}(3x_{+}^{4}+2kx_{+}^{2}-k^{2}-\frac{Q^{2}}{2})}{8\pi x_{+}(k+x_+^{2})}~
\label{t}
\end{equation}
and after the above mentioned re-scaling the entropy is found to be,
\begin{equation}
S=\int\frac{dM}{T}=\left(\frac{x_{+}^{2}}{4}+\frac{k}{2}lnx_{+} \right)+S_0~.
\label{s}
\end{equation}
In the above, $S_0$ is the integration constant which must be fixed by the physical consideration. To be precise, entropy is always determined upto some additive constant. But in all physical considerations the difference is important. Therefore in the present analysis we will consider only the difference. However, the integration constant $S_0$ could be determined using the usual thermodynamic prescription; i.e. determination of entropy in the $T\rightarrow 0$ limit, which might be the entropy of an extremal black hole.
\begin{figure}[h]
\centering
\includegraphics[angle=0,width=12cm,keepaspectratio]{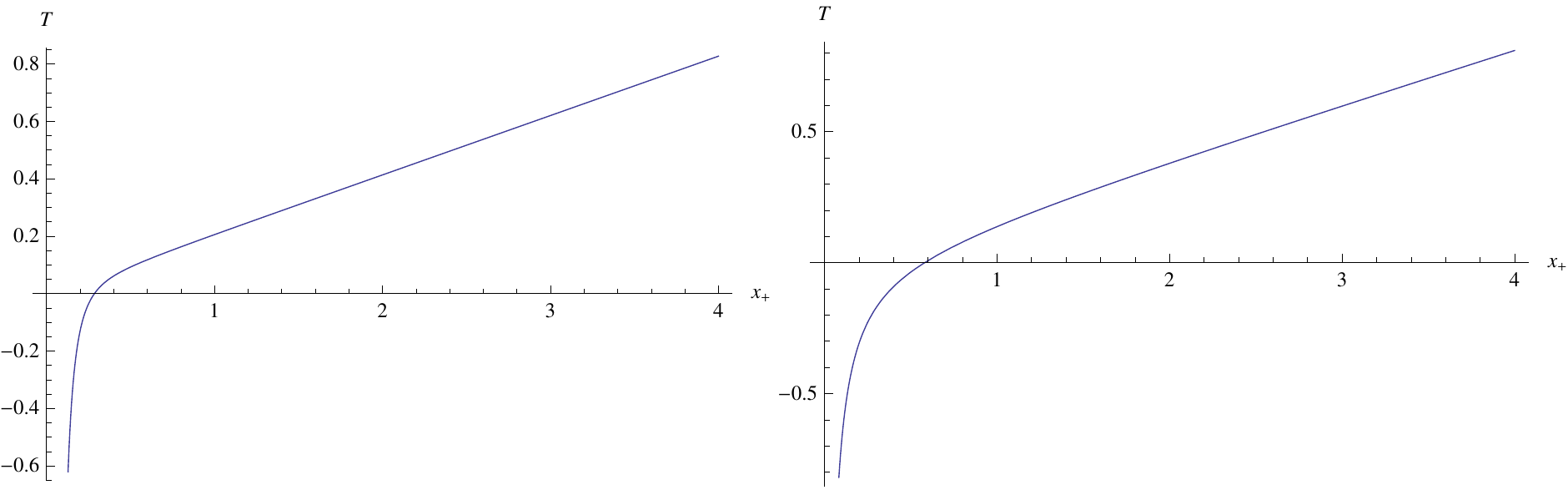}
\caption[]{\it Temperature ($ T $) plot of topological black holes for $ k=0,1 $ with respect to $x_{+}$ for $Q=5$ and $ l=1 $.}
\end{figure}

\begin{figure}[h]
\centering
\includegraphics[angle=0,width=12cm,keepaspectratio]{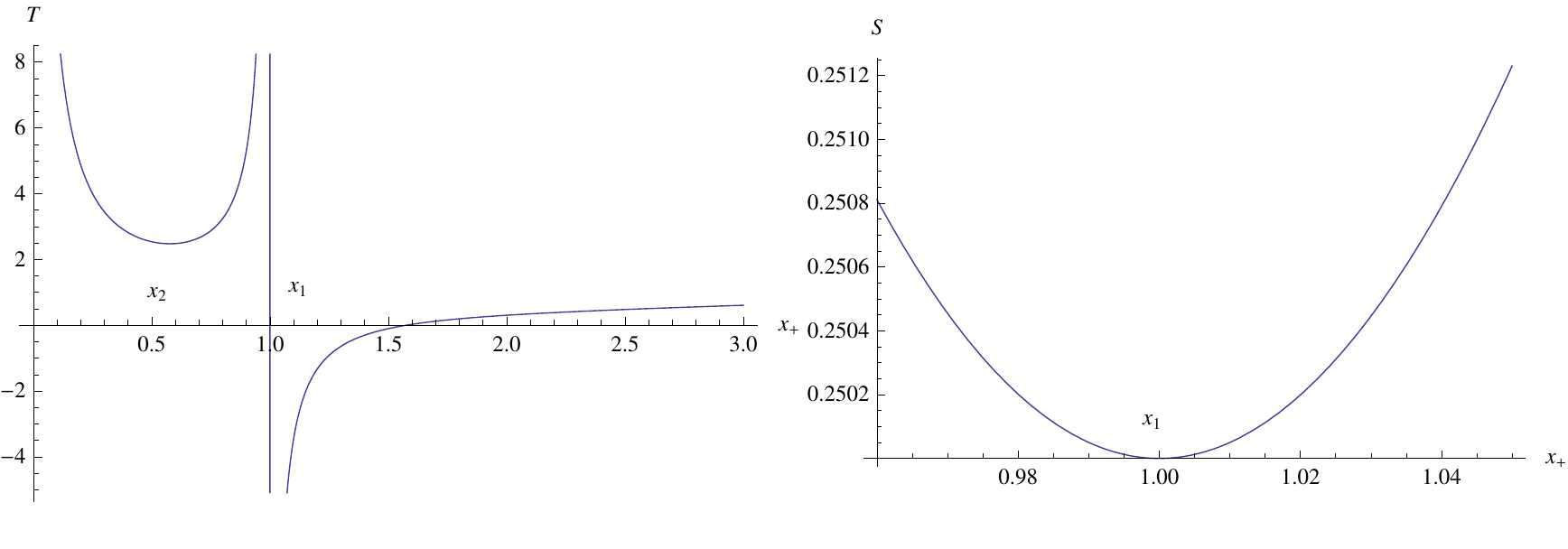}
\caption[]{\it Temperature ($ T $) and Entropy ($ S $) plot of topological black holes for $ k=-1 $ with respect to $x_{+}$ for $Q=5$ and $ l=1 $.}
\end{figure}

Before we proceed further, let us first try to analyze the behavior of Hawking temperature ($ T $) for different choices of $ k $. First of all, for the case $ k =0,1 $ we note that the Hawking temperature ($ T $) is a monotonically increasing function of horizon radius ($ x_+ $) (see fig. 3.12) which indicates that the corresponding heat capacities ($ C_Q $) are always positive definite. Therefore these black holes are globally stable and there is no (HP) phase transition.

Next consider the other case, $ k=-1 $. For this, the Hawking temperature (\ref{t}) reduces to the following form: 
\begin{equation}
T = \frac{\sqrt{-\Lambda}(3x_+^{2}+1)}{8\pi x_+} - \frac{\sqrt{-\Lambda}Q^{2}}{16 \pi x_+ (x_+^{2}-1)\label{tk}}.
\end{equation}
It it interesting to note that the Hawking temperature ($ T $) is always positive in the range $ 0<x_+<1 $. Furthermore, if we plot the entropy ($S$) as a function of $x_+$, using (\ref{s}), it shows that $S$ is also positive in this range (see fig. 3.13). Interestingly enough this condition is found to be valid for any value of the physical charge ($ Q $) of the black hole. On the other hand, we note that $ T\rightarrow -\infty $ as $ x_+\rightarrow 1 $. This is due to the fact that as $ x_+\rightarrow 1 $ the second term on the r.h.s of (\ref{tk}) dominates over the first one which ultimately produces a large negative temperature. This is clearly a nonphysical situation and the corresponding black hole solution does not exist for $ x_+\geq1 $.  Considering this fact, in the present paper we carry out our analysis in the {\it physical range} $ 0<x_+<1 $ where the temperature ($ T $) of the black hole is \textit{finite} as well as \textit{positive} definite. 

   In the above specified range ($ 0<x_+<1 $) we observe a change in slope at $ x_+=x_2 $ of the corresponding ($ T-x_+ $) plot (see fig. 3.13). This change in slope signals a discontinuity at $ x_+=x_2 $ in the corresponding heat capacity $ C_Q $. It gives a indication of \textit{second} order phase transition at $ x_+=x_2 $, which we shall refer as the critical point of phase transition\footnote{This situation is indeed quite different in the scenario of \textit{Einstein} gravity.  For \textit{hyperbolic} charged AdS black holes ($ k =-1 $) in the \textit{grand canonical} framework, the associated Gibbs free energy is always found to be negative which means that the black hole phase is \textit{globally} stable. However, using a \textit{canonical} ensemble it is found that the Helmholtz free energy could be positive which indicates that a HP transition might take place in this case \cite{ref26}.  A similar situation also holds for the Ricci flat AdS black holes ($ k = 0 $) in the \textit{Einstein} gravity where there is no HP phase transition between Ricci flat AdS black holes and thermal AdS in Poincare coordinates \cite{brimhm, ref30}. }. 
   
In order to make the discussion more transparent we  next compute the corresponding heat capacity ($ C_Q $).
Using (\ref{t}) and (\ref{s}) the heat capacity is determined as,
\begin{eqnarray}
C_Q&=&T\left(\frac{\partial S}{\partial T} \right)_{Q}=T\frac{\left(\frac{\partial S}{\partial x_+} \right)_{Q}}{\left(\frac{\partial T}{\partial x_+} \right)_{Q}}\nonumber\\
&=&\frac{(k+x_{+}^{2})^{2}(3x_{+}^{4}+2kx_+^{2}-k^{2}-\frac{Q^{2}}{2})}{6x_+^{6}+14kx_+^{4}+10k^{2}x_+^{2}+2k^{3}+Q^{2}(k+3x_+^{2})}~.
\label{cq} 
\end{eqnarray}
\begin{figure}[h]
\centering
\includegraphics[angle=0,width=8cm,keepaspectratio]{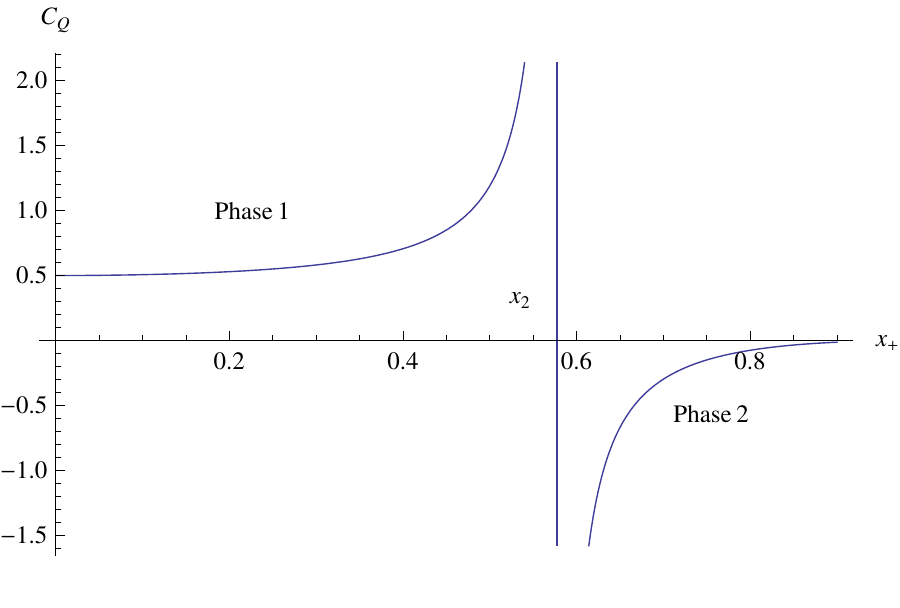}
\caption[]{\it Specific heat ($ C_{Q} $) plot of topological black holes for $ k=-1 $ with respect to $x_{+}$ for $Q=5$ and $ l=1 $.}
\end{figure}

From the above figure (Fig. 3.14), we see that specific heat ($ C_Q $) indeed suffers a discontinuity at $ x_+ = x_2 $. This shows that there is a genuine \textit{second} order phase transition at $ x_+ = x_2 $. For the phase 1 the heat capacity ($ C_Q $) is always found to be positive (see fig. 3.14) which means that this phase is thermodynamically {\it stable}. On the other hand, $ C_Q<0 $ for the phase 2 and therefore it is an {\it unstable phase}.  Given the above phase structure, as a second step, it will be quite natural to investigate the critical behavior of black holes in the \textit{Ho\v{r}ava-Lifshitz} theory of gravity near the critical point $x_2$, which we aim to discuss in the next section.

\section{Critical phenomena and scaling laws}

In this section, based on a thermodynamic approach, we explicitly calculate various critical exponents (that is associated with the \textit{second} order phase transition at $x_2$) for the \textit{hyperbolic} charged black holes (with $ k=-1 $) in the fixed charge ($ Q $) ensemble. To do that we shall basically try to extend our previous analytic methods (presented in sec 3.3) in the context of \textit{Ho\v{r}ava-Lifshitz} gravity. 

In order to calculate the critical exponents, as a first step, we write near the critical point,
\begin{eqnarray}
x_+=x_2(1+\Delta)
\label{eq1}
\end{eqnarray}
where $ |\Delta|<<1 $. Defining $T_2 \equiv T(x_2)$,
the Taylor expansion of the temperature ($ T $) about $ x_+=x_2 $ yields,
\begin{equation}
T=T_2+\Big[\left( \frac{\partial T}{\partial x_+}\right) _{Q}\Big]_{x_+=x_2}(x_+-x_2)+\frac{1}{2}\Big[\left(\frac{\partial^{2}T}{\partial x_+^{2}} \right)_{Q}\Big]_{x_+=x_2}(x_+-x_2)^{2}+ {\textrm{ higher order terms}}~.
\label{eq2} 
\end{equation}
It has been shown earlier that at the critical point $x_2$, $C_Q$ diverges (see Figure 3.14). Therefore (\ref{cq}) implies that the second term on the right hand side of (\ref{eq2}) vanishes. Hence neglecting the higher order terms and then using (\ref{eq1}) in (\ref{eq2}) we obtain,
\begin{equation}
\Delta = \sqrt{\frac{2}{D}}\frac{(T-T_2)^{1/2}}{x_2}
\label{eq3}
\end{equation}
where,
\begin{equation}
D=\left[ \left(\frac{\partial^{2}T}{\partial x_+^{2}} \right)_{Q}\right] _{x_+=x_2}=\frac{6x_2^{2}+3Q^{2}x_2^{2}-6x_2^{4}-6Q^{2}x_2^{4}-2-Q^{2}}{x_2^{3}(x_{2}^{2}-1)^{3}}.
\end{equation}

  Now to find the critical exponent $\alpha$ which is defined by the standard relation
\begin{equation}
C_Q\sim |T-T_2|^{-\alpha}~,
\label{alpha}
\end{equation}
we use the relation (\ref{eq1}) in (\ref{cq}) and then keep the terms only linear in $ \Delta $ to obtain
\begin{eqnarray}
C_Q \simeq \frac{\mathcal{N}(x_2,Q)}{\Delta (36x_2^{2}-56x_2^{4}+20x_2^{2}+6Q^{2}x_2^{2})}~,
\label{cq1}
\end{eqnarray}
with
\begin{equation}
\mathcal{N}(x_2,Q)=(x_2^{2}-1)^{2}(3x_2^{4}-2x_2^{2}-1-Q^{2}/2)~.
\end{equation}
Finally, using (\ref{eq3}) in the above we find the behavior of $ C_Q $ near the critical point $ x_+=x_2 $,
\begin{equation}
C_Q\simeq \frac{\mathcal{A}(x_2,Q)}{(T-T_2)^{1/2}}
\label{eq4}
\end{equation}
where,
\begin{equation}
\mathcal{A}(x_2,Q)=\frac{\mathcal{N}(x_2,Q)\sqrt{D}x_2}{\sqrt{2}(36x_2^{2}-56x_2^{4}+20x_2^{2}+6Q^{2}x_2^{2})}~.
\end{equation}
A comparison of (\ref{eq4}) with the standard relation (\ref{alpha}) yields $ \alpha=1/2 $.

  Next the determination of the critical exponent $ \beta $ which is defined through the relation (for a fixed value of charge $ Q $) as,
\begin{equation}
\Phi(x_+)-\Phi(x_2)\sim |T-T_2|^{\beta}
\label{eq5}
\end{equation}
will be done. Here the potential $\Phi(x)$ is given by
\begin{eqnarray}
\Phi(x) = \frac{Q}{x}~.
\label{potential}
\end{eqnarray}
To proceed, let us first Taylor expand $ \Phi(x_+)=\frac{Q}{x_+}$ about $x_+ = x_2$:
\begin{eqnarray}
\Phi(x_+) &=& \Phi(x_2) + \Big[\Big(\frac{\partial\Phi}{\partial x_+}\Big)_{Q}\Big]_{x_+=x_2} (x_+-x_2)+{\textrm{higher order terms}}
\nonumber
\\
&=&  \Phi(x_2)  - \frac{Q}{x_2^2} (x_+-x_2)+{\textrm{higher order terms}}~.
\label{phi1}
\end{eqnarray}
As before, neglecting the higher order terms and then using (\ref{eq1}) we obtain,
\begin{equation}
\Phi(x_+)-\Phi(x_2)=-\frac{Q}{x_2^{2}}\sqrt{\frac{2}{D}}(T-T_2)^{1/2}~,
\label{eq6}
\end{equation}
where in the final step (\ref{eq3}) also has been used.
This immediately determines $ \beta=1/2 $.

   To find out the exponent $ \gamma $, associated with the divergence of the inverse of the isothermal compressibility $ K_T^{-1} = \Big[Q\left(\frac{\partial \Phi}{\partial Q} \right)_{T}\Big]$, we first find the explicit expression of $K_T^{-1}$. The thermodynamic identity $\left(\frac{\partial \Phi}{\partial T} \right)_{Q} \left(\frac{\partial T}{\partial Q} \right)_{\Phi} \left(\frac{\partial Q}{\partial \Phi} \right)_{T}=-1$ yields
\begin{eqnarray}
\Big(\frac{\partial\Phi}{\partial Q}\Big)_{T} = -\Big(\frac{\partial\Phi}{\partial T}\Big)_{Q}\Big(\frac{\partial T}{\partial Q}\Big)_{\Phi}.
\label{KT1}
\end{eqnarray}
In order to evaluate the right hand side of (\ref{KT1}) first note that,
\begin{equation}
\Big(\frac{\partial\Phi}{\partial T}\Big)_{Q} = \frac{\Big(\frac{\partial\Phi}{\partial x_+}\Big)_{Q}}{\Big(\frac{\partial T}{\partial x_+}\Big)_{Q}}.\label{neweq1}
\end{equation}
Also, from the functional relation,
\begin{equation}
T=T(x_+,Q)
\end{equation}
we find,
\begin{equation}
\Big(\frac{\partial T}{\partial Q}\Big)_{\Phi} = \Big(\frac{\partial T}{\partial x_+}\Big)_{Q} \Big(\frac{\partial x_+}{\partial Q}\Big)_{\Phi} + \Big(\frac{\partial T}{\partial Q}\Big)_{x_+}.\label{neweq2}
\end{equation}
Right hand side of both (\ref{neweq1}) and (\ref{neweq2}) can be easily calculated using the relations (\ref{t}) and (\ref{potential}) which finally yields,
\begin{equation}
K_T^{-1}=Q\left(\frac{\partial \Phi}{\partial Q} \right)_{T} = \left( \frac{Q}{x_+}\right) \frac{6x_{+}^{6}+10x_+^{2}-14x_+^{4}-2+Q^{2}(x_+^{2}+1)}{6x_+^{6}+10x_+^{2}-14x_+^{4}-2+Q^{2}(3x_+^{2}-1)}.
\end{equation}
Then to obtain the near critical point expression, use (\ref{eq1}) and next (\ref{eq3}) in the above. This yields,
\begin{equation}
K_T^{-1}\simeq \frac{Q\sqrt{D}(6x_2^{6}+10x_2^{2}-14x_2^{4}-2+Q^{2}(x_2^{2}+1))}{\sqrt{2}(36x_2^{6}-56x_2^{4}+20x_2^{2}+6Q^{2}x_2^{2})}(T-T_2)^{-1/2}.
\end{equation}
Comparing with the standard relation 
$K_T^{-1}\sim |T-T_2|^{-\gamma}$,
defined for the fixed value of charge ($ Q $), we get $ \gamma =1/2 $.

   The critical exponent $ \delta $ is defined through the relation,
\begin{equation}
\Phi(x_+)-\Phi(x_2)\sim |Q-Q_2|^{1/\delta}
\label{delta}
\end{equation}
at constant temperature $ T $.
To find it, we first expand $ Q(x_+) $ in a sufficiently small neighborhood of $x_+= x_2 $ which yields,
\begin{eqnarray}
Q(x_+)&=& Q(x_2)+\left[ \left( \frac{\partial Q}{\partial x_+}\right)_{T}\right]_{x_+=x_2} (x_+-x_2)+\frac{1}{2} \left[ \left( \frac{\partial^{2} Q}{\partial x^{2}_+}\right)_{T}\right]_{x_+=x_2} (x_+-x_2)^{2}
\nonumber
\\
&+& {\textrm{higher order terms}}~.
\label{Q}
\end{eqnarray}
Since $ T=T(x_+,Q) $ (see Eq. (\ref{t})), one finds for the fixed $T$
\begin{equation}
\left( \frac{\partial Q}{\partial x_+}\right)_{T}=- \left( \frac{\partial T}{\partial x_+}\right)_{Q} \left( \frac{\partial Q}{\partial T}\right)_{x_+}~.
\label{eq7}
\end{equation}
Now at the critical point $x_+ = x_2$, $C_Q$ diverges and so, as earlier, $\Big[\left( \frac{\partial T}{\partial x_+}\right)_{Q}\Big]_{x_+=x_2}=0 $. Therefore, after neglecting the higher order terms, (\ref{Q}) reduces to
\begin{equation}
x_+ -x_2=\sqrt{\frac{2}{M}}(Q(x_+)-Q(x_2))^{1/2}\label{eq8}
\end{equation}
where,
\begin{equation}
M=\left[ \left( \frac{\partial^{2} Q}{\partial x^{2}_+}\right)_{T}\right]_{x_+=x_2}=\frac{18x_2^{8}+32x_2^{4}-44x_2^{6}-3Q^{2}x_2^{4}-4x_2^{2}-2-Q^{2}}{2Q^{2}x_2^{2}(x_2^{2}-1)^{2}}.
\end{equation}
Next since $\Phi(x_+) = \frac{Q}{x_+}$, we have the functional relation
$\Phi=\Phi(x_+,Q)$ and so
\begin{eqnarray}
\Big(\frac{\partial \Phi}{\partial x_+}\Big)_T = \Big(\frac{\partial\Phi}{\partial x_+}\Big)_Q + \Big(\frac{\partial\Phi}{\partial Q}\Big)_{x_+} \Big(\frac{\partial Q}{\partial x_+}\Big)_T~.
\label{phi2}
\end{eqnarray}
Therefore using (\ref{eq7}) we find,
\begin{equation}
\left[ \left( \frac{\partial \Phi}{\partial x_+}\right)_{T} \right]_{x_+=x_2} = \left[ \left( \frac{\partial \Phi}{\partial x_+}\right)_{Q} \right]_{x_+=x_2}=-\frac{Q}{x_2^{2}}.\label{eq9}
\end{equation}  
Finally, expanding $ \Phi(x_+) $ close to the critical point $ x_+\sim x_2 $ at constant $T$ and then using (\ref{eq8}) and (\ref{eq9}) we obtain,
\begin{equation}
\Phi(x_+)-\Phi(x_2)\simeq -\frac{Q}{x_2^{2}} \sqrt{\frac{2}{M}}\Big[Q(x_+)-Q(x_2)\Big]^{1/2}~.
\label{eq10}
\end{equation} 
Hence, the critical exponent is read off as $ \delta =2 $.

   Now re-expressing (\ref{eq4}) by using (\ref{eq3}) in the following form:
\begin{equation}
C_Q = \sqrt{\frac{2}{D}}\frac{{\cal{A}}}{x_2\Delta} =  \sqrt{\frac{2}{D}}\frac{{\cal{A}}}{(x_+-x_2)}~,
\end{equation}
and then substituting the value of $x_+-x_2$ from (\ref{eq8}) we obtain\footnote{$ Q(x_2)=Q_2 $.}
\begin{equation}
C_Q\sim \frac{1}{(Q(x_+)-Q_2)^{1/2}}~.
\label{eq11}
\end{equation}
Comparing this with the standard definition
$C_Q \sim {|Q(x_+)-Q_2|^{-\varphi}}$
we find $ \varphi=1/2 $.

 In the following the calculation, the critical exponent $ \psi $, defined by
\begin{equation}
S(x_+)-S(x_2)\sim |Q-Q_2|^{\psi}~,
\end{equation}
will be found out. Expansion of the entropy for fixed charge $Q$ about the critical point $x_2$ yields
\begin{eqnarray}
S(x_+) = S(x_2) + \Big[\Big(\frac{\partial S}{\partial x_+}\Big)_Q\Big]_{x_+ = x_2} (x_+ - x_2)+{\textrm{higher order terms}}~.
\label{new1}
\end{eqnarray}
Then following the identical steps as earlier and using (\ref{eq8}) we find
\begin{equation}
S(x_+)-S(x_2)\sim \frac{(x_2^{2}-1)}{2x_2}\sqrt{\frac{2}{M}}(Q-Q_2)^{1/2}~,
\end{equation}
which yields $ \psi = 1/2 $.

 In the following Table (3.7), we give the values of the critical exponents for the present example:
\begin{table}[h]
\caption{Various critical exponents and their values}   
\centering                          
\begin{tabular}{c c c c c c c c c}            
\hline\hline                        
$Critical~Exponents  $ & $\alpha$ & $\beta$  & $\gamma$ & $ \delta $ & $\varphi $ & $ \psi $ \\ [2.0ex]
\hline
Values & 1/2 & 1/2 & 1/2 & 2 & 1/2 & 1/2\\ [2.0ex]         
\hline                              
\end{tabular}
\label{E1}          
\end{table}

It is quite interesting to note that the critical exponents thus obtained corresponding to a \textit{second} order phase transition in the case of charged \textit{hyperbolic} black holes in \textit{Ho\v{r}ava-Lifshitz} theory of gravity are exactly equal to those obtained earlier for the \textit{Einstein-Born-Infeld} gravity with $k=1$. Therefore, as a natural consequence of this, it is quite trivial to show that black holes in the \textit{Ho\v{r}ava-Lifshitz} gravity also obey the \textit{thermodynamic scaling laws} (\ref{scaling laws}) in spite of differences in the nature of the black hole solutions as well as the phase structures. In this sense, it could be possible that the black holes in both type of theories fall under the same \textit{universality class}. 

Finally, we are in a position to check the additional scaling laws for the exponents $ \nu $ and $ \eta $ which are associated with the diverging nature of \textit{correlation length} and \textit{correlation function} near the critical point. Remember, like in the previous example, here also we do not explicitly compute these additional critical exponents. Instead we assume that the additional scaling relations,
\begin{equation}
\gamma = \nu (2-\eta),~~~ 2-\alpha = \nu d \label{add}
\end{equation}
to be valid in general. This immediately determines the other exponents. Taking $ d=3 $ and using the exponent values from Table 1, we finally obtain,
\begin{equation}
\nu =1/2,~~~ \eta =1.
\end{equation}
Like we mentioned earlier, in practical situations, the assumption that the scaling laws (\ref{add}) are valid, may not be true for black holes. In fact one should be able to find out a novel way to calculate the correlation between certain scalar modes in the curved back ground that eventually clarifies our findings. Therefore the relations (\ref{add}) are highly suggestive. Hence it must be checked that whether these are indeed valid or not. In other words, one should find the above exponents by some explicit computations. 

\section{Discussions}

In this chapter, based on a \textit{canonical} framework, we have provided an approach to study the issue of critical phenomena for charged black holes in AdS space. In order to make a comparative study between the \textit{scaling} behavior  of black holes in various theories of gravity, in the present analysis we consider black hole solutions in \textit{Einstein Born Infeld} gravity as well as in \textit{Ho\v{r}ava-Lifshitz} gravity. A number of interesting features have been observed in this regard. In the following we summarize our findings.

First of all, for \textit{Einstein Born Infeld} gravity, we note that the boundedness on the parameter space ($ b $ and $ Q $) that exists in ($3+1$) dimension eventually disappears in higher dimensions. Furthermore, based on a thermodynamic framework we have systematically addressed  various aspects of the critical phenomena in higher dimensional Born-Infeld AdS (BI AdS) black holes. Considering the appropriate limit ($ b\rightarrow\infty $) the critical behavior of higher dimensional Reissner Nordstrom AdS (RN AdS) black holes has also been studied simultaneously. During this exercise we have explicitly calculated all the \textit{static} critical exponents ($\alpha=1/2,\beta=1/2,\gamma=1/2,\delta=2,\varphi=1/2,\psi=1/2$) both for the BI AdS and RN AdS black holes. They are also found to satisfy the so called \textit{thermodynamic scaling relations} (\ref{scaling laws}) near the critical point which is characterized by the discontinuity in the heat capacity at constant charge ($ C_Q $).  We have also explored the \textit{static scaling hypothesis} which has been found to be compatible with the \textit{scaling} relations near the critical point. The \textit{scaling parameters} have been found to possess identical values ($ p=q=2/3 $). It is quite interesting to note that the \textit{static} critical exponents (Table 3.6) thus obtained are independent of the spatial dimensionality of the AdS space time. This indeed suggests the \textit{mean field} behavior in black holes as  thermodynamic systems and bolsters our confidence in applying \textit{Ehrenfest's} scheme to study phase transitions. On the top of it, from our analysis the \textit{thermodynamic scaling relations} for black holes are found to be valid in any dimensions. This is also compatible with a \textit{mean field} analysis that illustrates the dimension independence of the \textit{static} critical exponents in usual thermodynamic systems. As a matter of fact, from our analysis we find that both BI AdS and RN AdS black holes posses identical critical exponents and thereby similar \textit{scaling} behavior near the critical point. 

During the rest of our analysis, we adopted the standard thermodynamic approach to explore the phase structure of the topological charged black holes in \textit{Ho\v{r}ava-Lifshitz} gravity for the \textit{Lifshitz} point $z=3$. These black holes were introduced earlier  in \cite{ref114}. In \cite{ref114} the authors have computed the general expression for the Hawking temperature for these black holes. In spite of this  attempt, till date the issue regarding the nature of phase transition, particularly the issue of critical phenomena, was completely unexplored. In the present analysis, we attempted to provide an answer to all these questions. A number of interesting features have been observed in this regard which have never been explored so far to the best of our knowledge. These are as follows: 
\vskip 2mm
\noindent
$\bullet$  Black holes with spherical ($ k = 1 $) or planar ($ k = 0 $) topology always possess a positive specific heat and thereby corresponds to a \textit{globally} stable phase.
\vskip 2mm
\noindent
$\bullet$ For \textit{hyperbolic} charged black holes ($ k = -1 $), there exits a {\it physical range} $0<x_+<1$, where the black hole solution exists.
\vskip 2mm
\noindent 
$\bullet$  Within this \textit{physical} range, there happens to be a \textit{second} order phase transition from an unstable black hole to a locally stable phase with positive heat capacity.
\vskip 2mm
\noindent
$\bullet$  The corresponding situation is indeed \textit{opposite} in the \textit{Einstein} gravity. In the \textit{Einstein} gravity the (HP) phase transition always takes place for AdS black holes having \textit{spherical} topology ($ k = 1 $). On the other hand, the Ricci flat ($ k = 0 $) and the \textit{hyperbolic} ($ k = -1 $) AdS black holes are \textit{usually} found to posses positive heat capacity which means that these black holes are not only stable locally but also stable \textit{globally} \cite{ref26, brimhm, ref30}.
\vskip 2mm
\noindent
$\bullet$ Finally, the critical exponents near the critical point of the phase transition were derived. This was done again using the ordinary thermodynamic analogy. It may be noted that the critical point has been marked by the discontinuity of the heat capacity ($C_{Q}$) which suggests that it is a \textit{second} order phase transition. Moreover, as a point of conformation, we also found that the critical exponent, associated with the divergence in $C_Q$, is $1/2$ which simultaneously indicates a \textit{mean field} feature as well as a \textit{second} order nature of the phase transition. Interestingly, the critical exponents those are obtained in the context of HL gravity are found to be exactly identical to those of \textit{Einstein-Born-Infeld} theory and also they are found satisfy the \textit{thermodynamic scaling laws}. This in fact suggests two remarkable facts: (1) black holes in both the \textit{Einstein} as well as \textit{Ho\v{r}ava-Lifshitz} (HL) theory of gravity fall under the same \textit{universality class}, (2) there exists a universal \textit{mean field} behavior in both of these gravity theories. 

Before we conclude this chapter and move towards the next, it is customary to mention that for any gravitational theory constructed in an asymptotically AdS background, it is highly likely that there exists some dual description in terms of certain CFTs living on the boundary of the AdS.  Consequently the phase transition that we have been discussing so far, should have its natural interpretation in terms of boundary CFTs \footnote{According to Witten the HP transition could be realized as a confinement/deconfinement transition taking place in the dual strong coupling CFTs living on the boundary of the AdS \cite{ref17}.}. As a matter of fact, using the basic framework of \textit{AdS/CFT} duality \cite{ref54}-\cite{ref58}, it is in fact always possible to establish a mapping between the phase transition occurring in the dual CFTs to that with the phase transition occurring in the bulk gravitational theory in an asymptotically AdS space time \cite{ref26, brimhm,ref30}.      
  
Keeping the spirit of such an illuminating idea, in the next two chapters (Chapter 4 and Chapter 5), based on the basic principles of the \textit{AdS/CFT} duality, we aim to study gravity theories in an asymptotically AdS space, where due to instability occurring at low temperatures, the RN AdS (or, more specifically for our analysis the BI AdS) black holes develop \textit{scalar hair} below certain critical value of the temperature ($ T<T_c $) \cite{ref65,ref66}. Therefore in the bulk gravity theory we clearly have two distinct phases, one is the phase with the (planar) AdS black hole with no \textit{hair} and the other is the phase with AdS black holes developing a \textit{scalar hair} \cite{ref69}. Our aim will be to study the consequence of such a transition in the boundary CFTs where we shall try to carry out an explicit analytic computation of various quantities (like the order parameter, the critical temperature etc.) that characterize the phase transition, using the so called \textit{AdS/CFT} dictionary.

\chapter{AdS/CFT Duality And Phase Transition In Black Holes: Holographic Superconductors}
\section{Holographic $ s $- wave superconductors}
For the past sixty years, the BCS theory of superconductivity has been the most successful microscopic theory to describe weakly coupled superconductors with great accuracy \cite{bcs}. The basic principle that is responsible to exhibit superconductivity in these weakly coupled systems is the spontaneous breaking of $ U(1) $ symmetry at low temperatures \cite{wnbg}. However, it has been realized for quite some time that there are some materials, like heavy fermion compounds or high $ T_c $ cuprates where the understanding of the pairing mechanism remains incomplete \cite{ref87}. The failure of BCS theory in order to understand such strongly coupled systems invites new theoretical inputs. Surprisingly one such input comes from the so called AdS/CFT correspondence \cite{ref54}-\cite{ref58}.

The \textit{AdS/CFT} duality has been a powerful tool to deal with strongly coupled systems. It provides an exact correspondence between the gravity theory in a $ (d+1) $ dimensional AdS space time and a conformal field theory (CFT) residing on its $d$-dimensional boundary. Since the pioneering work of Witten \cite{ref17}, it has been known for quite some time that under the spell of \textit{AdS/CFT} correspondence, it is indeed possible to find a mapping between the phase structure of AdS black holes in the bulk gravitational theory to that with the dual CFTs living on the boundary of the AdS. Extending this idea further, in recent years the \textit{AdS/CFT} correspondence has been found to provide some meaningful theoretical insights in order to understand the physics of so called high $T_c$ superconductors, which are thereby termed as \textit{holographic superconductors}.

 The bulk gravitational description of holographic $s$-wave superconductors basically consists of a black hole and a complex scalar field minimally coupled to an abelian gauge field. The formation of \textit{scalar hair} below certain critical temperature ($T_c$) indicates the onset of a \textit{second} order phase transition in the bulk and a corresponding condensation in the dual CFTs. The mechanism that is responsible behind the formation of \textit{scalar hair} in the bulk is the breaking of a local $U(1)$ symmetry near the event horizon of the black hole at low temperatures \cite{ref65}-\cite{ref71}.
 
 According to the basic principles of \textit{AdS/CFT} duality, the gauge symmetries in the bulk gravitational theory corresponds to a global $ U(1) $ symmetry in the boundary CFTs. Therefore, the formation of scalar hair below $ T<T_c $ spontaneously breaks the global $ U(1) $ symmetry in the boundary CFTs. Therefore truly speaking the boundary theory should behave as a super fluid \cite{hsfluid} rather than a superconductor. When we talk about holographic superconductors then we mean that the boundary $ U(1) $ theory is weakly gauged, which is thereby broken due to the formation of scalar hair below $ T<T_c $ \cite{ref69}.
 
However considering the gauge theory sector, the onset of superconductivity in ($ 2+1 $) dimensions has always been a puzzle. This is because of the fact that, according to Mermin-Wagner theorem, due to the presence of large fluctuations there should not be any spontaneous symmetry breaking in ($ 2+1 $) dimensions.
The reason that we still have a holographic model of superconductors in ($ 2+1 $) dimensions is because of the fact that the fluctuations get suppressed in the large $ N $ limit where the bulk theory is described by the classical gravity formulated in an AdS space.

Till date, most of the investigations that have been performed on various holographic models superconductor are mostly based on the framework of usual Maxwell electrodynamics \cite{ref77}-\cite{ref85}.  
Recently, investigations have also been carried out in 
the framework of non-linear electrodynamics \cite{binwang}-\cite{ref86}. 
In particular, the effects of Born-Infeld (BI) corrections on the holographic superconductors has been
studied \textit{numerically} in \cite{binwang}. The analysis is important in its own right as  
BI electrodynamics is free from infinite self energies of charged point particles that arises in the Maxwell theory and it is also invariant under the electromagnetic duality transformations. All these features are sufficiently motivating to study \textit{Einstein} gravity in the frame work of BI electrodynamics.

The present chapter is based on the papers \cite{ref102,ref103}, where we explore the consequences of the phase transition occurring in the AdS black holes to that in the boundary CFTs. Using the \textit{AdS/CFT} dictionary, in the present analysis we explicitly compute various physical quantities (like, the critical temperature ($ T_c $), order parameter etc.) associated with this phase transition.  Based on the Sturm-Liouville (SL) eigenvalue problem\footnote{For details see the Apendix.} \cite{ref77}, we present a complete \textit{analytic} treatment in order to compute these quantities which finally resolves some of the major issues related to the holographic model of $s$-wave superconductors in the framework of BI electrodynamics. We perform our analysis both for the Schwarzschild AdS as well as Gauss Bonnet (GB) AdS back grounds. In the former case, we analytically establish (upto leading order in the BI coupling parameter ($ b $)) the relation between the critical temperature and the charge density and also the fact that at low temperatures $(T<T_c)$, the condensation is indeed affected due to the presence of BI coupling parameter ($ b $). The critical exponent for the condensation near the critical temperature also comes out naturally in our analysis.  Moreover, our results have been found to be in good agreement to that with the numerical results  existing in the literature \cite{binwang}. 

In the later case we present analytic studies of Gauss-Bonnet holographic superconductors in BI electrodynamics upto leading order in the Gauss-Bonnet (GB) as well as in the BI coupling parameter in the probe limit. This means that we study the effect of the leading possible higher derivative corrections to the onset of $ s $-wave order parameter condensation. The study of the effects of the higher curvature corrections on $ (3+1) $-dimensional holographic superconductors is motivated by the application of the Mermin-Wagner theorem (which forbids continuous symmetry breaking in $ (2+1) $-dimensions because of large fluctuations in lower dimensions) to the holographic superconductors \cite{ref78}.
The results that we obtain analytically are found to be in good agreement to that with the results obtained using numerical computations. At this stage, it is reassuring to note that all our calculations have been carried out in the \textit{probe} limit\footnote{By probe limit we mean that the back reaction of the matter fields on the metric could be ignored when the charge of the black hole is large enough \cite{ref69}.}. We feel that the analysis that is done is worthwhile since very little has been done in problems involving non-linear electrodynamics both analytically as well as numerically.       
 
Before we proceed further, let us first mention about the organization of the chapter. In section 4.2, we provide the basic holographic set up for the $s$-wave superconductors in the framework of BI electrodynamics, considering the background of a Schwarzschild-AdS space time.
In section 4.3, ignoring the back reaction of the dynamical matter field on the space time metric and using the perturbation technique, we compute the critical temperature in terms of a solution to the SL eigenvalue problem. In section 4.4,
we determine the temperature dependence of the $ s $- wave order parameter upto linear order in the BI coupling parameter ($ b $). In section 4.5, we give the basic setup for the holographic superconductors in Gauss-Bonnet (GB) AdS gravity. In section 4.6, we explicitly compute the relationship between the critical temperature and charge density for holographic superconductors in GB gravity.  In section 4.7, we compute the corresponding condensation operator near the critical temperature. Finally we draw our conclusion in section 4.8.

\section{Holographic superconductors in Schwarzschild AdS background}

We start with our construction of the holographic $s$-wave superconductors based on the fixed
background of Schwarzschild-AdS spacetime.
The metric of a planar Schwarzschild-AdS black hole reads,
\begin{eqnarray}
ds^2=-f(r)dt^2+\frac{1}{f(r)}dr^2+r^2(dx^2+dy^2)
\label{m1}
\end{eqnarray}
with,
\begin{eqnarray}
f(r)=r^{2}-\frac{r_{+}^3}{r}
\label{metric}
\end{eqnarray}
in units in which the AdS radius is unity, i.e. $l=1$.
The Hawking temperature is related to the horizon radius ($r_+$) as,
\begin{eqnarray}
T=\frac{3r_+}{4\pi}~.
\label{tempbisch}
\end{eqnarray}

Let us now consider an electric field and a charged complex scalar
field in this fixed background. The corresponding
Lagrangian density can be expressed as,
\begin{eqnarray}
\mathcal{L}=\mathcal{L}_{BI}-
|\nabla_{\mu} \psi- i A_{\mu}\psi|^2- m^2| \psi|^2
\label{m10}
\end{eqnarray}
where $\psi$ is a charged complex scalar field, $\mathcal{L}_{BI}$
is the Lagrangian density of the Born-Infeld electrodynamics \cite{binwang}
\begin{eqnarray}
\mathcal{L}_{BI}=\frac{1}{b}\bigg(1-\sqrt{1+\frac{b F}{2}}\bigg).
\label{m11}
\end{eqnarray}
Here $F\equiv F_{\mu\nu}F^{\mu\nu}$ and $F_{\mu\nu}$ is the
non-linear electromagnetic tensor which satisfies the BI
equation,
\begin{eqnarray}
\partial_{\mu}\bigg(\frac{\sqrt{-g}F^{\mu\nu}}{\sqrt{1+\frac{b F}{2}}}\bigg)=J^{\nu}
\label{m12}
\end{eqnarray}
with the BI coupling parameter $b$ indicating the difference between BI
and Maxwell electrodynamics. In the limit $b\rightarrow 0$, the Lagrangian
$\mathcal{L}_{BI}$ approaches to $-\frac{1}{4}F_{\mu\nu}F^{\mu\nu}$,
and one recovers the standard Einstein-Maxwell theory. 

 In order to solve the equations of motion both for the complex scalar field and the electromagnetic field,
we adopt the following ansatz \cite{ref70}
\begin{eqnarray}
A_{\mu}=(\phi(r),0,0,0),\;\;\;\;\psi=\psi(r)
\label{vector}
\end{eqnarray}
which finally yields the equations of motion for the complex scalar
field $\psi(r)$ and electrical scalar potential $\phi(r)$ as,
\begin{eqnarray}
\psi^{''}(r)+\left(\frac{f'}{f}+\frac{2}{r}\right)\psi'(r)
+\left(\frac{\phi^{2}(r)}{f^2}-\frac{m^2}{f}\right)\psi(r)=0\label{e1}
\end{eqnarray}
and
\begin{eqnarray}
\phi''(r)+\frac{2}{r}\phi'(r)\bigg(1-b\phi'^2 (r)\bigg)
-\frac{2\psi^2 (r)}{f}\phi(r)\bigg(1-b\phi'^2 (r)\bigg)^{3/2}=0\label{e2}
\end{eqnarray}
where prime denotes derivative with respect to $r$. In order to solve the
non-linear equations (\ref{e1}) and (\ref{e2}), we need
to know the boundary conditions for $\phi$ and $\psi$ near the black
hole horizon $r\sim r_+$ and at the spatial infinite
$r\rightarrow\infty$. The regularity condition at the horizon gives
the boundary conditions $\phi(r_+)=0$ and
$\psi=-\frac{3r_+}{2}\psi'$. 

 Under the change of coordinates $z=r_{+}/r$,  the field equations (\ref{e1}}) and (\ref{e2}}) turn out to be,
\begin{eqnarray}
z\psi''(z)-\frac{2+z^3}{1-z^3}\psi'(z)
+\left[z\frac{\phi^{2}(z)}{r_{+}^{2}(1-z^3)^2}-\frac{m^2}{z(1-z^3)}\right]\psi(z)=0\label{e1a}
\end{eqnarray}
\begin{eqnarray}
\phi''(z)+\frac{2bz^3}{r_{+}^2 }\phi'^{3}(z)-\frac{2\psi^2 (z)}{z^2 (1-z^3)}
\left(1-\frac{bz^4}{r_{+}^2}\phi'^{2}(z)\right)^{3/2}\phi(z) =0\label{e1aa}
\end{eqnarray}
where prime now denotes derivative with respect to $z$. These equations are to be solved in the
interval $(0, 1)$, where $z=1$ is the horizon and $z=0$ is the boundary.
The boundary condition $\phi(r_+)=0$ now becomes $\phi(z=1)=0$.

Setting $m^2$ close to BF bound \cite{bf1}-\cite{bf2}, the asymptotic boundary conditions for
the scalar potential $\phi(z)$ and the scalar field $\psi(z)$ turn out to be
\begin{eqnarray}
\phi\approx\mu-\frac{\rho}{r}=\mu-\frac{\rho}{r_{+}}z
\label{b2}
\end{eqnarray}
\begin{eqnarray}
\psi\approx\frac{\psi^{(-)}}{r^{\Delta_{-}}}+\frac{\psi^{(+)}}{r^{\Delta_{+}}}\label{b1}
\end{eqnarray}
where, 
\begin{eqnarray}
\Delta_{\pm}=\frac{3}{2}\pm\sqrt{\frac{9}{4}+m^2}
\label{dimension}
\end{eqnarray}
is the conformal dimension of the condensation operator $\mathcal{O}$ in the boundary
field theory. 
The coefficients $\psi^{(-)}$ and $\psi^{(+)}$
correspond to the vacuum expectation values of the condensation
operator $\mathcal{O}$ dual to the scalar field. Also $\mu$ and $\rho$ are 
interpreted as the chemical potential and the charge density of the dual theory
on the boundary.
Setting $m^2 =-2$ in eq.(\ref{dimension}), we have $\Delta_{-}=1$ and $\Delta_{+}=2$.
As in \cite{binwang}, we can impose the boundary condition that either $\psi^{(-)}$ or
$\psi^{(+)}$ vanish, so that the theory is stable in the asymptotic AdS region and we have a corresponding condensation without being sourced.
In our analysis we shall set $\psi^{(+)}=0$ and $\langle \mathcal{O}\rangle=\psi^{(-)}$. 

\section{Critical temperature for the condensation}
With the above set up in place, we are now in a position to investigate the relation between the critical temperature and the charge density. 

At the critical temperature $T_c$, $\psi=0$, so the field equation (\ref{e1aa})
for the electrostatic potential $\phi$ reduces to
\begin{eqnarray}
\phi''(z)+\frac{2bz^3}{r_{+(c)}^2}\phi'^{3}(z)=0.
\label{e1b}
\end{eqnarray}
To solve the above equation, we set $\phi'(z)=\xi(z)$ and obtain 
\begin{eqnarray}
\xi'(z)+\frac{2bz^3}{r_{+(c)}^2 }\xi^{3}(z)=0.
\label{eqn}
\end{eqnarray}
Integrating the above equation in the interval $[0, 1]$, we get 
\begin{eqnarray}
\frac{1}{\xi^{2}(1)}-\frac{1}{\xi^{2}(0)}=\frac{b}{r_{+(c)}^2}
\label{eqn1}
\end{eqnarray}
where we have used the fact that $\xi=\xi(0)$ at $z=0$ and $\xi=\xi(1)$ at $z=1$. 

At $z=0$, from eq.(\ref{b2}) we have
\begin{eqnarray}
\phi'|_{z=0}=\xi(0) &\approx&-\frac{\rho}{r_{+}}
=-\frac{\rho}{r_{+(c)}} \quad at ~ T=T_c ~.
\label{b22}
\end{eqnarray}
From eq(s)(\ref{eqn1}, \ref{b22}), we obtain
\begin{eqnarray}
\frac{1}{\xi^{2}(1)}=\frac{b}{r_{+(c)}^2}+ \left(\frac{r_{+(c)}}{\rho}\right)^2 ~.
\label{eqn1a}
\end{eqnarray}
Hence, integrating eq.(\ref{eqn}) in the interval $[1, z]$ and using eq.(\ref{eqn1a}) leads to
\begin{eqnarray}
\xi(z)=\phi'(z)=-\frac{\lambda r_{+(c)}}{\sqrt{1+b\lambda^2 z^4}}
\label{b20}
\end{eqnarray}
where 
\begin{eqnarray}
\lambda=\frac{\rho}{r_{+(c)}^2}
\label{lam}
\end{eqnarray}
and the negative sign has been taken before the square root in the expression for $\phi'(z)$ since $\phi'(0)$
is negative at $z=0$ (eq.\ref{b22}).

Integrating eq.(\ref{b20}) again from $z'=1$ to $z'=z$, we obtain
\begin{eqnarray}
\phi(z)=\int_{1}^{z}\frac{\lambda r_{+(c)}}{\sqrt{1+b\lambda^2 z'^{4}}}dz'
\label{lam1}
\end{eqnarray}
where we have used the fact that $\phi(z=1)=0$.

The above integral is not doable exactly and hence we shall expand the integrand binomially
and keep terms upto $\mathcal{O}(b)$ to get
\begin{eqnarray}
\phi(z)=\lambda r_{+(c)}(1-z)\left\{1-\frac{b\lambda^2}{10}(1+z+z^2 +z^3 +z^4)\right\}\quad,\quad b\lambda^2 < 1.
\label{sol}
\end{eqnarray}
Note that the above solution satisfies eq.(\ref{e1b}) upto $\mathcal{O}(b)$ along with the boundary condition $\phi(z=1)=0$.

 Using the above solution, we find that as $T\rightarrow T_c$, 
the field equation for the scalar field $\psi$ approaches the limit 
\begin{eqnarray}
-\psi'' +\frac{2+z^3}{z(1-z^3)}\psi'
+\frac{m^2}{z^2 (1-z^3)}\psi(z)
=\frac{\lambda^2 }{(1+z+z^2)^2}\left\{1-\frac{b\lambda^2 \zeta(z)}{5}\right\}\psi(z)
\label{e001}
\end{eqnarray}
where $\zeta(z)=(1+z+z^2 +z^3 +z^4)$.

 Near the boundary, we define \cite{ref77}
\begin{eqnarray}
\psi(z)=\frac{\langle \mathcal{O}\rangle}{\sqrt 2 r_+} zF(z)
\label{sol1}
\end{eqnarray}
where $F(0)=1$.
Substituting this form of $\psi(z)$ in eq.(\ref{e001}), we arrive at the following equation,
\begin{eqnarray}
- F''(z) + \frac{3z^2}{1-z^3}F'(z) + \frac{z}{1-z^3}F(z)
&=&\frac{\lambda^2 }{(1+z+z^2)^2}\left\{1-\frac{b\lambda^2}{5}\zeta(z)\right\}F(z) \nonumber\\
&\approx&\frac{\lambda^2 }{(1+z+z^2)^2}\left\{1-\frac{b(\lambda^2|_{b=0})}{5}\zeta(z)\right\}F(z)\nonumber\\
\label{eq5b}
\end{eqnarray}
which is to be solved subjected to the boundary condition $F' (0)=0$. Note that in the second line we have used the fact
that $b\lambda^2 =b(\lambda^2|_{b=0}) +\mathcal{O}(b^2)$, where $\lambda^2|_{b=0}$ is the value of $\lambda^2$ for $b=0$.

The above equation can be put in the standard Sturm-Liouville (SL) form (see eq.(\ref{app6}) in the Appendix)
with 
\begin{eqnarray}
p(z)=1-z^3~,~ q(z)=z~,~r(z)=\frac{1-z}{1+z+z^2}\left\{1-\frac{b(\lambda^2|_{b=0})}{5}\zeta(z)\right\}. 
\label{i1}
\end{eqnarray}
With the above identification, we now write down the eigenvalue $\lambda^2$ which minimizes the expression (see Appendix, eq. (\ref{app1}}))
\begin{eqnarray}
\lambda^2 [F(z)] = \frac{\int_0^1 dz\ \{ (1-z^3) [F'(z)]^2 + z [F(z)]^2 \} }{\int_0^1 dz \ \frac{1-z}{1+z+z^2}
\left\{1-\frac{b(\lambda^2|_{b=0})}{5}\zeta(z)\right\} [F(z)]^2}~.
 \label{eq5abc}
\end{eqnarray}

To estimate it, we use the following trial function \cite{ref77}
\begin{eqnarray}
F= F_\alpha (z) \equiv 1 - \alpha z^2
\label{eq50}
\end{eqnarray}
which satisfies the conditions $F(0)=1$ and $F'(0)=0$.

In the following we explicitly calculate the critical temperature ($ T_c $) corresponding to different choice in the BI parameter ($ b $). In the following we discuss a few of these cases separately.
 
 For $b=0$, we obtain
\begin{eqnarray}
\lambda_\alpha^2|_{b=0} = \frac{6-6\alpha + 10\alpha^2}{2\sqrt 3 \pi - 6\ln 3
+ 4 (\sqrt 3 \pi +3\ln 3 - 9)\alpha + (12\ln 3 - 13)\alpha^2} 
\end{eqnarray}
which attains its minimum at $\alpha \approx 0.2389$. Hence, we have
\begin{eqnarray}
\lambda^2|_{b=0} \approx \lambda_{0.2389}^2|_{b=0} \approx 1.268 
\end{eqnarray}
to be compared with the exact value $\lambda^2|_{b=0} = 1.245$.
The critical temperature therefore reads 
\begin{eqnarray}
T_c = \frac{3}{4\pi} r_{+(c)} = \frac{3}{4\pi} \sqrt{\frac{\rho}{\lambda|_{b=0}}}\approx 0.225\sqrt\rho 
\label{eqTc}
\end{eqnarray}
which is found to be matching exactly upto three decimal places to that with the value $T_c = 0.225\sqrt\rho$ \cite{binwang} obtained by numerical analysis.

For $b=0.1$, we obtain
\begin{eqnarray}
\lambda_\alpha^2|_{b=0.1} = \frac{\frac{1}{2}-\frac{\alpha}{2}+\frac{5\alpha^2}{6}}
{0.344-0.082\alpha +0.014\alpha^2} 
\end{eqnarray}
which attains its minimum at $\alpha \approx 0.2402$. Hence, we have
\begin{eqnarray}
\lambda^2|_{b=0.1} \approx \lambda_{0.2402}^2|_{b=0.1} \approx 1.317. 
\end{eqnarray}
The critical temperature therefore reads 
\begin{eqnarray}
T_c = \frac{3}{4\pi} r_{+(c)} = \frac{3}{4\pi} \sqrt{\frac{\rho}{\lambda|_{b=0.1}}}\approx 0.223\sqrt\rho 
\label{eqTc}
\end{eqnarray}
which again matches exactly to that with the result $T_c = 0.223\sqrt\rho$ obtained numerically in \cite{binwang}. Likewise one can obtain other values of $ T_c $ corresponding to different choices in $ b $.

From the above analysis it is indeed evident that the analytic scheme that we adopt here works quite nicely in the probe approximation. In the next section we extend this technique further in order to calculate the $ s $- wave order parameter ($ \langle \mathcal{O} \rangle $) near the critical point of the phase transition.


\section{$ s $ - wave order parameter}
In this section, employing our previous analytic technique, we shall explicitly compute the condensation operator ($\langle \mathcal{O}\rangle $) as a function of temperature. This will eventually help us to identify the critical exponent that is associated with the phase transition. 

 Away from (but close to) the critical temperature $T_c$, the field equation (\ref{e1aa}) for $\phi$ becomes (using eq.(\ref{sol1}))
\begin{eqnarray}
\phi''(z) +\frac{2bz^3}{r_{+}^2}\phi'^{3}(z)&=&\frac{\langle \mathcal{O}\rangle^2}{r_+^{2}}\mathcal{B}(z)\phi(z)\label{aw1} \\
\mathcal{B}(z)&=&\frac{F^{2}(z)}{1-z^3}\left(1-\frac{3bz^4}{2r_{+}^2}\phi'^{2}(z)\right)+ O(b^2)\nonumber
\end{eqnarray}
where the parameter $\langle \mathcal{O}\rangle^2/r_+^{2}$ is small.
We may now expand $\phi(z)$ in the small parameter $\langle \mathcal{O}\rangle^2/r_+^{2}$ as
\begin{eqnarray}
\frac{\phi}{r_+}=\lambda (1-z)\left\{1-\frac{b\lambda^2}{5}\zeta(z)\right\}+ \frac{\langle \mathcal{O}\rangle^2}{r_+^{2}} \chi(z) 
+\dots
\label{aw2} 
\end{eqnarray}
From eq(s)(\ref{aw1}, \ref{aw2}) (keeping terms upto $O(b)$), we obtain the equation for
the correction $\chi(z)$ near the critical temperature
\begin{eqnarray}
\chi''(z) +6b\lambda^2 z^3 \chi'(z) = \lambda \frac{F^{2}(z)}{1+z+z^2}\left\{1-\frac{b\lambda^2}{10}(\zeta(z)+15z^4)\right\}
\label{aw3}
\end{eqnarray}
with $\chi(1)=0=\chi'(1)$. Multiplying this equation by $e^{3b\lambda^2 z^{4}/2}$, we get
\begin{eqnarray}
\frac{d}{dz}\left(e^{3b\lambda^2 z^{4}/2}\chi'(z)\right) = \lambda e^{3b\lambda^2 z^{4}/2}
\frac{F^{2}(z)}{1+z+z^2}\left\{1-\frac{b\lambda^2}{10}(\zeta(z)+15z^4)\right\}~.
\label{aw4}
\end{eqnarray}
Integrating both sides of the above equation between $z=0$ to $z=1$, we obtain
\begin{eqnarray}
\chi'(0)=-\lambda\int_{0}^{1}dz~e^{3b\lambda^2 z^{4}/2}
\frac{F^{2}(z)}{1+z+z^2}\left\{1-\frac{b\lambda^2}{10}(\zeta(z)+15z^4)\right\}~.
\label{aw5}
\end{eqnarray}
Now from eq(s)(\ref{b2}, \ref{aw2}), we have
\begin{eqnarray}
\frac{\mu}{r_+}-\frac{\rho}{r_{+}^2}z&=&\lambda (1-z)\left\{1-\frac{b\lambda^2}{5}\zeta(z)\right\}+
\frac{\langle \mathcal{O}\rangle^2}{r_+^{2}}\chi(z) \nonumber\\
&=&\lambda (1-z)\left\{1-\frac{b\lambda^2}{5}\zeta(z)\right\}+
\frac{\langle \mathcal{O}\rangle^2}{r_+^{2}}(\chi(0)+z\chi'(0)+\dots)
\label{aw6}
\end{eqnarray}
where in the second line we have expanded $\chi(z)$ about $z=0$. Comparing the coefficient of $z$ on both sides
of the equation, we obtain
\begin{eqnarray}
\frac{\rho}{r_{+}^2}=\lambda-\frac{\langle \mathcal{O}\rangle^2}{r_+^{2}}\chi'(0).
\label{aw7}
\end{eqnarray}
Substituting $\chi'(0)$ from eq.(\ref{aw5}) in the above equation, we get
\begin{eqnarray}
\frac{\rho}{r_{+}^2}=\lambda\left\{1+\frac{\langle \mathcal{O}\rangle^2}{r_+^{2}}\mathcal{A}\right\}.
\label{aw8}
\end{eqnarray}
where,
\begin{eqnarray}
\mathcal{A}=\int_{0}^{1}dz~e^{3b\lambda^2 z^{4}/2}
\frac{F^{2}(z)}{1+z+z^2}\left(1-\frac{b\lambda^2}{10}(\zeta(z)+15z^4)\right).
\label{aw9}
\end{eqnarray}
Finally using eq(s)(\ref{tempbisch}, \ref{lam}), we get the following expression for $\langle \mathcal{O}\rangle$
\begin{eqnarray}
\langle \mathcal{O}\rangle = \gamma T_{c}\sqrt{1-\frac{T}{T_c}}
\label{aw10}
\end{eqnarray}
where,
\begin{eqnarray}
\gamma=\frac{4\pi\sqrt{2}}{3\sqrt{\mathcal{A}}}~.
\label{aw11}
\end{eqnarray}

The above relation (\ref{aw10}) reminds us about the standard \textit{square root} behavior of the condensation gap (near the critical temperature $ T_c $) that is predicted by the Landau - Ginzburg theory \cite{ref76}. This also shows that the critical exponent associated with the condensation is 1/2, which is the result of the \textit{mean field} theory. This also confirms that the phase transition we are talking about is a genuine \textit{second} order phase transition. From the above analysis it is also clear that a non zero condensate always corresponds to an AdS black hole associated with \textit{scalar hair} \cite{ref69}.

To see the effect of BI coupling ($ b $) on the condensation gap, one needs to calculate the coefficient $ \gamma $ for different choice in the BI coupling ($ b $). For example, taking $b=0.1$, computing $\mathcal{A}$ with $\alpha=0.2402$, we get $\gamma\approx8.19$
which is in good agreement to that with the exact result $\gamma\approx9.48$ \cite{binwang}. Likewise one can obtain other values of $ \gamma $ corresponding to different choice in the BI coupling ($ b $).
\begin{table}[htb]
\caption{A comparison of the analytical and numerical results for the critical temperature and the expectation value of the condensation operator}   
\centering                          
\begin{tabular}{c c c c c c c}            
\hline\hline                        
$b$ & $\zeta_{SL}(=\frac{3}{4\pi}\sqrt{\frac{1}{\lambda_{min}}})$  & $\zeta_{Numerical}$ &$\gamma_{SL}(=\frac{4\pi\sqrt{2}}{3\sqrt{\mathcal{A}}})$ & $\gamma_{Numerical} $ &  \\ [0.05ex]
\hline
0 & 0.225 & 0.225 & 8.07 & 9.31 \\
0.1 & 0.223 & 0.223 & 8.19 & 9.48 \\                              
0.2 & 0.221 & 0.221 & 8.33 & 9.62 \\
0.3 & 0.218 &0.219 & 8.54 & 9.74 \\ [0.05ex]  
\hline                              
\end{tabular}\label{E1}  
\end{table}

In Table 4.1 we eventually summarize the results obtained in sections 4.3 and 4.4. From the above Table 4.1
one can easily see that analytic results that are obtained using SL eigen value method are in excellent agreement to that with the results obtained numerically in \cite{binwang}.
 
From Table 4.1, a number of interesting facts could be extracted. First of all we note that the value of the coefficient $ \gamma $ increases with the increase in the BI coupling ($ b $). Which means that the BI coupling eventually increases the condensation gap and thereby making the formation of \textit{scalar hair} harder. This fact is also evident from the value of critical temperature ($ T_c $) where we find that $ T_c $ decreases as we increase the value of $ b $. All these facts together suggest that in the presence of BI correction to the usual Maxwell theory the onset of superconductivity turns out to be more difficult than in the ordinary Maxwell case. 

\section{Holographic superconductors in Gauss-Bonnet gravity}
One primary motivation behind the construction of holographic model of superconductivity in the context of higher curvature gravity is to examine whether higher curvature corrections suppress the condensation. In fact in order to check the validity of the Mermin-Wagner theorem, one should study the higher curvature gravity in four dimensions. Unfortunately apart from Gauss-Bonnet (GB) or Lovelock gravity \cite{kano1} and $ f(R) $ gravity \cite{kano2}, the other higher derivative terms in four dimensions give rise to \textit{ghost} degrees of freedom. It is interesting to note that in four dimensions the black hole solutions, that one obtain in the presence of these higher derivative terms are exactly identical to those obtained in the \textit{Einstein} gravity \cite{kano3}. Therefore these higher curvature terms play any significant role only for space time dimensions greater than four. 

Keeping this in mind, in the present analysis we construct the gravity dual of holographic superconductors in the fixed background of the Gauss-Bonnet (GB) AdS space time in five dimensions. The corresponding action for the Einstein Gauss-Bonnet gravity in five dimensions may be written as \cite{ref80},
\begin{equation}
S=\int d^{5}x \sqrt{-g}(R+\frac{12}{l^{2}}+\frac{\alpha}{2}(R^{2}-4 R^{\mu\nu}R_{\mu\nu}+R^{\mu\nu\rho\sigma}R_{\mu\nu\rho\sigma}))\label{eq1}
\end{equation} 
where $ \frac{12}{l^{2}} $ stands for the cosmological constant and $ \alpha $ is the Gauss-Bonnet (GB) coefficient. At this stage it is to be noted that for the rest of our analysis we set $ l=1 $. The Ricci flat solution for the action (\ref{eq1}) is given by,
\begin{equation}
ds^{2}=-f(r)dt^{2}+f^{-1}(r)dr^{2}+r^{2}(dx^{2}+dy^{2}+dz^{2})
\end{equation} 
where,
\begin{equation}
f(r)=\frac{r^{2}}{2\alpha}\left(1-\sqrt{1-4\alpha\left(1-\frac{M}{r^{4}}\right) } \right). 
\end{equation}
Here $ M $ is the mass of the black hole, which is related to the horizon radius ($ r_+ $) as,  $ r_+=M^{\frac{1}{4}} $. The Hawking temperature of the black hole is given by,
\begin{equation}
T=\frac{r_+}{\pi}\label{temp}.
\end{equation} 

It is interesting to note that in the limit $ r\rightarrow\infty $ we have
\begin{equation}
f(r)\sim\frac{r^{2}}{2\alpha}\left[1-\sqrt{1-4\alpha} \right] 
\end{equation}
which naturally sets an effective radius for the AdS space time as,
\begin{equation}
L_{eff}^{2}=\frac{2\alpha}{1-\sqrt{1-4\alpha} }.
\end{equation}
The corresponding upper bound $ \alpha=\frac{1}{4} $ is known as Chern-Simons limit.

Like in the previous example, in order to study the $ s $- wave superconductors in the frame work of BI electrodynamics we adopt the following Lagrangian density which includes the Maxwell field ($ A_{\mu} $) and a charged complex scalar field ($ \psi $) as,
\begin{equation}
\mathcal{L}=\frac{1}{b}\left(1-\sqrt{1+\frac{b}{2}F^{\mu\nu}F_{\mu\nu}} \right)-|\nabla_{\mu}\psi- iA_{\mu}\psi |^{2}-m^{2}\psi^{2} 
\end{equation}
where $ b $ is the Born-Infeld parameter and $ m^{2} $ is the mass square of the scalar field.

With the following gauge choices for the vector field and the scalar field,  
\begin{eqnarray}
A_{\mu}=(\phi(r),0,0,0),~~~~~\psi=\psi(r)
\end{eqnarray}
we eventually arrive at the following equations of motion for the scalar potential $ \phi(r) $ and the scalar field $ \psi(r) $
\begin{equation}
\partial_{r}^{2}\phi + \frac{3}{r}(1-b(\partial_{r} \phi)^{2})\partial_{r}\phi -\frac{2\psi^{2}\phi}{f}(1-b(\partial_{r}\phi)^{2})^{3/2}=0\label{eq2}
\end{equation}
and,
\begin{equation}
\partial_{r}^{2}\psi + \left(\frac{3}{r}+\frac{\partial_{r}f}{f} \right)\partial_{r}\psi + \frac{\phi^{2}\psi}{f^{2}}-\frac{m^{2}\psi}{f}=0\label{eq3}. 
\end{equation}
In order to solve (\ref{eq2}) and (\ref{eq3}) one needs the boundary conditions near the event horizon $ r\sim r_+ $ and at the spatial infinity $ r\rightarrow\infty $. Considering the former case we have,
\begin{equation}
\phi(r_+)=0,~~~~~\psi(r_+)=\frac{\partial_{r}f(r_+)}{m^{2}}\partial_{r}\psi(r_+)
\end{equation}
whereas in the later case we have,
\begin{eqnarray}
\phi(r)&=&\mu - \frac{\rho}{r^{2}}\label{phi}\\
\psi(r)&=&\frac{\psi^{-}}{r^{\Delta_{-}}}+\frac{\psi^{+}}{r^{\Delta_{+}}}
\end{eqnarray}
where $ \mu $ and $ \rho $ are respectively the chemical potential and the charge density and $ \Delta_{\pm} =2\pm\sqrt{4+m^{2}L_{eff}^{2}}$ is the conformal dimension of the dual operator $ \mathcal{O} $ in the boundary field theory. In the following analysis we shall set $ \psi^{-}=0 $ and $ \psi^{+}=<\mathcal{O}> $. 


\section{Relation between critical temperature and charge density}
In order to obtain an explicit relationship between the critical temperature and the charge density, we define $ z=\frac{r_+}{r} $. Under this choice of variables, we may express (\ref{eq2}) and (\ref{eq3}) as,
\begin{equation}
\partial_{z}^{2}\phi -\frac{1}{z}\partial_{z}\phi + \frac{3bz^{3}}{r_+^{2}}(\partial_{z}\phi)^{3}-\frac{2\psi^{2}\phi r_+^{2}}{f z^{4}}\left(1-\frac{b z^{4}(\partial_{z}\phi)^{2}}{r_+^{2}}\right)^{3/2}=0\label{eq4}
\end{equation}
and,
\begin{equation}
\partial_{z}^{2}\psi - \frac{1}{z}\partial_{z}\psi +\frac{\partial_{z}f}{f}\partial_{z}\psi + \frac{\phi^{2}\psi r_+^{2}}{z^{4}f^{2}} - \frac{m^{2}\psi r_+^{2}}{f z^{4}}=0\label{eq5}
\end{equation}
respectively.

At $ T=T_c $ , we have $ \psi=0 $. Therefore from (\ref{eq4}) we have
\begin{equation}
\partial_{z}^{2}\phi -\frac{1}{z}\partial_{z}\phi + \frac{3bz^{3}}{r_+^{2}}(\partial_{z}\phi)^{3}=0
\end{equation}
which has the solution,
\begin{equation}
\phi(z)=\lambda r_{+c}(1-z^{2})\left[1-\frac{b \lambda^{2}}{2}\xi(z) \right]+O(b^{2})\label{eq6} 
\end{equation}
with,
\begin{equation}
\xi(z)= (1+z^{2})(1+z^{4})~~~and,~~~\lambda = \frac{\rho}{r_{+c}^{3}}\label{lambda}.
\end{equation} 

 At this stage it is customary to mention that, since the BI coupling parameter ($ b $) is too small, therefore throughout the rest of our analysis we consider terms only linear in $ b $ and drop all the higher order terms in it.  

In order to investigate the boundary behavior of $ \psi $ (as $ T \rightarrow T_c $) we consider
\begin{equation}
\psi|_{z\rightarrow 0}\sim \frac{<\mathcal{O}>}{r_{+}^{3}}z^{3}F(z)\label{eq7}
\end{equation}
where $ F(0)=1 $ and $ F^{'}(0)=0 $.
Finally using (\ref{eq6}) and (\ref{eq7}), from (\ref{eq5}) we obtain
\begin{equation}
F^{''}+p(z)F^{'}+q(z)F+\lambda^{2}w(z)F=0\label{eq8}
\end{equation} 
where, the prime denotes the derivative w.r.t $ z $ and,
\begin{eqnarray}
p(z)&=&\frac{3(1-\sqrt{1-4\alpha + 4\alpha z^{4}})-12\alpha +20\alpha z^{4}}{z[1-4\alpha +4\alpha z^{4}-\sqrt{1-4\alpha + 4\alpha z^{4}}]}\nonumber\\
q(z)&=&\frac{1}{z^{2}}\left[\frac{3(1-4\alpha -4\alpha z^{4}-\sqrt{1-4\alpha + 4\alpha z^{4}})}{\sqrt{1-4\alpha + 4\alpha z^{4}}-1+4\alpha -4\alpha z^{4}}+\frac{2m^{2}\alpha}{\sqrt{1-4\alpha + 4\alpha z^{4}}-1} \right] \nonumber\\
w(z)&=&\frac{4\alpha^{2}(1-z^{2})^{2}(1-\frac{b}{2}\lambda^{2}\xi(z))^{2}}{(1-\sqrt{1-4\alpha + 4\alpha z^{4}})^{2}}.
\end{eqnarray}
It is now very much straightforward to convert (\ref{eq8}) into a standard Sturm-Liouville (SL) eigenvalue problem \cite{ref77} which reads,
\begin{equation}
(T(z)F(z))^{'}-Q(z)F(z)+\lambda^{2}P(z)F(z)=0
\end{equation} 
where,
\begin{eqnarray}
T(z)&=&\frac{z^{3}}{2\sqrt{\alpha}}(\sqrt{1-4\alpha + 4\alpha z^{4}}-1)\approx z^{3}\sqrt{\alpha}(z^{4}-1)[1-\alpha(z^{4}-1)]\nonumber\\
Q(z)&=&-T(z)q(z)\approx -3z\sqrt{\alpha}(3z^{4}+6\alpha z^{4}-7\alpha z^{8})\nonumber\\
P(z)&=&T(z)w(z)\approx \frac{\sqrt{\alpha}z^{3}(z^{2}-1)(1+\alpha(z^{4}-1))(1-\frac{b}{2}\lambda^{2}\xi(z))^{2}}{z^{2}+1}.\label{eq9}
\end{eqnarray}
Here we have retained terms only upto an order $ \alpha^{3/2} $ while computing all the above expressions. It is quite interesting to note that one can further simplify $ P(z) $ using the fact that,
\begin{equation}
b\lambda^{2}=b(\lambda^{2}|_{b=0})+O(b^{2}).
\end{equation} 
This finally yields,
\begin{equation}
P(z)\approx \frac{\sqrt{\alpha}z^{3}(z^{2}-1)(1+\alpha(z^{4}-1))(1-b(\lambda^{2}|_{b=0})\xi(z))}{z^{2}+1}.
\end{equation}
\begin{figure}[h]
\centering
\includegraphics[angle=0,width=12cm,keepaspectratio]{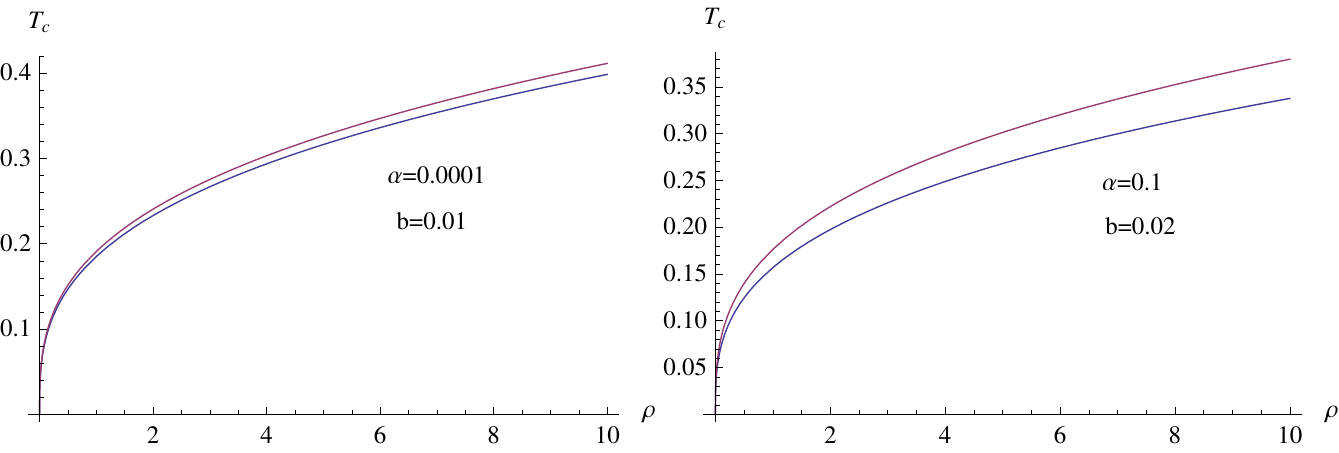}
\caption[]{\it Critical temperature ($ T_{c}-\rho $) plot for Gauss Bonnet (GB) holographic superconductors for different choice of parameters $ \alpha $ and $ b$. The red (upper) curve corresponds to the numerical value whereas the blue (lower) one corresponds to analytical value. From the plots it is evident that the critical temperature is indeed suppressed due to the higher curvature effects.}
\label{figure 2a}
\end{figure}
\begin{table}[htb]
\caption{A comparison of the analytical and numerical results for the critical temperature for $ \alpha=0.0001 $}   
\centering                          
\begin{tabular}{c c c c c c c}            
\hline\hline                        
$b$ & $ a $& $ \lambda_{min}^{2} $ & $\zeta_{SL}\left(=\frac{1}{\pi \lambda_{min}^{1/3}}\right)$  & $\zeta_{Numerical}$ &  \\ [0.05ex]
\hline
0 & 0.721772 & 18.2331 & 0.196204& 0.196204 \\
0.01 &0.754014 & 25.9147 & 0.185037&0.191012  \\
0.02 & 0.821082 &44.0958& 0.169349 &0.185073  \\ [0.5ex] 
\hline                              
\end{tabular}\label{E1}  
\end{table}
\begin{table}[htb]
\caption{A comparison of the analytical and numerical results for the critical temperature for $ \alpha=0.1 $}   
\centering                          
\begin{tabular}{c c c c c c c}            
\hline\hline                        
$b$ & $ a $& $ \lambda_{min}^{2} $ & $\zeta_{SL}\left(=\frac{1}{\pi \lambda_{min}^{1/3}}\right)$  & $\zeta_{Numerical}$ &  \\ [0.05ex]
\hline
0 & 0.709061 & 21.5679 & 0.19078& 0.189939 \\
0.01 &0.754297 & 33.3478 &0.177421&0.184679  \\
0.02 &0.868304 &70.2223& 0.156713 &0.17616  \\ [0.5ex] 
\hline                              
\end{tabular}\label{E1}  
\end{table}

With all the above expressions in hand, the minimum value of $ \lambda^{2} $ may be obtained considering the variation of the following functional,
\begin{equation}
\lambda^{2}[F(z)]=\frac{\int_{0}^{1}dz(T(z)(F^{'}(z))^{2}+Q(z)F^{2}(z))}{\int_{0}^{1}dz P(z) F^{2}(z)}
\end{equation}
with the choice $ F(z)=1-a z^{2} $ and $ m^{2}=-3/L_{eff}^{2} $. It is reassuring to note that this choice of mass is well above the BF bound \cite{bf1}-\cite{bf2}.

Finally using (\ref{temp}) and (\ref{lambda}) we obtain for $ T \sim T_c  $
\begin{equation}
T_c=\zeta \rho^{1/3}
\end{equation}
where $ \zeta=\frac{1}{\pi \lambda_{min}^{1/3}} $.

Let us now tabulate various values of the coefficient ($ \zeta $) for different choice of parameters $ \alpha $  and $ b $. Looking at the results enumerated in Tables (4.2 and 4.3) it is indeed evident that for $ b=0 $ case the results are in good agreement to that with the results existing in the literature \cite{ref78}, \cite{ref80}. Furthermore, we note that the critical temperature ($ T_c $) decreases with the increase in the value of the BI coupling ($ b $), which means that the condensation gets harder as the BI coupling ($ b $) becomes larger. This result is consistent with the earlier findings for the Schwarzschild AdS case. In Figure 4.1, we show the effect of Gauss Bonnet as well as BI coupling parameters on the critical temperature $ (T_c) $.  

Finally, as an effect of the Gauss-Bonnet (GB) coupling ($ \alpha $) on the condensation operator, we note that, for a given value of $ b $, the condensation gap becomes larger for the higher values of $ \alpha $ \cite{ref78}, \cite{ref80}. This indeed suggests that the condensation gets suppressed due to the presence of higher derivative corrections in the bulk gravitational theory.


\section{Critical exponent and condensation values}

In this section we aim to examine the effect of BI coupling parameter on the condensation operator near the critical point. We will be basically following the same approach as that of sec. 4.4. 
In order to do that we first note that close to the critical temperature ($ T_c $) equation (\ref{eq4}) may be written as,
\begin{equation}
\partial_{z}^{2}\phi -\frac{1}{z}\partial_{z}\phi + \frac{3bz^{3}}{r_+^{2}}(\partial_{z}\phi)^{3}=\frac{<\mathcal{O}>^{2}}{r_+^{4}}\mathcal{B}\phi \label{eq10}
\end{equation}
where, $ \mathcal{B}=\frac{2z^{2}}{f}\left(1-\frac{b z^{4}(\partial_{z}\phi)^{2}}{r_+^{2}}\right)^{3/2}F^{2}(z) $.

Since $ \frac{<\mathcal{O}>^{2}}{r_+^{4}} $ is a very small parameter (as we are very close to the critical temperature), therefore it will be natural to expand $ \phi(z) $ as,

\begin{equation}
\frac{\phi(z)}{r_+}=\lambda (1-z^{2})\left[1-\frac{b \lambda^{2}}{2}\xi(z) \right]+ \frac{<\mathcal{O}>^{2}}{r_+^{4}} \chi(z)\label{eq11}
\end{equation}
with $ \chi(1)=\chi^{'}(1)=0 $.

Using (\ref{eq11}), from (\ref{eq10}) we obtain,
\begin{equation}
\chi^{''}(z)-\frac{\chi^{'}}{z}+36b\lambda^{2}z^{5}\chi^{'}=\lambda \mathcal{B}(1-z^{2})\left( 1-\frac{b}{2}\lambda^{2}\xi(z)\right)\label{eq12}. 
\end{equation}
From (\ref{eq12}) we note that, in the limit  $ z\rightarrow0 $
\begin{equation}
\chi^{''}(0)=\frac{\chi^{'}(z)}{z}|_{z\rightarrow 0}\label{eq13}.
\end{equation} 

In order to obtain the r.h.s. of (\ref{eq13}), we note that equation (\ref{eq12}) may be written as,
\begin{equation}
\frac{d}{dz}\left(e^{6b\lambda^{2}z^{6}}\frac{\chi^{'}}{z} \right)=\lambda \frac{2z^{3}}{r_+^{2}}\frac{e^{6b\lambda^{2}z^{6}}(1-\frac{b}{2}\lambda^{2}\Gamma(z))}{(1+z^{2})(1+\alpha(1-z^{4}))}F^{2}(z)\label{eq14} 
\end{equation}
where, $ \Gamma(z)=1+z^{2}+z^{4}+13z^{6} $.
Finally integrating (\ref{eq14})  between the limits $ z=0 $ and $ z=1 $ and keeping terms only linear in $ b $, we obtain
\begin{eqnarray}
\frac{\chi^{'}}{z}|_{z\rightarrow 0}&=&-\frac{\lambda}{r_+^{2}}\mathcal{A}\label{eq15}\\
with,~~~~\mathcal{A}&\approx &\int_{0}^{1}\frac{2z^{3}F^{2}(z)(1-\frac{b}{2}\lambda^{2}(1+z^{2}+z^{4}+z^{6}))(1-\alpha(1-z^{4}))}{1+z^{2}}dz\nonumber.
\end{eqnarray}
\begin{table}[htb]
\caption{A comparison of the analytical and numerical results for the condensation operator for $ \alpha=0.0001 $}   
\centering                          
\begin{tabular}{c c c c c c c}            
\hline\hline                        
$b$ & $ a $& $ \lambda_{min}^{2} $ & $\gamma_{SL}$  & $\gamma_{Numerical}$ &  \\ [0.05ex]
\hline
0 & 0.721772 & 18.2331 & 7.70525& 7.70677 \\
0.01 &0.754014 & 25.9147 & 8.73543&8.7599  \\
0.02 & 0.821082 &44.0958& 10.36 &11.1853  \\ [0.5ex] 
\hline                              
\end{tabular}\label{E1}  
\end{table}
\begin{table}[htb]
\caption{A comparison of the analytical and numerical results for the condensation operator for $ \alpha=0.1 $}   
\centering                          
\begin{tabular}{c c c c c c c}            
\hline\hline                        
$b$ & $ a $& $ \lambda_{min}^{2} $ & $\gamma_{SL}$  & $\gamma_{Numerical}$ &  \\ [0.05ex]
\hline
0 & 0.709061 & 21.5679 & 7.90303& 7.86604 \\
0.01 &0.754297 & 33.3478 &9.26635&9.3564  \\
0.02 &0.868304 &70.2223& 11.7452 &14.4142 \\ [0.5ex] 
\hline                              
\end{tabular}\label{E1}  
\end{table}
 
 Again from (\ref{phi}) and (\ref{eq11}) one can have near $ z=0 $
\begin{equation}
\mu - \frac{\rho}{r_+^{2}}z^{2}=\lambda r_{+}(1-z^{2})\left[1-\frac{b \lambda^{2}}{2}\xi(z) \right]+\frac{<\mathcal{O}>^{2}}{r_+^{3}}(\chi(0)+z\chi^{'}(0)+\frac{z^{2}}{2}\chi^{''}(0)+...).
\end{equation}
Comparing the coefficients of $ z^{2} $ from both sides and using (\ref{temp}), (\ref{lambda}), (\ref{eq13}) and (\ref{eq15}) for $ T \rightarrow T_c $ we finally obtain,
\begin{eqnarray}
<\mathcal{O}> &\approx & \gamma \pi^{3}T_c^{3}\sqrt{1-\frac{T}{T_c}}\\
with,~~~~\gamma &=& \sqrt{\frac{6}{\mathcal{A}}}.
\end{eqnarray}
 
In order to discuss the effect of higher curvature corrections on the holographic $ s $- wave condensate, let us first tabulate various values of the coefficient $ \gamma $ those obtained numerically as well as analytically using SL eigenvalue problem. From the above Tables 4.4 and 4.5 we note that for a given value of the BI coupling ($ b $), the value of $ \gamma $ increases with the increase in the GB coupling ($ \alpha $), which shows that higher curvature corrections increases the condensation gap and thereby making it harder to form at low temperatures.

\section{Discussions}
In this chapter, considering the probe limit and based on the Sturm-Liouville (SL) eigenvalue problem, we have explored several crucial aspects of holographic $ s $-wave superconductors in the framework of Born-Infeld (BI) electrodynamics. The \textit{analytic} results that have been obtained during this exercise are valid upto leading order in the Born-Infeld (BI) coupling parameter ($ b $). The entire content of this chapter could be divided mainly into two parts. The first part of it has been devoted towards the study of holographic $ s $- wave condensate in the background of a planar Schwarzschild AdS space time. The relation between the critical temperature ($ T_c $) and the charge density has been obtained through an iterative procedure. It has been observed that the Born-Infeld coupling parameter ($ b $) decreases the critical temperature ($ T_c $) of the condensate. Moreover, we have also found that the condensation gap increases due to the presence of BI coupling in the theory. All these facts eventually suggest that it is comparatively harder for the scalar condensation to form in Born-Infeld electrodynamics than that in the usual Maxwell case. The results that have been obtained analytically are found to be in very good agreement with the existing numerical results \cite{binwang}. On top of it, the critical exponent of the condensation has been found to be $ 1/2 $  which is the universal value in the \textit{mean field} theory.

During the rest of this chapter, we have made an explicit analysis on holographic $s$-wave superconductors in the background of Gauss-Bonnet (GB) AdS spacetime. In order to do that we have performed our analysis both analytically as well as numerically. Using Sturm-Liouville (SL) eigenvalue problem we have established the relationship between the critical temperature ($ T_c $) and the charge density in the probe limit. It has been observed that both the BI coupling parameter ($b$) as well the GB coupling parameter ($\alpha$) (at the leading order) affect the formation of \textit{scalar hair} at low temperatures ($ T<T_c $). The physical meaning of the leading order expansions in $b$ and $\alpha$ is that what has been studied is the effect of leading higher derivative corrections of the gauge fields as well as the space time curvature to the onset of the $s$-wave order parameter condensation. It has been observed that the condensation is harder to form for the higher values of BI/GB coupling parameters. The critical exponent associated with the condensation again comes out to be $ 1/2 $ which is in good agreement to that with the universal \textit{mean field} value. 

The primary motivation behind the analysis of the present chapter was to carry out an explicit analytic computation that parallels the existing numerical results regarding various properties of holographic $ s $- wave superconductors coupled to nonlinear electrodynamics. Of course we have not yet explored all the properties. For example, we have not so far explored the onset of superconductivity in presence of an external magnetic field. It would be quite interesting to explore how the application of an external magnetic field affects the formation of \textit{scalar hair} in the bulk. Regarding the magnetic response in holographic superconductors, it is interesting to note that most of the analysis that have been carried out so far are based on numerical techniques and also these are performed in the usual framework of Maxwell electrodynamics \cite{ref88}-\cite{ref95}. Therefore it would be quite natural to ask whether it is indeed possible to carry out an explicit analytic computation regarding the onset of holographic $ s $- wave condensate in the presence of nonlinear corrections to the usual Maxwell action. We are going to explore these issues in the next chapter.

\chapter*{Appendix}
\section*{ Sturm-Liouville problem and calculus of variations}

Let us consider the determination of stationary values of the quantity $\lambda$
defined by the ratio
\begin{eqnarray}
\lambda=\frac{\int_{a}^{b}\left\{p(x)(y'(x))^2 -q(x)y^2 (x)\right\}dx}{\int_{a}^{b} r(x)y^2 (x) dx}\equiv
\frac{I_1}{I_2}
\label{app1}
\end{eqnarray}
where $p(x)$, $q(x)$ and $r(x)$ are known functions of $x$ and prime denotes derivative with respect to $x$.

 Varying $\lambda$ with respect to $y(x)$, we get
\begin{eqnarray}
\delta\lambda=\frac{1}{I_2}(\delta I_1 -\lambda\delta I_2)~.
\label{app2}
\end{eqnarray}
Computation of $\delta I_1$ and $\delta I_2$ yield
\begin{eqnarray}
\delta I_1=2\left\{p(x)y'(x)\delta y(x)\right\}|_{a}^{b} - 2\int_{a}^{b}\left\{(p(x)y'(x))' +q(x)y(x)\right\}\delta y(x) dx
\label{app3}
\end{eqnarray}
\begin{eqnarray}
\delta I_2=2\int_{a}^{b}r(x)y(x)\delta y(x) dx~.
\label{app4}
\end{eqnarray}
Substituting eq(s)(\ref{app3}, \ref{app4}) in eq.(\ref{app2}), we get
\begin{eqnarray}
\delta\lambda=\frac{2}{I_2}\left\{(p(x)y'(x)\delta y(x))|_{a}^{b} -\int_{a}^{b}\left[(p(x)y'(x))' +q(x)y(x)+\lambda r(x) y(x)
\right]\delta y(x) dx\right\}~.
\label{app5}
\end{eqnarray}
Hence, the condition $\delta\lambda=0$ for the stationary values of $\lambda$ leads to the Euler equation
\begin{eqnarray}
\frac{d}{dx}\left(p(x)\frac{dy(x)}{dx}\right)+q(x)y(x)+\lambda r(x) y(x)=0
\label{app6}
\end{eqnarray}
and to the following boundary conditions:
\begin{eqnarray}
(p(x)y'(x))|_{x=a}=0 \quad or \quad y(a)~ prescribed\nonumber\\
(p(x)y'(x))|_{x=b}=0 \quad or \quad y(b)~ prescribed.
\label{app7}
\end{eqnarray}

\chapter{Magnetic Response In Holographic Superconductors}
\section{An overview}
One of the major characteristic properties of ordinary superconductors is that they exhibit perfect \textit{diamagnetism} as the temperature is lowered through $ T_c $ in the presence of an external magnetic field. In other words, at low temperature superconductors expel magnetic field lines. This is known as \textit{Meissner} effect which could be put into the form of a following parabolic law \cite{ref76},
\begin{equation}
B_c(T)\approx B_c(0)\left[1-\left(\frac{T}{T_c} \right)^{2}  \right]. 
\end{equation}
 
 Depending on their behavior in the presence of an external magnetic field, ordinary superconductors are classified into two categories, namely type I and type II. In type I superconductors, for $ B>B_c $ there exists a \textit{first order} phase transition from the superconducting phase to the normal phase where $ B_c $ is the value of the critical field strength. Whereas, on the other hand, in type II superconductors for $ B>B_{c1} $ (lower critical field strength) the magnetic field starts to penetrate the superconducting sample in the form of quantized flux known as \textit{vortices}. These vortices become more dense with the increase in the magnetic field strength, and finally at $ B = B_{c2} $, where where $ B_{c2} $ is the upper critical field strength, there happens to be a gradual \textit{second order} phase transition and as a result the material ceases to super conduct for $ B>B_{c2} $. 

Inspired from all these facts, till date a number of attempts have been made in order to investigate the effects of applying an external magnetic field to  holographic superconductors \cite{ref88}-\cite{ref95}. From these various analysis, it is more or less confirmed that holographic superconductors are of type II rather than type I, which is in agreement with their so called high $ T_c $ behavior. 
The observations that have been made so far could be summarized as follows:

 In order to study the effect of an external magnetic field on the holographic condensate, a magnetic field is added in the bulk gravitational description. In the presence of this external magnetic field, the dual of the conducting/normal phase of the boundary CFTs could be described by a black hole in AdS with both electric and magnetic charges and no \textit{scalar hair}. According to the \textit{AdS/CFT} dictionary, the boundary value of the magnetic field added in the bulk acts as a back ground magnetic field in the dual CFTs. It is found that as the magnetic field is increased, it shrinks the condensate away completely \cite{ref90,ref91}. Using the \textit{AdS/CFT} dictionary one can in fact show that the (upper) critical field strength ($ B_{c2} $) decreases with the increase in temperature and eventually vanishes at $ T = T_c $. This establishes the fact that, like ordinary superconducting materials, holographic superconductors can also sustain a larger magnetic field at low temperatures. Regarding the formation of vortex states, in \cite{ref88} the authors have found that one can in fact have a holographic realization of \textit{Abrikosov lattice} near the critical point of the \textit{second} order transition. Moreover they have also observed that the condition of the minimum free energy essentially leads to a \textit{triangular lattice} configuration, which is according to Ginzburg - Landau theory is the thermodynamically most stable configuration for ordinary type II superconductors.  
 
All the above results are indeed enough encouraging to carry out further analysis in this particular direction. Therefore motivated from the above observations, in the present analysis we wish to explore several other intriguing issues regarding the formation of holographic condensate in the presence of an external magnetic field. In the following we give the motivation behind the analysis of the present chapter.
\vskip 1mm
\noindent
$\bullet$  Since all the above attempts are mostly concerned with\textit{ numerical} techniques, the question that naturally arises is that whether it is possible, in general, to have an \textit{analytic} scheme which could be employed to investigate the behavior of holographic superconductors even in the presence of an external magnetic field. Furthermore, it is also to be noted that all the above analysis are mostly performed for ($ 2+1 $) dimensional holographic superconductors. Therefore, it will be quite interesting to extend this analysis for the ($ 3+1 $) dimensional case. 
\vskip 1mm
\noindent
$\bullet$ Most importantly, one may note that, so far all the attempts to study the effect of an external magnetic field on holographic superconductors are made in the framework of usual Maxwell electrodynamics. Therefore, it would be quite interesting to explore whether the higher derivative corrections of the gauge fields can affect the behavior of holographic superconductors in the presence of an external magnetic field. As it is evident from the earlier chapter that these higher derivative corrections could be incorporated in the theory of superconductors by replacing the Maxwell action in \cite{ref70} with a suitable non linear action for classical electrodynamics\footnote{In the previous chapter we have already incorporated the effect of higher derivative corrections of $ U(1) $ gauge fields by replacing the Maxwell action with the Born-Infeld (BI) action and eventually studied the effect of these higher derivative corrections on the holographic $ s $- wave condensate in the absence of external magnetic field. }. 
 
The analysis of the present chapter is based on \cite{ref104}, where using a new \textit{analytic} scheme\footnote{This new analytic scheme is known as \textit{matching technique} which is based on the matching of the solutions to the field equations near the horizon and near the asymptotic AdS region \cite{ref78}.}, we investigate the effect of adding an external magnetic field on holographic $ s $- wave condensate in the presence of both (i) Born-Infeld (BI) \cite{binwang} as well as the (ii) Weyl curvature corrections \cite{nref4} to the usual Maxwell action. Considering the probe limit the entire analysis has been carried out in the back ground of a planar Schwarzschild AdS space time. At this stage it is reassuring to note that these non linear generalizations essentially correspond to the higher derivative corrections of the gauge fields. In both the cases the analytic expressions for the critical temperature ($ T_c $) as well as the condensation values have been obtained, where the computations are performed to the leading order in the coupling  parameter(s). It has been observed that the coupling parameter(s) of the theory indeed affect the formation of the \textit{scalar hair} at low temperatures. Finally, the effect of applying an external magnetic field on the formation of holographic $ s $- wave condensate has been studied by adding a static magnetic field in the bulk theory. Interestingly, we find that the superconducting phase disappears for $ B>B_c $, where $ B_c $ is the critical field strength. Moreover, we note that the value of this critical field strength ($ B_{c} $) indeed gets affected due to the presence of non linearity in the original theory, which also in turn affects the formation of \textit{scalar hair} at low temperatures.
 
Before going further, let us briefly mention about the organization of the chapter. In section 5.2, we investigate the effect of adding an external magnetic field on the $ s $- wave holographic superconductors in the presence of BI corrections to the usual Maxwell action. In section 5.3, similar analytic computations have been carried out incorporating the effect of Weyl curvature corrections in the original Maxwell theory.  Finally, the chapter is concluded in section 5.4.  
 
\section{The Meissner effect with Born-Infeld corrections}
In the present section, considering the probe limit, we aim to discuss the effect of an external magnetic field on the holographic $ s $- wave condensate in the presence of Born-Infeld (BI) corrections to the usual Maxwell action.
In order to do that, as a first step, using the matching technique \cite{ref78}, we explicitly derive the expression for the critical temperature ($ T_c $) as well the order parameter near the critical point of the phase transition. In the next step, using these results we explicitly derive the expression for the critical magnetic field ($ B_c $) close to the critical point.

We begin with the metric of a planar Schwarzschild AdS black hole, which may be written as \cite{ref70},
\begin{eqnarray}
ds^2=-f(r)dt^2+\frac{1}{f(r)}dr^2+r^2(dx^2+dy^2)
\label{m1}
\end{eqnarray}
where,
\begin{eqnarray}
f(r)=r^{2}-\frac{r_{+}^3}{r}.
\label{metric}
\end{eqnarray}
Note that we have set the AdS radius to unity, i.e. $l=1$.
The Hawking temperature is related to the horizon radius ($r_+$) as
\begin{eqnarray}
T=\frac{3r_+}{4\pi}~.
\label{ntemp}
\end{eqnarray}
The entire analysis have been performed over this fixed back ground.

In order to study the holographic dual of this theory we adopt the following action which includes a complex scalar field minimally coupled to the Maxwell field $ A_{\mu} $ as,
\begin{equation}
S= \frac{1}{16\pi G_4}\int d^{4}x\sqrt{-g}\left[R-2\Lambda +\frac{1}{b}\bigg(1-\sqrt{1+\frac{b F}{2}}\bigg)-|\nabla_{\mu}\psi-iA_{\mu}\psi|^2-m^2|\psi|^2 \right], 
\end{equation}  
where, $ F= F_{\mu\nu}F^{\mu\nu}$, $ b $ is the BI coupling parameter and $ \Lambda\left( =-\frac{3}{l^{2}}\right)  $ is the cosmological constant. It is reassuring to note that in the limit $ b\rightarrow 0 $ one recovers the usual Maxwell action. It is to be noted that the higher order terms in the coupling parameter $ b $ essentially implies the higher derivative corrections of the gauge fields. 

In the probe limit, the Maxwell and scalar field equations may be found as,
\begin{eqnarray}
\frac{1}{\sqrt{-g}}\partial_{\mu}\left( \frac{\sqrt{-g}F^{\mu\nu}}{\sqrt{1+\frac{bF}{2}}}\right)-i\left( \psi^{*}\partial^{\nu}\psi - \psi(\partial^{\nu}\psi)^{*}\right)-2A^{\nu}|\psi|^{2} = 0 \label{neq1}
\end{eqnarray} 
and,
\begin{equation}
\partial_{\mu}\left(\sqrt{-g}\partial^{\mu}\psi \right) -i\sqrt{-g}A^{\mu}\partial_{\mu}\psi -i\partial_{\mu}\left(\sqrt{-g}A^{\mu}\psi \right)-\sqrt{-g}A^{2}\psi -\sqrt{-g}m^{2}\psi =0\label{neq2} 
\end{equation}
respectively.

Considering the following anstaz,
\begin{eqnarray}
A_{\mu}=(\phi(r),0,0,0),\;\;\;\;\psi=\psi(r)
\label{vector}
\end{eqnarray}
and setting $ m^{2}=-2 $ along with a change in the variable from $ r $ to $ z(=\frac{r_+}{r}) $ the above set of equations (\ref{neq1}, \ref{neq2}) turn out to be,
\begin{eqnarray}
\phi''(z)+\frac{2bz^3}{r_{+}^2 }\phi'^{3}(z)-\frac{2\psi^2 (z)\phi(z)r_+^{2}}{z^4 f(z)}
\left(1-\frac{bz^4}{r_{+}^2}\phi'^{2}(z)\right)^{3/2} =0\label{neq5}
\end{eqnarray}
and,
\begin{eqnarray}
\psi''(z)+\frac{f^{'}(z)}{f(z)}\psi'(z)+ \frac{r_+^{2}\phi^{2}(z)\psi(z)}{z^{4}f^{2}(z)}+\frac{2r_+^{2}\psi(z)}{z^{4}f(z)}=0\label{neq6}
\end{eqnarray}
respectively.

Let us now talk about the boundary conditions:
\vskip 1mm
\noindent
$\bullet$ Regularity at the horizon $ z=1 $ implies,
\begin{eqnarray}
\phi (1)=0,~~~\psi^{'}(1)=\frac{2}{3}\psi(1) .\label{neq7}
\end{eqnarray}

\vskip 1mm
\noindent
$\bullet$ In the asymptotic region ($ z\rightarrow 0 $) the solutions may be written as,
\begin{eqnarray}
\phi(z) = \mu - \frac{\rho}{r_+}z,~~~~\psi(z)= J_{-} z + J_{+}z^{2},\label{neq8}
\end{eqnarray}  
where $ \mu $ and $ \rho $ are the chemical potential and the charge density of the dual field theory. In the following analysis we set $ J_{-}=0 $ .

\subsection{Matching method and condensation values}
With the above expressions in hand, as a next step, we aim to derive an analytic expression for the critical temperature and the condensation values in the presence of above non linear (BI) corrections to the usual Maxwell action. In order to do that, we first Taylor expand both $ \phi(z) $ and $ \psi(z) $ near the horizon as,
\begin{eqnarray}
\phi (z)=\phi (1) - \phi^{'}(1)(1-z) + \frac{1}{2}\phi^{''}(1)(1-z)^{2} + .. ..\label{neq9}
\end{eqnarray}
and,
\begin{eqnarray}
\psi(z)=\psi(1)-\psi^{'}(1)(1-z)+\frac{1}{2} \psi^{''}(1)(1-z)^{2} + ....\label{neq10}
\end{eqnarray}
respectively, where without loss of generality we choose $ \phi'(1)<0 $ and $ \psi(1)>0 $.

On the other hand, near $ z=1 $ from (\ref{neq5}) we obtain,
\begin{equation}
\phi^{''}(1)= -\frac{2b \phi'^{3}(1)}{r_+^{2}}-\frac{2\psi^2 (1)\phi'(1)}{3}\left( 1-\frac{3b}{2r_+^{2}}\phi'^{2}(1)\right) 
+ O(b^{2})\label{neq11},
\end{equation}
where we have used the fact that near the event horizon ($ z=1 $) the function $ f(z) $ could be Taylor expanded as in (\ref{neq9}, \ref{neq10}). Finally, substituting (\ref{neq11}) into (\ref{neq9}) we obtain,
\begin{equation}
\phi(z)= -\phi'(1)(1-z)-\left[\frac{b \phi'^{2}(1)}{r_+^{2}}+\frac{\psi^{2}(1)}{3}\left( 1-\frac{3b}{2r_+^{2}}\phi'^{2}(1)\right)  \right] \phi'(1)(1-z)^{2} + O(b^{2})\label{neq12}.
\end{equation}
Similarly, from (\ref{neq6}) and using (\ref{neq7}), near $ z=1 $ we obtain,
\begin{equation}
\psi''(1)= \frac{8}{9}\psi(1) -\frac{\psi(1)\phi'^{2}(1)}{18r_+^{2}}\label{neq13}.
\end{equation} 
Substituting (\ref{neq13}) into (\ref{neq10}) we finally obtain,
\begin{equation}
\psi (z) = \frac{1}{3}\psi(1)+\frac{2}{3}\psi(1)z+\left( \frac{4}{9}-\frac{\phi'^{2}(1)}{36r_+^{2}}\right)\psi(1)(1-z)^{2}\label{neq14}. 
\end{equation}

Following the methodology developed in \cite{ref78}, one can obtain an analytic expression for the critical temperature ($ T_c $) by matching the solutions (\ref{neq8}), (\ref{neq12}) and (\ref{neq14}) at some intermediate point $ z=z_m $.

In order to match these two sets of asymptotic solutions smoothly at $ z=z_m $ we need the following four conditions,
\begin{eqnarray}
\mu -\frac{\rho z_m}{r_+} = \beta(1-z_m) + \beta\left[\frac{b \beta^{2}}{r_+^{2}} +\frac{\alpha^{2}}{3}\left(1-\frac{3b\beta^{2}}{2r_+^{2}} \right)  \right] (1-z_m)^{2}\label{neq15}
\end{eqnarray}
\begin{equation}
-\frac{\rho}{r_+} = -\beta - 2\beta (1-z_m)\left[\frac{b \beta^{2}}{r_+^{2}} +\frac{\alpha^{2}}{3}\left(1-\frac{3b\beta^{2}}{2r_+^{2}} \right)  \right] \label{neq16}
\end{equation}
\begin{eqnarray}
 J_{+}z_m^{2} = \frac{\alpha}{3}+\frac{2\alpha z_m}{3} + \alpha \left(\frac{4}{9}-\frac{{\tilde{\beta^{2}}}}{36} \right)(1-z_m)^{2}\label{neq17} 
\end{eqnarray}
and,
\begin{equation}
J_+z_m = \frac{\alpha}{3} - \alpha \left(\frac{4}{9}-\frac{{\tilde{\beta^{2}}}}{36} \right)(1-z_m) \label{neq18}
\end{equation}
where we have set $ \beta=-\phi'(1) $, $ \alpha = \psi(1) $ and $ \tilde{\beta}=\frac{\beta}{r_+} $.

As a first step, from (\ref{neq16}) we obtain,
\begin{equation}
\alpha^{2}= \frac{3}{2(1-z_m)}\left[\left(\frac{\rho}{\beta r_+} -1\right)+ \frac{b\beta^{2}}{r_+^{2}} \left(\frac{3\rho}{2\beta r_+}-\frac{(7-4z_m)}{2} \right)  \right] + O(b^{2})\label{neq19}.
\end{equation}
Using (\ref{ntemp}), from (\ref{neq19}) we finally obtain,
\begin{equation}
\alpha^{2}=\frac{3}{2(1-z_m)} \left(\frac{T_c}{T}\right)^{2}\left(1+\frac{b\tilde{\beta^{2}}}{2}(7-4z_m)\right) \left(1-\frac{T^{2}}{T_c^{2}} \right) + O(b^{2})\label{neq20},  
\end{equation}
where,
\begin{equation}
T_c = \frac{3\sqrt{\rho}}{4\pi\sqrt{\tilde{\beta}}}\sqrt{1-2b\tilde{\beta^{2}}(1-z_m)}\label{neq21}.
\end{equation}
Finally, for $ T\sim T_c $ i.e, very close to the critical temperature, from (\ref{neq20}) we obtain,
\begin{equation}
\alpha = \psi(1) = \sqrt{\frac{3}{1-z_m}} \left(1+\frac{b\tilde{\beta^{2}}}{4}(7-4z_m)\right) \sqrt{1-\frac{T}{T_c}}+O(b^{2})\label{neq22}.
\end{equation}

From (\ref{neq17}) and (\ref{neq18}) one can further obtain,
\begin{equation}
J_+ = \frac{\alpha(2+z_m)}{3z_m} ~~ and,~~~ \tilde{\beta}= 2\sqrt{\frac{7-z_m}{1-z_m}}\label{neq23}.
\end{equation}

Finally, using (\ref{ntemp}), (\ref{neq22}) and (\ref{neq23}), near the critical temperature ($ T\sim T_c $) the condensation operator may be calculated as \footnote{Where, $ <\mathcal{O}> = \sqrt{2} J_+ r_+^{2}. $} \cite{ref78},
\begin{eqnarray}
 <\mathcal{O}> = \frac{16\sqrt{2} \pi^{2}}{9}\left(\frac{2+z_m}{3z_m} \right)\sqrt{\frac{3}{1-z_m}} T_c^{2}  \left(1+\frac{b\tilde{\beta^{2}}}{4}(7-4z_m)\right) \sqrt{1-\frac{T}{T_c}}\nonumber\\
 + higher~~ order~~ terms.
\end{eqnarray}

Before we proceed further, let us make some comments on the results that have been obtained so far. First of all, from (\ref{neq21}) we note that, in order to have a meaningful notion for the critical temperature ($ T_c $) we must have the following upper bound 
\begin{equation}
b\leq \frac{1}{2\tilde{\beta}^{2}(1-z_m)} = \frac{1}{8(7-z_m)}
\end{equation} 
on the BI coupling parameter. Also from (\ref{neq21}) we note that, in the presence of non linear (BI) corrections the critical temperature ($ T_c $) for the condensate eventually decreases which makes it harder for the \textit{scalar hair} to form at low temperatures. At this stage it is interesting to note that similar features have already been observed in the previous chapter but using a different analytic scheme.

\subsection{The Meissner effect}
 
In order to study the effect of an external static magnetic field on a holographic superconductor we add a magnetic field in the bulk. According to the gauge/gravity duality, the asymptotic value of this magnetic field corresponds to a magnetic field added to the boundary field theory, i.e, $ B({\bf{x}})=F_{xy}({\bf{x}},z\rightarrow 0) $ . Since the condensate is small everywhere near the upper critical value ($ B_{c} $) of the magnetic field, therefore we may regard the scalar field $ \psi $ as a perturbation near the critical field strength $ B\sim B_{c} $. Based on the above physical arguments, we adopt the following ansatz \cite{ref90},
\begin{equation}
A_t = \phi(z),~~~A_y = B x,~~~and~~~\psi = \psi (x,z)\label{neq24}.
\end{equation}

With the above choice, the scalar field equation for $ \psi $ turns out to be,
\begin{eqnarray}
\psi''(x,z)+\frac{f^{'}(z)}{f(z)}\psi'(x,z)+ \frac{r_+^{2}\phi^{2}\psi}{z^{4}f^{2}(z)}+\frac{2r_+^{2}\psi}{z^{4}f(z)}+\frac{(\partial_{x}^{2}\psi - B^{2}x^{2}\psi)}{z^{2}f(z)}=0\label{neq25}.
\end{eqnarray}
 
In order to solve the solve (\ref{neq25}), we take the following separable form
\begin{equation}
\psi (x,z) = X(x) R(z)\label{neq26}.
\end{equation}
 
 Substituting (\ref{neq26}) into (\ref{neq25}) we finally obtain,
 \begin{eqnarray}
 z^{2}f(z)\left[\frac{R''}{R}+ \frac{f'}{f}\frac{R'}{R}+\frac{\phi^{2}r_+^{2}}{z^{4}f^{2}}+\frac{2r_{+}^{2}}{z^{4}f} \right]-\left[ -\frac{X''}{X} + B^{2} x^{2}\right]=0.  
 \end{eqnarray}
 
 The equation for $ X(x)$ could be identified as the Schrodinger equation for a simple harmonic oscillator localized in one dimension with frequency determined by $ B $ \cite{ref90},
 \begin{equation}
 -X^{''}(x) + B^{2}x^{2}X(x) = \lambda_{n} B X(x)
 \end{equation}
 where, $ \lambda_{n}= 2n+1 $ denotes the separation constant. We will be considering the lowest mode ($ n=0 $) solution which is expected to be most stable\cite{ref71,ref90}.  
 
 With this particular choice, the equation of $ R(z) $ turns out to be,
 \begin{equation}
 R''(z)+\frac{f^{'}(z)}{f(z)}R'(z)+ \frac{r_+^{2}\phi^{2}(z)R(z)}{z^{4}f^{2}(z)}+\frac{2r_+^{2}R(z)}{z^{4}f(z)}=\frac{B R(z)}{z^{2}f(z)}\label{neq27}.
 \end{equation}
 
 At the horizon ($ z=1 $), from (\ref{neq27}) one can obtain the following relation,
 \begin{equation}
 R'(1) = \left(\frac{2}{3} - \frac{B}{3r_+^{2}}\right) R(1)\label{neq28}. 
 \end{equation}
 On the other hand, the asymptotic solution ($ z\rightarrow 0 $) for (\ref{neq27}) may be written as,
 \begin{equation}
 R(z) = J_- z + J_+ z^{2} \label{neq29}
 \end{equation}
 where according to our previous choice $ J_- = 0 $.
 
 Near $ z=1 $, we may Taylor expand $ R(z) $ as, 
 \begin{equation}
 R(z) = R (1) - R^{'}(1)(1-z) + \frac{1}{2} R^{''}(1)(1-z)^{2} + .. ..\label{neq30}
 \end{equation}
 
 In order to calculate $ R'' (1) $ we will be using (\ref{neq27}). Considering (\ref{neq27}) close the horizon ($ z=1 $) we find,
 \begin{eqnarray}
 R''(1) &=& - \left[ \frac{f' R''+f'' R'}{f'(z)}\right]_{z=1}-\left[ \frac{r_+^{2}R(z)(-\phi'(1)(1-z)+..)^{2}}{(1-z)^{2}f'^{2}(z)}\right]_{z=1} -\frac{2r_+^{2}R'}{f'(1)} +\frac{B R'}{f'(1)}\nonumber\\
&=& -R''(1) - \frac{f''(1)}{f'(1)}R'(1)-\frac{r_+^{2}R(1)\phi'^{2}(1)}{f'^{2}(1)}-\frac{2r_+^{2}R'(1)}{f'(1)} +\frac{B R'(1)}{f'(1)}. 
 \end{eqnarray}
After some simple algebraic steps and using (\ref{neq28}) we finally obtain,
\begin{equation}
R''(1) = \left[\frac{8}{9}-\frac{5B}{9r_+^{2}}-\frac{\phi'^{2}(1)}{18 r_+^{2}} +\frac{B^{2}}{18r_+^{4}}\right]R(1)\label{neq31}. 
\end{equation} 
 Substituting (\ref{neq31}) into (\ref{neq30}) and using (\ref{neq28}) we find,
 \begin{equation}
 R(z) = \frac{1}{3}R(1) + \frac{2}{3}R(1)z+\frac{B R(1)}{3r_+^{2}}(1-z)+\frac{1}{2}\left[\frac{8}{9}-\frac{5B}{9r_+^{2}}-\frac{\phi'^{2}(1)}{18 r_+^{2}} +\frac{B^{2}}{18r_+^{4}}\right]R(1-z)^{2}\label{neq32}. 
 \end{equation}

Using the previous technique, viz, matching the above two solutions (\ref{neq29}, \ref{neq32}) for some intermediate point $ z=z_m $, we find the following set of equations,
\begin{equation}
J_+z_m^{2} = \frac{1}{3} R(1)+\frac{2}{3} R(1)z_m + \frac{B}{3r_+^{2}} R(1)(1-z_m)+ \frac{1}{2}\left[\frac{8}{9}-\frac{5B}{9r_+^{2}}-\frac{\phi'^{2}(1)}{18 r_+^{2}} +\frac{B^{2}}{18r_+^{4}}\right]R(1-z_m)^{2}\label{neq33}
\end{equation}
and, 
\begin{equation}
2J_+z_m = \frac{2}{3}R(1) - \frac{B}{3r_+^{2}} R(1) - \left[\frac{8}{9}-\frac{5B}{9r_+^{2}}-\frac{\phi'^{2}(1)}{18 r_+^{2}} +\frac{B^{2}}{18r_+^{4}}\right]R(1)(1-z_m)\label{neq34}.
\end{equation}
\begin{figure}[h]
\centering
\includegraphics[angle=0,width=12cm,keepaspectratio]{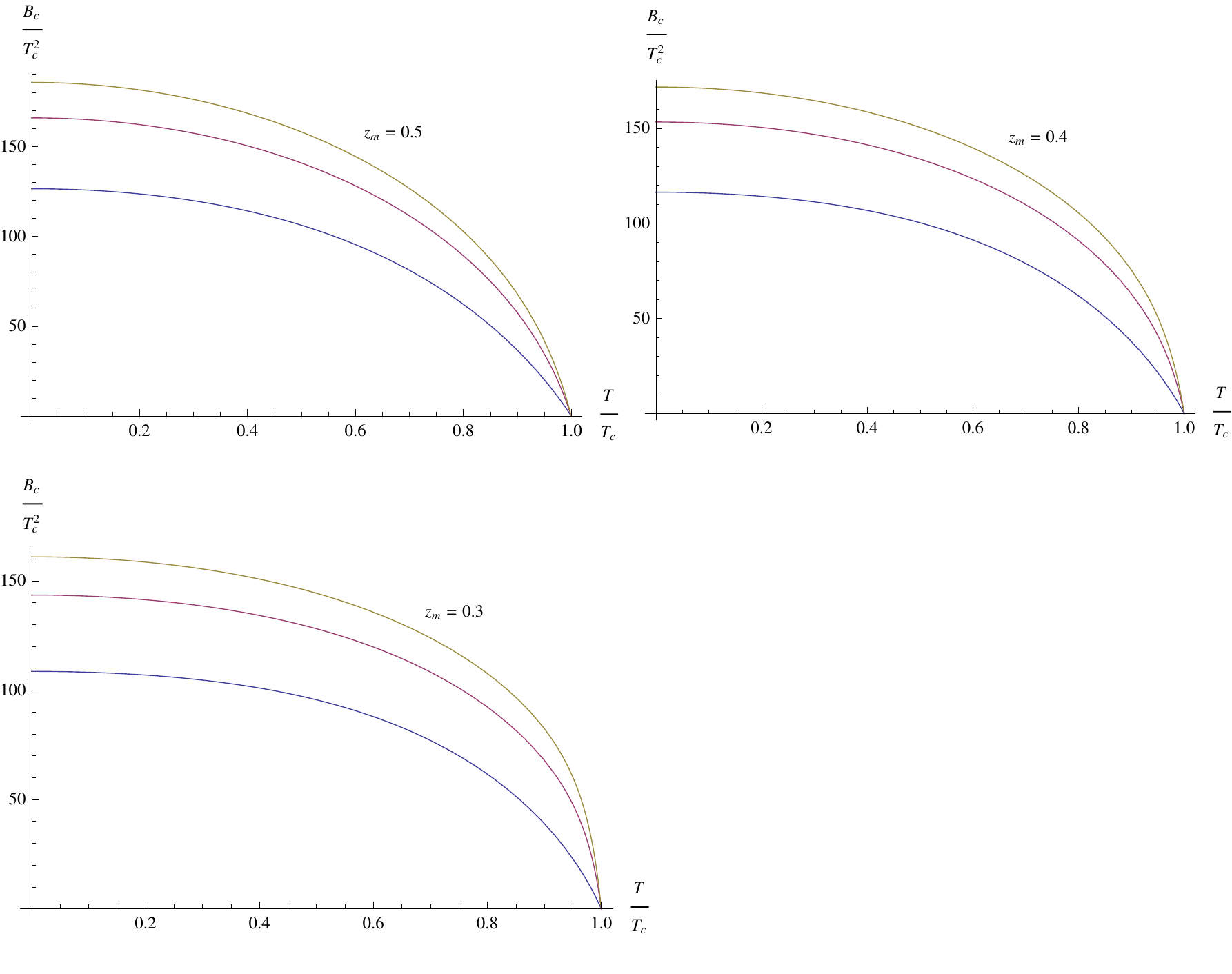}
\caption[]{\it Critical magnetic field ($ B_{c}$) plots for $ s $-wave holographic superconductors for different choice of BI parameters ($ b$), $ b=0 $ (lower curve), $ b=0.006 $ (middle curve) and $ b=0.009 $ (upper curve).}
\label{figure 2a}
\end{figure}
 \begin{figure}[h]
\centering
\includegraphics[angle=0,width=16cm,keepaspectratio]{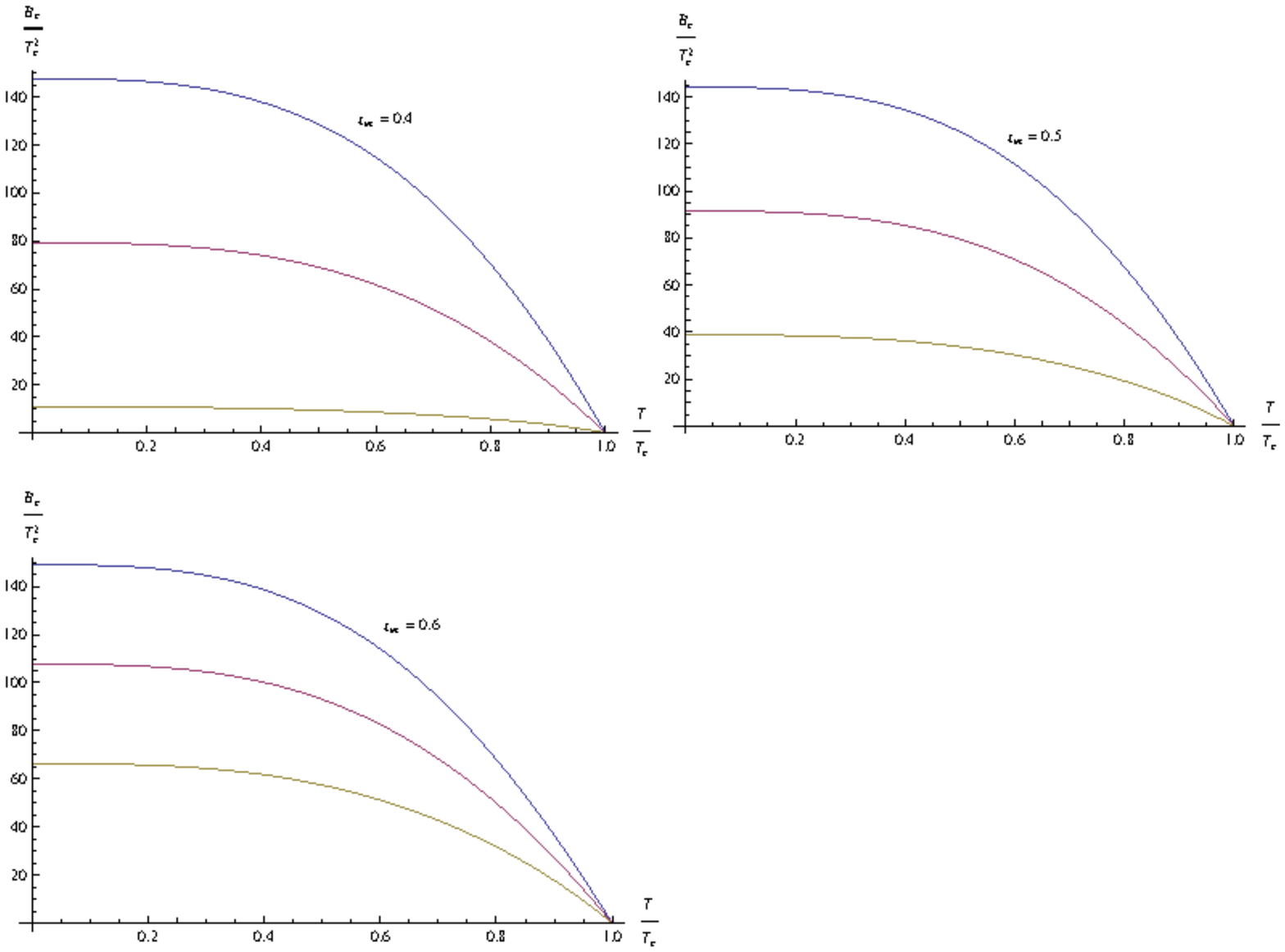}
\caption[]{\it Critical magnetic field strength ($ B_{c}-\frac{T}{T_c} $) plot for $ s $-wave holographic superconductors for different choice of Weyl coupling parameters $ \gamma $. The upper curve corresponds to $ \gamma=-0.006 $, middle curve corresponds to $ \gamma = 0 $ and the lower curve corresponds to $ \gamma=0.006 $.}
\label{figure 2a}
\end{figure}
 From the above set of equations (\ref{neq33}, \ref{neq34}) it is quite trivial to find out the following quadratic equation in $ B $,
 \begin{equation}
 B^{2}+2Br_+^{2}\left( \frac{1+2z_m}{1-z_m}\right) +\frac{4r_+^{4}(7-z_m)}{(1-z_m)}-\phi'^{2}(1)r_+^{2} = 0
 \end{equation}
which has a solution,
\begin{equation}
B= \sqrt{\phi'^{2}(1)r_+^{2}-\frac{9r_+^{4}(3-4z_m)}{(1-z_m)^{2}}}-r_+^{2}\left(\frac{1+2z_m}{1-z_m} \right) \label{nbc}.
\end{equation} 
 
 Now consider the case for where the value of the external magnetic field ($ B $) is very close to the upper critical value i.e, $ B\sim B_{c} $. This implies a vanishingly small condensation and therefore we may ignore all the quadratic terms in $ \psi $. With all these approximations from (\ref{neq5}) we obtain, 
 \begin{eqnarray}
\phi''(z)+\frac{2bz^3}{r_{+}^2 }\phi'^{3}(z) = 0\label{neq35}
\end{eqnarray}
which has a solution,
\begin{equation}
\phi (z) = \frac{\rho}{r_+}(1-z)\left(1-\frac{b\rho^{2}}{10r_+^{4}}\zeta(z) \right)+ O(b^{2})\label{neq36} 
\end{equation} 
with, $ \zeta(z) = 1+z+z^{2}+z^{3}+z^{4} $. 
 
Near the boundary ($ z\rightarrow 0 $) of the AdS space, we can approximately write (\ref{neq36}) as,
\begin{equation}
\phi (z) \simeq \frac{\rho}{r_+}\left(1-\frac{b\rho^{2}}{10r_+^{4}}\right) - \frac{\rho z}{r_+}\label{neq37}.
\end{equation} 
 Comparing (\ref{neq37}) with the first relation in (\ref{neq8}) one finds,
 \begin{equation}
 \mu = \frac{\rho}{r_+}\left(1-\frac{b\rho^{2}}{10r_+^{4}}\right)\label{eq38}.
 \end{equation}
 
 Near the horizon ($ z=1 $), from (\ref{neq35}) we obtain,
 \begin{equation}
 \phi''(1) = - \frac{2b}{r_+^{2}}\phi'^{3}(1)\label{eq39}.
\end{equation}  
 Substituting (\ref{eq39}) into (\ref{neq9}) we obtain,
 \begin{equation}
 \phi (z) = -\phi'(1)(1-z) - \frac{b}{r_+^{2}}\phi'^{3}(1)(1-z)^{2}\label{eq40}.
 \end{equation}
 
 Matching the solutions (\ref{neq8}) and (\ref{eq40}) at $ z=z_m $, we obtain the following set of equations
 \begin{equation}
 \mu - \frac{\rho z_m}{r_+} = -\phi'(1)(1-z_m) - \frac{b\phi'^{3}(1)}{r_+^{2}}(1-z_m)^{2}\label{eq41}
\end{equation}  
 and,
\begin{equation}
-\frac{\rho}{r_+} = \phi'(1)+ \frac{2b\phi'^{3}(1)}{r_+^{2}}(1-z_m)\label{eq42}.
\end{equation} 
 
 Finally, using (\ref{eq38}) the above set of equations (\ref{eq41}, \ref{eq42}) may be written as,
 \begin{equation}
 \frac{\rho}{r_+}(1-z_m) + \phi'(1)(1-z_m) = \frac{b\rho^{3}}{10r_+^{5}}-\frac{b\phi'^{3}(1)}{r_+^{2}}(1-z_m)^{2}\label{eq43}
 \end{equation}
 and,
 \begin{equation}
 -\frac{\rho}{r_+} -\phi'(1) = \frac{2b\phi'^{3}(1)}{r_+^{2}}(1-z_m)\label{eq44}
 \end{equation}
 respectively.
Using the above equations (\ref{eq43}, \ref{eq44}) the following relation could be easily obtained as,
\begin{equation}
\left(\beta^{4}+\beta^{3}\eta \right)(1-z_m)^{2} + \frac{1}{10} (\beta\eta^{3}+\eta^{4}) = 0 \label{eq45}
\end{equation}
where, $ \beta = \phi'(1) $ and $ \eta = \frac{\rho}{r_+} $. This is a quartic equation in $ \beta $.
One of the solutions for (\ref{eq45}) may be written as,
\begin{eqnarray}
\beta = -\eta ~~~~ \Rightarrow \phi'(1) = -\frac{\rho}{r_+}\label{neq46}.
\end{eqnarray}

Substituting (\ref{neq46}) into (\ref{nbc}) and using (\ref{ntemp}, \ref{neq21}), after a few algebraic steps we finally arrive at the following expression for the critical value of the magnetic field strength,

\begin{eqnarray}
B_{c} \simeq \frac{16\pi^{2}T_c^{2}(0)}{9}\left(1 + \frac{2 b\tilde{\beta}^{2}(1-z_m)}{1-\frac{9(3-4z_m)}{4(1-z_m)(7-z_m)}\left(\frac{T}{T_c} \right)^{4} }\right)\times \nonumber\\
\left[\tilde{\beta}\sqrt{1-\frac{9(3-4z_m)}{4(1-z_m)(7-z_m)}\left(\frac{T}{T_c} \right)^{4} } - \left(\frac{1+2z_m}{1-z_m} \right)  \left( \frac{T}{T_c}\right)^{2}\right] + O(b^{2}).   
\end{eqnarray}   
 
In Figure 5.1, the variation of the critical field strength ($ B_c $) has been depicted corresponding to different choices of the matching points ($ z_m=0.5, 0.4, 0.3 $). From these plots it is evident that the qualitative feature of individual plots does not alter due to different choices in the values of $ z_m $. The only quantitative change (for a given value of $ b $) that one can notice is that the value of the critical field strength ($ B_c $) is lowered for the smaller values of $ z_m $. Moreover, considering a particular case i.e; for a given $ z_m $, from these plots (Fig. 5.1) one can note that as the temperature of the $ s $-wave superconductor is lowered through the critical temperature $ T_c(0) $ (which is the critical temperature corresponding to zero magnetic field) the critical field strength gradually increases from zero to it's maximum value. At this stage, one should note that the critical value ($ B_{c} $) of the magnetic field strength is higher in the presence of non linear corrections than in the usual Maxwell case. This upper critical value increase as we increase the value of the BI coupling parameter ($ b $). The increasing field strength will try to reduce the condensate away completely, which implies that in the presence of higher derivative corrections to the usual Maxwell action the condensation gets harder.

 \section{Magnetic field effect in presence of Weyl corrections}
 
 In this section we aim to carry out our analysis in presence of Weyl corrections to the original Maxwell action. In order to do that we begin with the action for a ($ 3+1 $) Weyl corrected holographic superconductor \cite{nref4},
\begin{equation}
S= \int d^{5}x \sqrt{-g}\left[ \frac{1}{16 \pi G_N}\left( R + \frac{12}{L^{2}} \right) +  \mathcal{L}_{m}\right] 
\end{equation}
with,
\begin{equation}
\mathcal{L}_{m} = - \frac{1}{4}\left( F^{\mu\nu}F_{\mu\nu} - 4\gamma C^{\mu\nu\rho\sigma}F_{\mu\nu}F_{\rho\sigma}\right)- \frac{1}{L^{2}} |\nabla_{\mu}\psi -iA_{\mu}\psi |^{2} - \frac{m^{2}}{L^{4}}|\psi |^{2},
\end{equation}
where $ G_N $ is the gravitational Newton's constant, $ \frac{12}{L^{2}} $ corresponds to a negative cosmological constant, $ \gamma $ ($-\frac{L^{2}}{16} < \gamma < \frac{L^{2}}{24} $) is a dimensionful constant and $ C_{\mu\nu\rho\sigma} $ is the Weyl tensor. In the following analysis, without loss of any generality we set $ L = 1 $.

The metric for the planar Schwarzschild AdS black hole in ($ 4 + 1 $) dimensions may be written as,
\begin{equation}
ds_{5}^{2} = r^{2}(-f(r)dt^{2} + dx_i dx^{i}) + \frac{dr^{2}}{r^{2}f(r)}
\end{equation}
with,
\begin{equation}
f(r) = 1 - \frac{r_+^{4}}{r^{4}}
\end{equation}
where $ r_+ $ is the radius of the outer event horizon.

The Hawking temperature of the black hole may be written as,
\begin{equation}
T = \frac{r_+}{\pi}\label{temp}.
\end{equation}

Equations of motion for the Maxwell field ($ A_{\mu} $) and the complex scalar field ($ \psi $) may be written as,
\begin{equation}
\frac{1}{\sqrt{-g}} \partial_{\mu}(\sqrt{-g}F^{\mu\nu}) - \frac{4 \gamma}{\sqrt{-g}} \partial_{\mu} (\sqrt{-g}C^{\mu\nu\rho\sigma} F_{\rho\sigma})-i\left( \psi^{*}\partial^{\nu}\psi - \psi(\partial^{\nu}\psi)^{*}\right)-A^{\nu} |\psi |^{2} = 0 \label{eq1}
\end{equation} 
and,
\begin{equation}
\partial_{\mu}\left(\sqrt{-g}\partial^{\mu}\psi \right) -i\sqrt{-g}A^{\mu}\partial_{\mu}\psi -i\partial_{\mu}\left(\sqrt{-g}A^{\mu}\psi \right)-\sqrt{-g}A^{2}\psi -\sqrt{-g}m^{2}\psi = 0 \label{eq2}.
\end{equation}

Considering the following ansatz,
\begin{eqnarray}
A_{\mu}=(\phi(r),0,0,0,0),\;\;\;\;\psi=\psi(r)
\label{vector}
\end{eqnarray}
the above set of equations (\ref{eq1}, \ref{eq2}) turns out to be,
\begin{equation}
\left(1 - \frac{24 \gamma r_+^{4}}{r^{4}} \right) \phi''(r) + \left( \frac{3}{r} + \frac{24 \gamma r_+^{4}}{r^{5}}\right) \phi'(r) - \frac{2 \phi \psi^{2}}{r^{2}f} = 0 \label{eq3}  
\end{equation}
and,
\begin{equation}
\psi''(r) + \left( \frac{5}{r} + \frac{f^{'}}{f}\right) \psi'(r) + \frac{\phi^{2}\psi}{r^{4}f^{2}} - \frac{m^{2}\psi}{r^{2}f} = 0\label{eq4} 
\end{equation}
respectively.

In order to express the above set of equations in a more suitable way, we set $ m^{2} = -3 $ and choose $ z=\frac{r_+}{r} $. With this choice of variable, (\ref{eq3}) and (\ref{eq4}) turn out to be,
\begin{equation}
\phi''(z) - \frac{1}{z} \left(\frac{1 + 72 \gamma z^{4}}{1 - 24 \gamma z^{4}} \right) \phi'(z) - \frac{2\psi^{2}(z)\phi(z)}{z^{2}(1 - 24 \gamma z^{4})f(z)} = 0 \label{eq5}
\end{equation}
and,
\begin{equation}
\psi''(z) + \left( \frac{f'(z)}{f(z)} - \frac{3}{z}\right)\psi'(z) + \frac{\phi^{2}(z)\psi(z)}{r_+^{2} f^{2}(z)} + \frac{3\psi (z)}{z^{2}f(z)} = 0 \label{eq6}
\end{equation}
respectively.

Before going further, it is customary to mention about the boundary conditions.
\vskip 1mm
\noindent
$\bullet$ Boundary conditions at the horizon ($ z=1 $) may be written as,
\begin{equation}
\phi (1) = 0,~~~ ~~~\psi'(1) = \frac{3}{4} \psi (1)\label{eq7}.
\end{equation}
\vskip 1mm
\noindent
$\bullet$ The asymptotic ($ z\rightarrow 0 $) boundary conditions may be expressed as,
\begin{eqnarray}
\phi (z) = \mu - \frac{\rho z^{2}}{r_+^{2}}\label{eq8}\\
\psi (z) = J_- z +  J_+ z^{3} \label{eq9}
\end{eqnarray}  
where $ \mu $ and $ \rho $ are the chemical potential and the charge density of the boundary field theory. On the other hand $ J_+ $ is related to the vacuum expectation of the condensation operator ($ <\mathcal{O}_+> $) in the dual field theory and $ J_- $ acts as a source. Since we want condensation with out being sourced, therefore we set $ J_- = 0 $.

\subsection{Matching method and condensation values}
 In order to investigate the effect of an external magnetic field on $ s $- wave condensation, as a first step, it is essential to compute the critical temperature ($ T_c(0) $) in the absence of external magnetic field. To do that, we first Taylor expand $ \phi (z) $
and $ \psi (z) $ near the horizon ($ z = 1 $) as, 
\begin{eqnarray}
\phi (z)=\phi (1) - \phi^{'}(1)(1-z) + \frac{1}{2}\phi^{''}(1)(1-z)^{2} + .. ..\label{eq10}
\end{eqnarray}
and,
\begin{eqnarray}
\psi(z)=\psi(1)-\psi^{'}(1)(1-z)+\frac{1}{2} \psi^{''}(1)(1-z)^{2} + ....\label{eq11}
\end{eqnarray}
respectively, where without loss of generality we choose $ \phi'(1)<0 $ and $ \psi(1)>0 $.

From (\ref{eq5}) we note that,
\begin{equation}
\phi''(1) = \frac{\phi'(1)}{1-24 \gamma}\left[1 + 72 \gamma - \frac{\psi^{2}(1)}{2} \right]\label{eq12}. 
\end{equation}
Substituting (\ref{eq12}) into (\ref{eq10}) we finally obtain,
\begin{equation}
\phi (z) = -\phi'(1) (1-z) + \frac{\phi'(1)}{2(1-24 \gamma)}\left[1 + 72 \gamma - \frac{\psi^{2}(1)}{2} \right] (1-z)^{2}\label{eq13}.
\end{equation}

On the other hand, from (\ref{eq6}) and using (\ref{eq7}) we obtain,
\begin{equation}
\psi'' (1) = \frac{9}{32} \psi (1) - \frac{\phi'^{2}(1)\psi (1)}{32 r_+^{2}}\label{eq14}.
\end{equation}
Finally, substituting (\ref{eq14}) into (\ref{eq11}) we find,
\begin{equation}
\psi (z) = \psi (1) - \frac{3}{4} \psi (1) (1-z) + \frac{1}{2} \left(\frac{9}{32} -  \frac{\phi'^{2}(1)}{32 r_+^{2}} \right) \psi (1) (1-z)^{2}\label{eq15}.
\end{equation}

As a next step, following the matching method \cite{ref78}, we match the solutions (\ref{eq8}, \ref{eq13}) and (\ref{eq9}, \ref{eq15}) at some intermediate point $ z = z_m $, which yields the following set of equations,
\begin{eqnarray}
\mu - \frac{\rho z_m^{2}}{ r_+^{2}} &=& \beta (1-z_m) - \frac{\beta}{2(1-24\gamma)}\left[1 + 72\gamma - \frac{\alpha^{2}}{2} \right](1-z_m)^{2}\\
- \frac{2\rho z_m}{r_+^{2}} &=& -\beta + \frac{\beta(1 + 72\gamma)}{(1-24\gamma)}\left[1 - \frac{\alpha^{2}}{2(1 + 72\gamma)} \right](1-z_m)\label{eq16}\\
J_+z_m^{3} &= &\alpha - \frac{3}{4}\alpha (1-z_m) + \frac{\alpha}{2} \left(\frac{9}{32} - \frac{\tilde{\beta}^{2}}{32}\right)(1-z_m)^{2}\label{eq17}\\
J_+z_m^{2} &=& \frac{\alpha}{4} - \frac{\alpha}{3} \left(\frac{9}{32} - \frac{\tilde{\beta}^{2}}{32}\right)(1-z_m)\label{eq18}  
\end{eqnarray}
where, $ \beta = -\phi'(1) $, $ \psi (1) = \alpha $ and $ \tilde{\beta} = \beta/r_+ $.

After some algebraic steps, from (\ref{eq16}) we obtain,
\begin{equation}
\alpha^{2} =  \frac{4 z_m \rho (1 - 24\gamma)}{\pi^{3} T^{3} \tilde{\beta}(1-z_m)} \left(1 - \frac{T^{3}}{T_c^{3}} \right)\label{eq19} 
\end{equation}
with the identification of the critical temperature 
\begin{equation}
T_c = \left[\frac{2 z_m}{\pi^{3} \tilde{\beta}}\left( \frac{1 - 24\gamma}{z_m - 24\gamma(4-3z_m)}\right) \right]^{\frac{1}{3}} \rho^{1/3}\label{eq20}
\end{equation}
where,  $ \gamma < \frac{z_m}{24 (4-3 z_m)} $.

From (\ref{eq17}) and (\ref{eq18}) it is easy to show that,
\begin{equation}
\tilde{\beta} = \frac{\sqrt{9z_m^{2}+60z_m+75}}{\sqrt{(1-z_m)(3-z_m)}}.
\end{equation}

On the other hand, using (\ref{eq17}), (\ref{eq18}), (\ref{eq19}) and (\ref{eq20}) one can easily find,
\begin{equation}
J_+ = \frac{(3z_m + 5)}{4z_m^{2}(3-z_m)} \alpha =\frac{(3z_m + 5)}{4z_m^{2}(3-z_m)}\sqrt{\frac{2}{1-z_m}} \sqrt{z_m- 24\gamma(4-3z_m)} \left( \frac{T_c}{T}\right)^{\frac{3}{2}}\sqrt{1 - \frac{T^{3}}{T_c^{3}}} \label{eq21}. 
\end{equation}   

Finally, using (\ref{temp}) the condensation operator for $ T\sim T_c $ may be found as,
\begin{equation}
<\mathcal{O}> = J_+ r_+^{3} \simeq \frac{\sqrt{3}(3z_m + 5)\pi^{3}T_c^{3}}{4z_m^{2}(3-z_m)}\sqrt{\frac{2}{1-z_m}} \sqrt{z_m- 24\gamma(4-3z_m)} \sqrt{1 - \frac{T}{T_c}}.
\end{equation} 

Therefore from the above analysis it is evident that the critical exponent associated with the $ s $- wave order parameter is $ 1/2 $, which is the universal result of the \textit{mean field} theory.  This further confirms the fact that the phase transition occurring at $ T = T_c $ is indeed a \textit{second order} phase transition.

\subsection{The Meissner effect}

With the above expressions in hand, we are now in a position to study the effect of an external static magnetic field on holographic superconductors in presence Weyl corrections. In order to study the consequences of applying an external magnetic field ($ B $) in the dual field theory we adopt the following ansatz \cite{ref90},
\begin{equation}
A_t = \phi (z),~~~A_y = B x ~~~ and~~~ \psi = \psi (x,z).
\end{equation}

With the above choice, the equation for $ \psi (x,z) $ turns out to be,
\begin{equation}
\psi''(x,z) + \left(\frac{f'}{f} - \frac{3}{z} \right) \psi'(x,z) + \frac{\phi^{2}(z) \psi (x,z)}{r_+^{2} f^{2}(z)} + \frac{3 \psi (x,z)}{z^{2} f(z)} + \frac{1}{r_+^{2}f(z)} (\partial_x^{2} \psi - B^{2}x^{2}\psi) = 0 \label{eq21}. 
\end{equation}
In order to solve the above equation (\ref{eq21}) we take the solution of the following form,
\begin{equation}
\psi (x,z) = X(x) R(z).\label{eq22}
\end{equation}
Substituting (\ref{eq22}) into (\ref{eq21}) we find,
\begin{eqnarray}
r_+^{2}f\left[ \frac{R''(z)}{R(z)} + \left(\frac{f'}{f} - \frac{3}{z} \right) \frac{R'(z)}{R(z)} + \frac{\phi^{2}(z)}{r_+^{2} f^{2}} + \frac{3}{z^{2}f}\right]- \left[-\frac{X''(x)}{X(x)} + B^{2}x^{2}\right]= 0. 
\end{eqnarray}
It is now straightforward to show that the equation corresponding to $ X(x) $ basically represents the motion of a one dimensional harmonic oscillator with frequency determined by $ B $ \cite{ref90}.
\begin{equation}
- X'' (x) + B^{2} x^{2} X(x) = \lambda_n B X(x)
\end{equation} 
 where $ \lambda_n = 2n+1 $ is the separation constant. In the following analysis we shall set $ n =0 $ thereby paying attention to the lowest mode of solutions.
 
 With this particular choice, the equation for $ R(z) $ turns out to be
 \begin{equation}
 R''(z) + \left(\frac{f'}{f} - \frac{3}{z} \right) R'(z) + \frac{\phi^{2} R(z)}{r_+^{2}f^{2}(z)} + \frac{3 R (z)}{z^{2} f(z)} = \frac{B R(z)}{r_+^{2} f(z)}\label{eq23}.
 \end{equation}
 
 Before we proceed further, let us first discuss the boundary conditions for (\ref{eq23}).
\vskip 1mm
\noindent
$\bullet$   
For $ z = 1 $ one can easily obtain,
\begin{equation}
R'(1) = \left( \frac{3}{4} - \frac{B}{4r_+^{2}}\right) R(1)\label{r}.
\end{equation}
\vskip 1mm
\noindent
$\bullet$ The asymptotic ($ z\rightarrow 0 $) behavior of (\ref{eq23}) may be written as
\begin{equation}
R (z) = J_- z + J_+ z^{3}\label{1}
\end{equation} 
with $ J_- = 0 $.

On the other hand, following our previous approach, we Taylor expand $ R(z) $ close to the horizon as,
\begin{equation}
R(z) = R(1) - R'(1) (1-z) + \frac{1}{2} R''(1) (1-z)^{2} + .. \label{eq24}
\end{equation} 
In order to compute $ R''(1) $ let us consider (\ref{eq23}) close to the horizon ($ z = 1 $), which yields,
\begin{eqnarray}
R''(1)&=&-\left[ \frac{f'' R'+f' R''}{f'}\right]_{z=1}+3R'-\left[ \frac{(-\phi'(1)(1-z)+..)^{2} R(z)}{r_+^{2}(-f'(1)(1-z))^{2}}\right]_{z=1}-\frac{3R'}{f'}+\frac{B R'}{r_+^{2}f'}\nonumber\\
&=& \left[9 - \frac{6B}{r_+^{2}} + \frac{B^{2}}{r_+^{4}} - \frac{\phi'^{2}(1)}{r_+^{2}} \right]\frac{R(1)}{32}\label{eq25}. 
\end{eqnarray}
Substituting (\ref{eq25}) into (\ref{eq24}) and using (\ref{r})  we get,
\begin{equation}
R (z) = R(1) - \left[\frac{3}{4} - \frac{B}{4r_+^{2}} \right] R(1) (1-z) + \left[9 - \frac{6B}{r_+^{2}} + \frac{B^{2}}{r_+^{4}} - \frac{\phi'^{2}(1)}{r_+^{2}} \right]\frac{R(1)}{64}(1-z)^{2}.\label{eq26} 
\end{equation}

Following the basic arguments of matching method \cite{ref78}, we match the equations (\ref{1}) and (\ref{eq26}) at some intermediate point $ z = z_m $ which finally yields the following set of equations,  
\begin{eqnarray}
J_+z_m^{3} = R - \left[\frac{3}{4} - \frac{B}{4r_+^{2}} \right] R (1-z_m) + \left[9 - \frac{6B}{r_+^{2}} + \frac{B^{2}}{r_+^{4}} - \frac{\phi'^{2}}{r_+^{2}} \right]\frac{R}{64}(1-z_m)^{2}\label{eq27}
\end{eqnarray}
and,
\begin{equation}
3z_m^{2}J_+ = \left[\frac{3}{4} - \frac{B}{4r_+^{2}} \right] R(1) -\left[9 - \frac{6B}{r_+^{2}} + \frac{B^{2}}{r_+^{4}} - \frac{\phi'^{2}(1)}{r_+^{2}} \right]\frac{R(1)}{32}(1-z_m) \label{eq28}.
\end{equation}

Using (\ref{eq27}) and (\ref{eq28}) it is now straightforward to obtain the following quadratic equation in $ B $, 
\begin{equation}
\zeta_1 B^{2} + \zeta_2 Br_+^{2} - \zeta_3 \phi'^{2}(1)r_+^{2} + \zeta_4 r_+^{4} = 0\label{eq29}
\end{equation}
where,
\begin{eqnarray}
\zeta_1 &=& (1-z_m)(3-z_m)=\zeta_3 \nonumber\\
\zeta_2 &=& 30-8z_m-6z_m^{2}\nonumber\\
\zeta_4 &=& 75+60z_m+9z_m^{2}.
\end{eqnarray}
Solution of (\ref{eq29}) may be written as,
\begin{equation}
B = \frac{1}{2\zeta_1}\left[\sqrt{(\zeta_2^{2}-4\zeta_1 \zeta_4)r_+^{4} + 4\zeta_1\zeta_3\phi'^{2}(1)r_+^{2}} - \zeta_2 r_+^{2} \right]\label{b}. 
\end{equation}

Considering $ B\sim B_{c} $, we may ignore all the quadratic and other higher order terms in $ \psi $ as the condensation is very small near the critical field strength. With this argument (\ref{eq5}) becomes,
\begin{equation}
\phi''(z) - \frac{1}{z} \left(\frac{1 + 72 \gamma z^{4}}{1 - 24 \gamma z^{4}} \right) \phi'(z) = 0\label{eq30}.
\end{equation}
It is now trivial to note that for $ z\rightarrow 0 $ the solution for (\ref{eq30}) may be written as,
\begin{equation}
\phi (z) = \frac{\rho}{r_+^{2}} (1- z^{2})\label{eq31}
\end{equation}
where comparing with (\ref{eq8}) one can identify the chemical potential as $ \mu = \frac{\rho}{r_+^{2}} $.

As a next step, we consider (\ref{eq30}) close to the horizon ($ z = 1 $) which yields,
\begin{equation}
\phi''(1) = \left(\frac{1 + 72 \gamma}{1 - 24 \gamma} \right) \phi'(1)\label{eq32}.
\end{equation}
 Substituting (\ref{eq32}) into (\ref{eq10}) and then matching with (\ref{eq31}) for some intermediate value of $ z=z_m $,  we eventually find the following set of equations,
\begin{equation}
\frac{\rho}{r_+^{2}} (1-z_m^{2}) = -\phi'(1) (1-z_m) + \frac{1}{2} \phi'(1) \left( \frac{1+72\gamma}{1-24\gamma}\right) (1-z)^{2}\label{eq34}
\end{equation}
and,
\begin{equation}
-\frac{2\rho z_m}{r_+^{2}} = \phi'(1) - \phi'(1)\left( \frac{1+72\gamma}{1-24\gamma}\right)(1-z_m)\label{eq35}.
\end{equation}
Finally, using (\ref{eq34}, \ref{eq35}) it is quite trivial to show that,
\begin{equation}
\phi'(1) = -\frac{2\rho}{r_+^{2}}~~~\Rightarrow \phi'^{2}(1)r_+^{2} = \frac{4\rho^{2}}{r_+^{2}}\label{eq36}.
\end{equation}
Substituting (\ref{eq36}) into (\ref{b}) and using (\ref{temp},  \ref{eq20}) we finally obtain 
\begin{eqnarray}
B_{c} \simeq \pi^{2} \tilde{\beta}T_c^{2}(0) \left(1-\frac{96\gamma (1-z_m)}{z_m\left( 1+\frac{(\zeta_2^{2}-4\zeta_1\zeta_4)}{4\tilde{\beta}^{2}\zeta_1\zeta_3}\left(\frac{T}{T_c} \right)^{6}\right) } \right)\nonumber\\
\times\left[ {\sqrt{1+\frac{(\zeta_2^{2}-4\zeta_1\zeta_4)}{4\tilde{\beta}^{2}\zeta_1\zeta_3}\left(\frac{T}{T_c} \right)^{6}}} - \frac{\zeta_2}{2\tilde{\beta}\sqrt{\zeta_1\zeta_3}}\left(\frac{T}{T_c} \right)^{3} \right] + \mathcal{O}(\gamma^{2})\label{bc}.
\end{eqnarray}

From the Figure 5.2 one can note that the critical field strength ($ B_c $) increases as the temperature is lowered through ($ T_c $). This resembles the basic qualitative features of type II superconductors where the upper critical field strength ($ B_{c2} $) shows identical behavior at low temperatures\cite{poole}. Furthermore, from (\ref{bc}) we note that for $ T \sim T_c $
\begin{equation}
B_c \varpropto \left(1 - \frac{T}{T_c} \right) 
\end{equation}
which agrees well with the standard expression for the upper critical field strength as predicted by Ginzburg- Landau theory \cite{ref76},\cite{poole}. Moreover, for a given value of $ z_m $, the value of the critical field strength ($ B_{c} $) has been found to be increasing for $ \gamma<0 $, which indicates the onset of a harder condensation. One may also note that (for a given value of $ \gamma $) the critical field strength is higher for higher values of $ z_m $.

\section{Discussions}
 
 In this chapter, based on the \textit{matching technique} \cite{ref78}, various properties of holographic $ s $- wave superconductors have been investigated in the the probe limit. The present analysis was dedicated towards investigating the behavior of holographic superconductors immersed in an external magnetic field in the presence of various non linear corrections to the usual Maxwell action. These non linear corrections basically correspond to higher derivative corrections of the gauge fields. In the present work two such non linear corrections have been considered, namely (1) Born-Infeld (BI) correction and (2) Weyl correction. In both the cases it has been observed (upto \textit{leading} order in the coupling parameters ) that the properties of holographic superconductors are indeed affected due to the presence of these non linear effects.
The observations may be put as follows:
\vskip 1mm
\noindent
$\bullet$ In the appropriate limits ($ b\rightarrow 0 $, $ \gamma\rightarrow 0 $) our results smoothly goes over to that of the results for the usual Maxwell case.
 \vskip 1mm
\noindent
$\bullet$  In both the examples, it has been observed that the critical temperature ($ T_c $) indeed gets affected due to the presence of these higher derivative corrections. In the first example, $ T_c $ has been found to be decreasing as we increase the strength of the BI coupling parameter ($b$), whereas in the second case it decreases for $ \gamma<0 $. These indicate that the condensation gets harder due to these non linear effects.
\vskip 1mm
\noindent
$\bullet$  In both the cases an upper bound in these coupling parameters have been found above which the analysis can not be carried out.
\vskip 1mm
\noindent
$\bullet$ In both the cases the critical exponent associated with the condensation value near the critical point has been found to be equal to $ 1/2 $ which is in good agreement with the universal \textit{mean field} value and indeed suggests the onset of a \textit{second order} phase transition near the critical temperature.  
\vskip 1mm
\noindent
$\bullet$ Finally, and most importantly it has been observed that the $ s $-wave condensate exists below certain critical value ($ B_c $) of the external magnetic field. In the first example it is observed that the value of the critical field strength ($ B_{c} $) increases as we increase the strength of the BI coupling  parameter ($ b $), whereas in the second example it is found that the critical field strength increases for $ \gamma<0 $. These in fact suggest that as the magnetic field strength increases it will try to reduce the condensate away completely and thereby making it harder to form the scalar hair at low temperatures. Moreover one can identify the critical field strength to that with the upper critical value of the magnetic field in type II superconductors which also satisfies the standard relation predicted by Ginzburg-Landau theory.

\chapter{Summary and Outlook}

This is the concluding chapter of the thesis where we would like to give a summary of the entire analysis presented so far along with mentioning some of its future prospects. To do that, at the very beginning it is customary to mention that the motivation of the present thesis was to analyze the phase transition as well as the critical phenomena occurring in AdS black holes from various perspectives, particularly exploiting the \textit{AdS/CFT} correspondence. Such analyses are important since it could open a gate to a new realm of black hole physics as well as the theory of \textit{Quantum Gravity}. 

We start summarizing our results from Chapter 2. In this chapter, based on the fundamental concepts of thermodynamics we have examined phase transition in black holes defined in Anti-de Sitter (AdS) spaces.
Although there had been a number of investigations \cite{ref27},\cite{ref28,ref29} in order to analyze the phase transition phenomena in higher dimensional (charged) AdS black holes, most of these were confined to the canonical ensemble. A detailed study based on a grand canonical frame work was particularly lacking in the literature. Moreover, explicit expressions for the critical temperature ($ T_0 $) and/or critical mass ($ M_0 $), were in general, unavailable. In our work we have tried to fill up these gaps by performing an explicit computation in the grand canonical framework.

The key idea of our analysis was to consider the AdS black holes as grand-canonical ensembles and study phase transition defined by the discontinuity of \textit{second} order derivatives of Gibbs potential. In order to generalize our calculations we have carried out a detailed analysis of the phase transition phenomena occurring in $ (n+1) $ dimensional Reissner Nordstrom (RN)-AdS black holes using a grand canonical ensemble. We have \textit{analytically} checked that this phase transition between the `smaller' and `larger' mass black holes obey \textit{Ehrenfest} relations defined at the critical point and hence confirm a \textit{second} order phase transition. Similar results were also obtained for the rotating (Kerr AdS) black holes in $ (3 + 1) $ dimensions. Explicit expressions for the critical temperature ($ T_0 $) as well as the critical mass ($ M_0 $) were also derived. At this stage, it is also reassuring to note that, in the appropriate limit our results for the RN AdS case smoothly goes over to that of the results for $ (n+1) $ dimensional Schwarzschild AdS black holes. 

In Chapter 3, we have further extended our analysis by performing a detailed study on the \textit{scaling} behavior of charged Born Infeld (BI) AdS black holes using a canonical framework\footnote{The notion of considering a canonical ensemble in the context of black hole thermodynamics implies that the electric charge ($ Q $) of the black hole has been kept constant \cite{ref28,ref29}}. The critical behavior has been studied near the critical point(s) which were characterized by the discontinuity in the corresponding heat capacity at constant charge ($ C_Q $). We have explicitly calculated the critical exponents of the relevant thermodynamic quantities. Interestingly enough these exponents have been found to satisfy all the \textit{thermodynamic scaling} laws. We have also checked the Generalized Homogeneous Function (GHF) hypothesis (or the \textit{scaling hypothesis}) which has been shown to be compatible with the \textit{thermodynamic scaling} laws. In the appropriate limit our results also provide the corresponding expressions for the RN AdS case.    

Apart from studying the \textit{scaling} behavior of black holes in the usual \textit{Einstein gravity}, we have also explored similar situations going beyond the usual framework of \textit{Einstein} gravity. In this regard, we have made a detailed analysis of the phase transition for  the static, \textit{topological} charged black hole solutions with arbitrary scalar curvature $2k$ in \textit{Ho\v{r}ava-Lifshitz} gravity at the Lifshitz point $z=3$. Such analyses are important in the sense that from these analyses it is easy to make a comparative study between the \textit{scaling} behavior of black holes in various theories of gravity. Surprisingly in our analysis we have noticed that there exists certain class of \textit{topological} black hole solutions ($ k = -1 $) in \textit{Ho\v{r}ava-Lifshitz} gravity which falls under the same \textit{universality class} to that with the black hole solutions for $ k = 1 $ in the usual \textit{Einstein} gravity.

The remainder of the thesis has been devoted towards the study of the phase transition phenomena associated with \textit{hairy} black holes considered in an asymptotically AdS space. The motivation of our analyses was to explore the consequences of such an instability in the dual field theory living on the boundary of the AdS. Using the \textit{AdS/CFT} dictionary it has been found that such an instability essentially leads to a spontaneous breaking of $ U(1) $ symmetry in the dual field theory, thereby triggering a superconductivity in the boundary. These are thus termed as \textit{holographic superconductors}.

The majority of the analyses that have been performed on these holographic description of superconductors are mostly based on numerical techniques, where the computation of the critical temperature, the order parameter etc. have been performed numerically. However, these numerical approaches have their own limitations, for example, it is indeed quite difficult to predict the fate of the condensation at low temperatures based on numerical computations \cite{ref67}. Therefore in order to have a better understanding regarding the low temperature behavior of the holographic superconductors, a search for alternative approaches became inevitable.  

In the present thesis we have tried to find out answers to some of these questions by adopting suitable analytic techniques and considering a \textit{probe} limit. In our analysis we have considered a dual gravitational description for $ s $- wave superconductors coupled with Born Infeld (BI) electrodynamics, which could be viewed as a non linear generalization of the usual Maxwell electrodynamics. The motivation of our analysis was to explore the effect of various higher derivative corrections of the gauge fields on the holographic $ s $- wave condensate based on analytic computations. Furthermore we have also investigated the effect of adding an external magnetic field on the holographic $ s $- wave condensate in presence of various non linear corrections to the usual Maxwell action.
In the following we summarize our findings.
\vskip 1mm
\noindent
$\bullet$ In Chapter 4, based on Sturm-Liouville (SL) eigenvalue problem, we have explicitly computed the critical temperature as well as the order parameter for the $ s $- wave superconductors both in the usual \textit{Einstein} gravity as well as in the Gauss Bonnet (GB) gravity. From our analysis, we have found that the higher derivative corrections of the gauge fields indeed affect the condensation at low temperatures. It has been observed that the condensation gap increases as the BI coupling becomes stronger. This ensures that the higher derivative corrections make the condensation harder. Moreover, from our analysis we have been able to find out the critical exponent associated with the order parameter near the critical point of the phase transition to be 1/2, which is the universal feature of a \textit{mean field} theory. This also confirms the \textit{second} order nature of the phase transition. As a final remark, we would like to mention that, using our analytic scheme we have been able to establish the earlier results that were obtained numerically \cite{binwang}.  
\vskip 1mm
\noindent
$\bullet$ In Chapter 5, we have explored the effect of adding an external magnetic field on holographic $ s $- wave condensate in the presence of higher derivative corrections of the gauge fields. Surprisingly we have noticed a close similarity between our findings and that with the results obtained for the ordinary superconductors. It was observed that the holographic superconductors can sustain a large magnetic field at low temperatures and there exists a critical field strength ($ B_c $), above which the superconductivity gets destroyed. This is reminiscent of the famous \textit{Meissner} effect for the ordinary superconductors, where the magnetic field expels the condensate rather than the condensate expels it. The reason for this is that the scalar fields cannot back react on the back ground magnetic fields \cite{ref90}. Our analysis revealed the fact that the critical field strength ($ B_c $) increases with the increase in the value of the nonlinear couplings, which thereby make the condensation harder.

 Finally, as concluding remarks, we would like to make the following comments.
\vskip 1mm
\noindent
$\bullet$ From our analyses it is perhaps evident that there exists a connection between the laws black hole physics and that of the \textit{mean field} theory to certain extent. In each sector of our analysis from \textit{Ehrenfest's} equations to the critical phenomena, we have noticed a close connection between the laws of black hole physics and that with the ordinary thermodynamic systems which in some sense could be described by sort of \textit{mean field} approximation. The analysis based on \textit{Ehrenfest's} equations were particularly illuminating in the sense that it defined a completely new way to deal with the phase transition in black holes.
\vskip 1mm
\noindent
$\bullet$Another interesting point in this context would be to explore the underlying \textit{renormalization group} scheme to study the critical phenomena in black holes which could explain the \textit{scaling} relations in a better way. We want to put all these issues as a future perspective in order to make a further probe into the subject of critical phenomena in black holes. We believe that our approach could illuminate these and other related issues regarding the underlying microscopic structure of black holes.
\vskip 1mm
\noindent
$\bullet$ Regarding our analyses on holographic superconductors, we would like to stress that the notion of the \textit{Gauge/gravity} duality is no way a coincident, rather it provides a deeper and remarkable connection between the gravitational physics and that with other non gravitational areas of physics. The fact that one should appreciate is that, it is the \textit{AdS/CFT} dictionary that puts some light on the mysterious behavior of high $ T_c $ superconductors, while on the other hand there is no satisfactory explanation for such a phenomenon in the condensed matter physics. Perhaps one day we will be able to find materials whose behavior could be described by some suitable dual gravitational theory.


\newpage
\thispagestyle{empty}

~~~~~~~~~~~~~~~~~ ~~~~~~~~~~~~~~~~~~~~ ~~~~~~~~~~~~~

~~~~~~~~~~~~~~~~~~~~~~~~ ~~~~~~~~~~~~~~~~~~~~~~~~~~~~~~

~~~~~~~~~~~~~~~~~~~~~~~~~~~~~~~ ~~~~~~~~~~~~~~~~~~~~~~~~

~~~~~~~~~~~~~~~~~~~~~~~~ ~~~~~~~~~~~~~~~~~~~~~~~~~~~~~~~~~

~~~~~~~~~~~~~~~~~~~~~~~~~~~~~~~~~~~~~~~~~~~~ ~~~~~~~~~~~~~~~~~~

~~~~~~~~~~~~~~~~~~~~~~~~~~~~~~~~~~~~~~~~~~~~~~~~~~~~~~~~~~~~~~

~~~~~~~~~~~~~~~~~~~~~~~~~~~~~~~~~~~~~~~~~~~ ~~~~~~~~~~~~~~~~~~~

~~~~~~~~~~~~~~~~~~~~~~~~~~~~~~~~~~~~~ ~~~~~~~~~~~~~~~~~~~~~~~~

\begin{center}

 {\bf \uppercase{\huge R\Large EPRINTS
}}
\end{center}


\end{document}